\documentclass{aa}
\usepackage{natbib}
\bibpunct{(}{)}{;}{a}{}{,} 
\usepackage{txfonts}
\usepackage{graphicx}
\usepackage{soul}
\usepackage{amssymb,amsmath}
\usepackage{mathtools}
\usepackage{color}
\usepackage{bm}
\usepackage{url}
\usepackage{hyperref} 
\DeclarePairedDelimiter\bra{\langle}{\rvert}
\DeclarePairedDelimiter\ket{\lvert}{\rangle}
\DeclarePairedDelimiterX\braket[2]{\langle}{\rangle}{#1 \delimsize\vert #2}
\DeclareMathOperator{\Tjs}{3-\!\mathnormal{j}}
\DeclareMathOperator{\Sjs}{6-\!\mathnormal{j}}
\definecolor{darkgreen}{rgb}{0,0.60,0}
\begin{document} 

\title{The transfer of polarized radiation in resonance lines with \\
partial frequency redistribution, $J$-state interference, \\
and arbitrary magnetic fields.}
\subtitle{A radiative transfer code and useful approximations} 
\titlerunning{Transfer of polarized radiation with PRD, $J$-state interference and magnetic fields}

\author{
E. Alsina Ballester\inst{1,2,3}
\and
L. Belluzzi\inst{1,4,5}
\and
J. Trujillo Bueno\inst{2,3,6}
}

\institute{
{IRSOL Istituto Ricerche Solari ``Aldo e Cele Dacc\`o''},  
   Universit\`a della Svizzera italiana (USI), CH-6605 Locarno-Monti, Switzerland
\and
Instituto de Astrof\'{i}sica de Canarias (IAC), 
E-38205 La Laguna, Tenerife, Spain
\and
Departamento de Astrof\'{i}sica, Universidad de La Laguna {(ULL)}, E-38206 La Laguna, Tenerife, Spain
\and
Leibniz-Institut f\"{u}r Sonnenphysik (KIS), D-79104 Freiburg, Germany
\and
Euler Institute, Universit\`a della Svizzera italiana (USI), 
CH-6900 Lugano, Switzerland
\and
Consejo Superior de Investigaciones Cient\'{i}ficas (CSIC), 
Spain\\
\email{ernest.alsina@iac.es}
}

\abstract
{}
%
{We present the theoretical framework and numerical methods we have implemented to solve the problem of the generation and transfer of polarized radiation in spectral lines without assuming local thermodynamical equilibrium, while accounting for scattering polarization, partial frequency redistribution (due to both the Doppler effect and elastic collisions), $J$-state interference, and {hyperfine} structure. 
The resulting radiative transfer code allows one to model the impact of magnetic fields of an arbitrary strength and orientation through the Hanle, incomplete Paschen-Back, and magneto-optical effects. 
We also evaluate the suitability of a series of approximations for modeling the scattering polarization in the wings of strong resonance lines at a much lower computational cost, which is particularly valuable for the numerically intensive case of three-dimensional radiative transfer.}
{We examine the suitability of the considered approximations by using our radiative transfer code to model the Stokes profiles of the Mg~{\sc ii} $h$ \& $k$ lines and of the H~{\sc i} Lyman-$\alpha$ line in magnetized one-dimensional models of the solar atmosphere. }
{Neglecting Doppler redistribution in the scattering processes that are unperturbed by elastic collisions (i.e., treating them as coherent in the observer's frame) produces a negligible error in the scattering polarization wings of the Mg~{\sc{ii}} resonance lines and a minor one in the Lyman-$\alpha$ wings, although it is unsuitable to model the cores of these lines. 
For both lines, the scattering processes that are perturbed by elastic collisions only give a significant contribution to the intensity component of the emissivity. 
Neglecting collisional as well as Doppler redistribution (so that all scattering processes are coherent) represents a rough but suitable approximation for the wings of the Mg~{\sc{ii}} resonance lines, but a very poor one for the Lyman-$\alpha$ wings.  
The magnetic sensitivity in the scattering polarization wings of the considered lines can be modeled by accounting for the magnetic field in only the $\eta_I$ and $\rho_V$ coefficients of the Stokes-vector transfer equation (i.e., using the zero-field expression for the emissivity).} 
{}

 \date{}
 \keywords{Radiative transfer --
                Scattering -- 
                Polarization --
                Line: profiles --
                Methods: numerical --
                Sun: chromosphere}
 \maketitle     
 \section[Sect1]{Introduction}
 \label{sec::introduction}
The intensity and polarization profiles of strong resonance lines are valuable observables for exploring the outer layers of the solar atmosphere. In particular, strong ultraviolet lines, such as 
the Lyman-$\alpha$ lines of H {\sc i} and He {\sc ii} and the Mg~{\sc{ii}} $h$ \& $k$ doublet, were predicted to show significant linear polarization signals that arise from the scattering of anisotropic radiation, especially when observed close to the solar limb \citep{TrujilloBueno+11,TrujilloBueno+12,BelluzziTrujilloBueno12,Belluzzi+12}. 
These scattering polarization signals encode information on the thermodynamic structure of the solar regions from which the radiation emerges. In the line core, they also carry the fingerprints of the strength and orientation of the magnetic fields due to the action of the Hanle effect \citep[{e.g.,}][hereafter LL04]{BStenflo94,TrujilloBueno01,BLandiLandolfi04}. The scattering polarization profiles of strong resonance lines often show broad lobes that extend far beyond the Doppler core, thus allowing 
the conditions in deeper layers of the solar atmosphere {to be probed}. Interestingly, via the so-called magneto-optical (MO) effects, these wing signals are sensitive to the presence of longitudinal magnetic fields with strengths comparable to those that characterize the onset of the Hanle effect \citep{AlsinaBallester+16, AlsinaBallester+19,delPinoAleman+16,delPinoAleman+20}. 
 
Modeling the intensity and polarization patterns observed in such lines through radiative transfer (RT) calculations is crucial to understanding the physical processes that shape them and to developing reliable techniques to extract information on the solar atmosphere. Strong resonance lines, and especially their scattering polarization signals, must be modeled without assuming local thermodynamic equilibrium 
{(LTE).} 
In order to suitably model both the intensity and scattering polarization in the line wings, one must also account for the frequency correlations between incoming and outgoing photons in scattering processes, or partial frequency redistribution (PRD) phenomena. 
In the case of multiplets, their wing scattering polarization signals are often also strongly impacted by the quantum interference between different fine structure (FS) levels (i.e., $J$-state interference), provided that the energy separation between them is not too large \citep{BelluzziTrujilloBueno11}. In order to account for such interference, we need to consider a two-term or multi-term atomic model (e.g., Sects.~7.5 and 7.6 of LL04). 

The H~{\sc{i}} Lyman-$\alpha$ line and the Mg~{\sc{ii}} $h$ \& $k$ doublet in the ultraviolet range of the solar spectrum are strong resonance lines of particular scientific interest \citep[e.g., the review by][]{TrujilloBueno+17}. 
They present magnetically sensitive scattering polarization signals that must be modeled considering the abovementioned physical mechanisms. Following the theoretical predictions summarized in the abovementioned review paper, the intensity and polarization profiles of the H~{\sc{i}} Lyman-$\alpha$ and Mg~{\sc{ii}} $h$ \& $k$ lines were observed by the Chromospheric Lyman Alpha SpectroPolarimater (CLASP) and the Chromospheric Layer SpectroPolarimeter (CLASP2) sounding rocket experiments, respectively. 
Among a number of valuable findings \citep[see][]{Kano+17,TrujilloBueno+18,Ishikawa+21}, they provided observational confirmation that both the H~{\sc{i}} Lyman-$\alpha$ and the Mg~{\sc{ii}} $k$ lines present broad, {large-amplitude,} scattering polarization {signals in the wings}, 
which is in agreement with theoretical predictions \citep[see][]{BelluzziTrujilloBueno12,Belluzzi+12,AlsinaBallester+16,delPinoAleman+16,delPinoAleman+20}. 

The modeling of the intensity of these spectral lines, accounting for PRD effects, has been a subject of theoretical research for many decades; for an in-depth discussion, with a particular focus on the hydrogen lines including Lyman-$\alpha$, the reader is referred to Sect.~15.6 of \cite{BHubenyMihalas15}, and the references therein \citep[see][]{Vernazza+73,MilkeyMihalas73b, MilkeyMihalas73,Cooper+89,HubenyLites95}. 
Because a suitable modeling of the polarization of these lines requires accounting for 
{all the} 
phenomena mentioned above, it was approached in steps of increasing complexity.  First, the limit of complete frequency redistribution (CRD; see LL04) was considered \citep{TrujilloBueno+11,TrujilloBueno+12}. This limit is suitable for modeling the scattering polarization signals observed in the core of these lines, where the Hanle effect operates, but not for their large and broad wing signals. 
Subsequently, making use of a theoretical framework based on the concept of metalevels \citep[see][]{LandiDeglInnocenti+97}, \cite{BelluzziTrujilloBueno12} and \cite{Belluzzi+12} modeled the aforementioned lines taking PRD effects and $J$-state interference into account. Within this framework, the impact of magnetic fields can also be taken into account, but the redistribution effects of collisions cannot be included in a self-consistent manner. 
The approach based on the Kramers-Heisenberg formula for scattering amplitudes \citep[see][]{BStenflo94} allows one to account for the same physics \citep{Sowmya+14} and is subject to the same limitation regarding collisional redistribution. 
The same lines were later modeled by \cite{AlsinaBallester+16,AlsinaBallester+19}, making use of the theoretical framework presented by \cite{Bommier97a,Bommier97b}. This theory, based on a perturbative expansion of atom-photon interaction{s}, allows one to account for PRD effects (including collisional as well as Doppler redistribution) and the impact of magnetic fields. 
However, it is only suitable for a two-level atomic system with an unpolarized\footnote{An atomic level or term is said to be polarized when the states composing it present population imbalances or quantum coherence.} and infinitely sharp lower level, and thus $J$-state interference cannot be taken into account. 
A recent theoretical framework developed by \cite{Casini+14}, based on a diagrammatic treatment of atom-photon interaction{s}, allowed PRD effects, magnetic fields, and $J$-state interference to be accounted for in the collisionless regime. Shortly thereafter, \citet{Casini+17a,Casini+17b} extended it to account for the redistribution effects of collisions, and \cite{delPinoAleman+16} applied it to model the Mg~{\sc{ii}} $h$ \& $k$ lines. 
Finally, \cite{Bommier17,Bommier18} presented a generalization of her 1997 theoretical framework, which is suitable for a two-term atomic system with an infinitely sharp and unpolarized lower term\footnote{Throughout this work, we refer to a term as infinitely sharp when all the $J$ levels pertaining to it are infinitely sharp.} and, furthermore, allows for the inclusion of  {hyperfine} structure (HFS). 
In this paper, we present the numerical scheme we have used to develop a novel 
{non-}LTE 
RT code, making use of the redistribution matrices presented in \cite{Bommier17,Bommier18}. It represents an extension to the one presented in \cite{AlsinaBallester+17}, which was valid for the simpler case of a two-level atom. We have already applied this novel code to carry out research in other spectral lines, most notably to model the scattering polarization signals of the Na~{\sc{i}} D lines \citep[see][]{AlsinaBallester+21}. 

When considering RT problems with PRD effects, the line scattering emissivity must generally be computed accounting for its detailed dependency on the frequency and direction of propagation of the incident radiation. Even when introducing simplifying approximations, such as removing the angle-frequency coupling induced by the Doppler effect in the reference frame of the observer (i.e., the observer's frame) through the so-called angle-averaged approximation \citep[e.g.,][]{ReesSaliba82}, 
{this generally remains the most computationally demanding step} in the numerical solution of {non-}LTE RT problems, especially when considering complex atomic models or accounting for the energy shifts induced by the presence of an external magnetic field \citep[e.g.,][]{Paganini+21}. 
For PRD calculations in realistic three-dimensional (3D) models of the solar atmosphere, which may contain more than $500^3$ spatial points, the numerical requirements become harsh enough to render them unfeasible, even with current high-performance computational techniques. It is thus of interest to devise and test approximations that allow one to quickly and reliably compute the scattering emissivity, even if their suitability is restricted to specific spectral ranges.

In this article we introduce the {non-}LTE RT code that we have developed and we then apply it to numerically evaluate a series of approximations for modeling the wings of the intensity and scattering polarization profiles of strong resonance lines. The investigation shown in the main text is focused on the Mg~{\sc{ii}} doublet near $2800$~\AA , whereas the analogous study for the H~{\sc{i}} Lyman-$\alpha$ line is found in the Appendix. 
We first investigate the approximation of neglecting Doppler redistribution for the fraction of scattering processes that are coherent in the atomic frame (i.e., for these processes, we assume that coherence occurs in the observer's frame). This approximation was already investigated by \cite{BelluzziTrujilloBueno12} for the magnesium doublet, but in the absence of magnetic fields and neglecting collisional redistribution. 
We then analyze the impact of additionally neglecting the contribution to scattering polarization from the scattering processes that are rendered noncoherent by collisions and, subsequently, of neglecting both Doppler and collisional redistribution (i.e., we assume that all scattering processes are coherent in the observer's frame). Finally, we study the suitability of neglecting the impact of the magnetic field in the emissivity, taking it into account in all or only some of the coefficients of the propagation matrix. 

This article is structured as follows. In Sect.~\ref{sec::formulation}, we show the main expressions involved in the considered RT problem, {which are} suitable for the case without HFS, as well as the detailed numerical scheme used to develop the transfer code. 
In Sect.~\ref{sec::Illustrative}, the abovementioned approximations are explained in detail and their suitability is evaluated through numerical calculations. The main conclusions are presented in Sect.~\ref{sec::Conclusions}. 
The atomic model parameters used in the calculations for the Mg~{\sc{ii}} $h$ \& $k$ lines, and the details of this modeling, are presented in Appendix~\ref{sec::AppAtomMod}. 
Appendix~\ref{sec::AppLymanAlpha} presents an analogous analysis to that of the main text but considering the H~{\sc{i}} Lyman-$\alpha$ line, in which the differences and similarities with the magnesium doublet are highlighted. 
Appendix~\ref{sec::AppendixExpression} contains the relevant expressions for the numerical scheme, including those suitable for the general case in which HFS is included and those obtained under the approximations investigated in this paper.  

\section[Sect2]{Formulation of the problem}
\label{sec::formulation} 
In this section, we present the relevant equations entering the considered non-LTE RT problem, the methods applied to solve it, and the details of its numerical implementation. The ensuing code is suitable for calculations considering one-dimensional (1D) models of the solar atmosphere. We consider a two-term atomic model whose lower term is infinitely sharp and unpolarized; this assumption is generally justified for strong resonance lines. 
We point out that suitably modeling the polarization signals of certain spectral lines of interest, such as the sodium doublet, requires considering a two-term atom with HFS \citep[see][]{AlsinaBallester+21}. The generalization of the present approach to include HFS is found in Appendix~\ref{sec::AppendixExpression}. 

\subsection[Sect2p1]{The radiative transfer equation}
\label{sec::sRTE} 
The intensity and polarization properties of a given radiation beam are characterized by its Stokes vector $\boldsymbol{I}$, whose components $I_i$ ($i=\{0,1,2,3\}$) correspond to the Stokes parameters $I$, $Q$, $U$, and $V$, respectively. 
The variation of the Stokes parameters of a radiation beam as it propagates through a medium is described by the radiative transfer equation (RTE), a first-order inhomogeneous ordinary differential equation which can be written in vectorial form as 
\begin{equation}
 \frac{\mathrm{d}}{\mathrm{d}s} \boldsymbol{I} = \boldsymbol{\varepsilon} - \mathbf{K} \boldsymbol{I} \, ,
 \label{eq::RTEq}
\end{equation} 
where $s$ is the coordinate along the considered ray path, $\mathbf{K}$ is the propagation matrix, characterizing the attenuation of the Stokes parameters and the couplings between them, and $\boldsymbol{\varepsilon}$ is the emission vector, whose four components 
quantify the emission in the four Stokes parameters. 
All quantities appearing in the RTE depend in general on the spatial point $\mathbf{r}$ and on the frequency $\nu$ and propagation direction $\mathbf{\Omega}$ of the considered radiation beam. These dependencies have been omitted in the expressions shown in this subsection for notational simplicity. 
From the numerical point of view, it is often convenient to solve the RTE on optical depth scale rather than geometrical depth \citep[e.g.,][]{Janett+17}. Indicating with $\tau$ the optical depth coordinate, the relation between the two scales is given by 
$\mathrm{d}\tau = - \eta_I \, \mathrm{d}s$, 
where $\eta_I$ is the absorption coefficient, which quantifies the attenuation of the four Stokes parameters as the radiation propagates through the medium (i.e., without modifying its degree of polarization). 
The propagation matrix has the form 
\begin{equation}
 \mathbf{K} = \left(\begin{array}{c c c c}
                     \eta_I & \eta_Q & \eta_U & \eta_V \\
                     \eta_Q & \eta_I & \rho_V & -\rho_U \\
                     \eta_U & -\rho_V & \eta_I & \rho_Q \\
                     \eta_V & \rho_U & -\rho_Q & \eta_I
                    \end{array} \right) \, .
 \label{eq::Kmat}
\end{equation} 
The coefficients $\eta_Q$, $\eta_U$, and $\eta_V$ quantify the differential attenuation of Stokes parameters $Q$, $U$, and $V$, respectively (i.e., dichroism). The anomalous dispersion coefficients $\rho_Q$, $\rho_U$, and $\rho_V$ describe the couplings between Stokes parameters other than $I$. 
The $\rho_V$ coefficient, in particular, quantifies the coupling between Stokes $Q$ and $U$, resulting in a rotation of the plane of linear polarization. Under the assumption of an unpolarized lower term, this coupling is exclusively due to the presence of a magnetic field, and we thus refer to it as Faraday rotation. In the line wings, $\rho_V$ reaches values comparable to those of $\eta_I$ in the presence of longitudinal magnetic fields with strengths similar to those that characterize the onset of the Hanle effect \citep[see][]{AlsinaBallester+16,AlsinaBallester+19}. Thus, as noted in the introduction, these MO effects give rise to an appreciable magnetic sensitivity in the scattering polarization wings of strong resonance lines, even in the presence of relatively weak magnetic fields.  

All the elements of the propagation matrix and components of the emission vector can in general be given as a sum of two contributions, one due to line processes and one due to continuum processes. These contributions will hereafter be labeled with $\ell$ and $c$, respectively. 
The analytical expressions for the line contributions to the elements of the propagation matrix and to the emission vector, for the atomic system considered in this work, can be found in Sects.~\ref{sec::sPropMatExp} and \ref{sec::sEmis}, respectively. The expressions for the continuum contribution to all coefficients of the RTE are discussed in Sect.~\ref{sec::sContinuum}. 

\subsection[Sect2p2]{The two-term atomic model}
\label{sec::sTwoterm}  
We consider a two-term atomic system in which each term is specified by $(\beta L S\!)$, 
where $\beta$ is a set of inner quantum numbers, $L$ is the quantum number for the orbital angular momentum operator $\vec{L}$, and $S$ is the quantum number for the electronic spin operator $\vec{S}$. 
In the absence of magnetic fields, the energy eigenstates of this atomic system coincide with those of the total angular momentum operator $\vec{J}$ and are given by $\ket{\beta L S J M}$, where $J$ is the total angular momentum quantum number and is $M$ the magnetic quantum number. When an external magnetic field is present and the induced splitting between the magnetic sublevels is comparable to the separation between the different $J$-levels of the considered term (i.e., the Paschen-Back effect regime), $J$ ceases to be a good quantum number for the atomic Hamiltonian. Its eigenstates can thus be given as linear combinations of the eigenstates of $\vec{J}$, 
\begin{align}
 \ket{\beta L S j M} = \sum_J {C}^{j}_J(M) \ket{\beta L S J M} \, ,  
 \label{eq::EigenCouplingNoHFS}
\end{align}
where the label $j$ is introduced to distinguish between different states with the same quantum numbers $\beta, L, S$, and $M$. The sum over $J$ ranges from $|L-S|$ to $L + S$, but for simplicity of notation these limits are not given explicitly. The coupling coefficients ${C}^{j}_J(M)$ can be obtained by diagonalizing the atomic Hamiltonian (see Appendix~\ref{sec::AppendixExpression}). Of course, in the absence of a magnetic field, $C^{j}_J(M) = \delta_{J,j}$. 

We consider the spectral lines arising from radiative transitions (under the electric dipole approximation) between states of the upper and lower term, for which we hereafter use the shorthand notation $\ket{j_u M_u}$ and $\ket{j_\ell M_\ell}$, respectively.  
We introduce the normalized strength (e.g., Sect.~3.4 of LL04) for the transition between states $\ket{j_\ell M_\ell}$ and $\ket{j_u M_u}$   
\begin{align}
 S^{j_\ell M_\ell, j_u M_u}_q & = \frac{3}{2 S + 1} \sum_{J_u J^\prime_u} C^{j_u}_{J_u}(M_u) \, C^{j_u}_{J^\prime_u}(M_u) \notag \\
 & \times \sum_{J_\ell J^\prime_\ell} C^{j_\ell}_{J_\ell}(M_\ell) \, C^{j_\ell}_{J^\prime_\ell}(M_\ell) \notag \\
 & \times \sqrt{(2 J_u + 1) (2 J^\prime_u + 1) (2 J_\ell + 1) (2 J^\prime_\ell + 1)} \notag \\
 & \times 
 \left\{\begin{array}{c c c}
         J_u  & J_\ell & 1 \\
         L_\ell & L_u  & S 
        \end{array}
  \right\}
 \left\{
 \begin{array}{c c c}
          J^\prime_u  & J^\prime_\ell & 1 \\
         L_\ell & L_u  & S 
 \end{array}
 \right\} \notag \\
 & \times 
 \left(\begin{array}{c c c}
 J_u  & J_\ell  & 1 \\
 - M_u  & M_\ell & -q
 \end{array}\right)
 \left(\begin{array}{c c c}
 J^\prime_u  & J^\prime_\ell  & 1 \\
 - M_u  & M_\ell & -q
 \end{array} \right) \, ,
 \label{eq::TransitionnHFS}
 \end{align}
 where $q$ is an integer which can take values $-1$, $0$, and $1$. The quantities given in the brackets and curly brackets are the so-called $\Tjs$ and $\Sjs$ symbols, respectively, quantifying the couplings between various angular momenta (see Sects.~2.2 and 2.3 of LL04). 
 The transition frequency is 
 \begin{equation}
   \nu_{j_u M_u, j_\ell M_\ell} = \bigl(E_{j_u M_u} - E_{j_\ell M_\ell} \bigr)/h \, , 
 \end{equation} 
where $h$ is the Planck constant and $E_{j_u M_u}$ and $E_{j_\ell M_\ell}$ are the energies of the upper and lower states, respectively, which correspond to eigenvalues of the atomic Hamiltonian (see Appendix~\ref{sec::AppendixHamilton}). 
  
 \subsection[Sect2p3]{Propagation matrix elements in a two-term atom}
 \label{sec::sPropMatExp} 
The line contribution to the elements of the propagation matrix can be derived as detailed in Sect.~7.6 of LL04. In the reference frame of the observer, and under the assumption that the lower term is unpolarized, this contribution can be written in terms of the transition strengths introduced above \footnote{These expressions were already introduced in Eqs.~(9-12) of \cite{AlsinaBallester+19}, where the order of the $j_u M_u$ and $j_\ell M_\ell$ was unintentionally inverted in the definition of the transition strength. Despite this change in definition, all the expressions therein remain correct.} 
\begin{subequations}
  \begin{align}
  \eta^\ell_i(\nu,\boldsymbol{\Omega}) 
    = \frac{k_M}{\sqrt{\pi}\Delta\nu_D} & \sum_{K = 0}^2 \sum_{Q=-K}^K \sqrt{\frac{2 K + 1}{3}}
    \mathcal{D}^K_{0 Q}(R_B)^\ast \, 
    \mathcal{T}^K_Q(i,\boldsymbol{\Omega}) \notag \\
    & \times \sum_{q=-1}^1 (-1)^{1+q} 
      \left(\begin{array}{c c c}
            1 & 1 & K\\
            q & -q & 0
         \end{array} \right) 
    \notag \\
   & \times \!\!\!\!\! \sum_{j_u M_u j_\ell M_\ell} 
     S^{j_\ell M_\ell, j_u M_u}_q \, %
    H\bigl(a,u + u_{j_u M_u, j_\ell M_\ell} \bigr) \, , \label{eq::EtaNoHFS} \\
   \rho^\ell_i(\nu,\boldsymbol{\Omega}) = \frac{k_M}{\sqrt{\pi} \Delta\nu_D} & \sum_{K = 0}^2 \sum_{Q=-K}^K \sqrt{\frac{2 K + 1}{3}} \mathcal{D}^K_{0 Q}(R_B)^\ast \, 
   \mathcal{T}^K_Q(i,\boldsymbol{\Omega}) \notag \\
   & \times \sum_{q=-1}^1 (-1)^{1+q} 
      \left(\begin{array}{c c c}
            1 & 1 & K\\
            q & -q & 0
         \end{array} \right) 
    \notag \\
   & \times \!\!\!\!\! \sum_{j_u M_u j_\ell M_\ell} 
    S^{j_\ell M_\ell, j_u M_u}_q \, %
    L\bigl(a, u + u_{j_u M_u, j_\ell M_\ell} \bigr) \, . \label{eq::RhoNoHFS}  
 \end{align}
  \label{eq::PropLineTransformed}
\end{subequations}
\noindent Throughout this work, the maximum and minimum values of $M$ in the sums depend on the quantum numbers $S$ and $L$ of the considered term, but for simplicity of notation the summation limits are not given explicitly. For the same reason we do not show the limits for the $j$ labels, which are determined by the numbers $S\!$, $L$, and $M$ (see Sect.~3.4 of LL04). 

The quantities $D^K_{Q Q^\prime}(R_B)$ are the so-called rotation matrices (e.g., Sect.~2.4 of LL04). They quantify the rotation from a reference system in which the $Z$-axis, which corresponds to the quantization axis for total angular momentum, is parallel to the magnetic field vector (i.e., the magnetic reference system) into the system in which it is parallel to the local vertical (i.e., the reference system of the problem). 
 \begin{figure}[]
  \centering
  \includegraphics[width=0.34\textwidth]{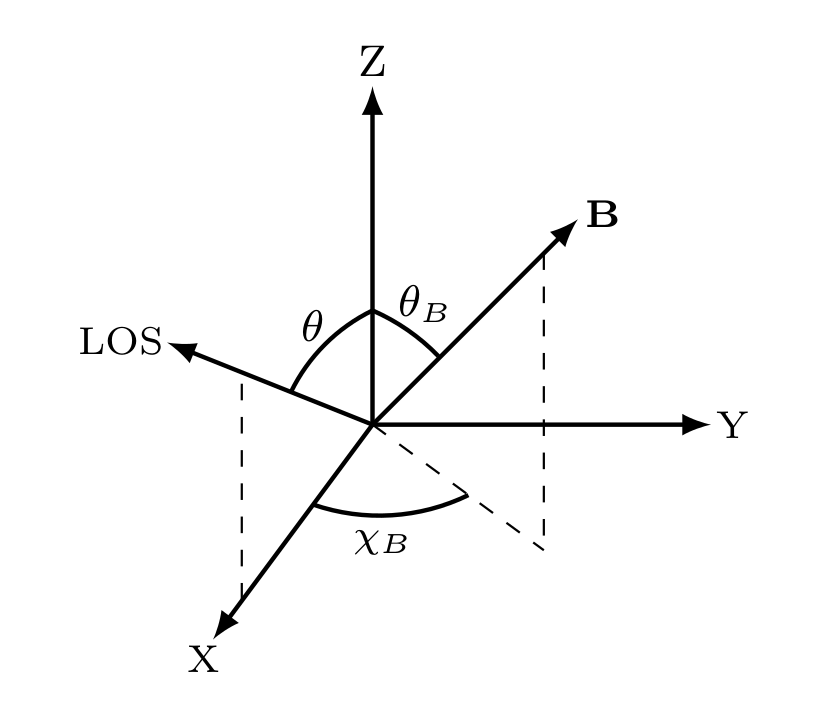}
        \caption{{Right-handed Cartesian coordinate system considered in the problem. The Z-axis is taken parallel to the local vertical.}  
 The direction toward the observer, or line of sight (LOS), is contained in the $X$--$Z$ plane and is therefore determined solely by its inclination with the local vertical $\theta$. The direction of the magnetic field is specified by its inclination $\theta_B$ and azimuth $\chi_B$.}  
        \label{Fig::geometry} 
 \end{figure} 
This rotation is given by $R_B = (0,-\theta_B, -\chi_B)$, where $\theta_B$ and $\chi_B$ are the Euler angles, illustrated in Fig.~\ref{Fig::geometry}. The expressions for the geometrical tensors $\mathcal{T}^K_Q(i,\mathbf{\Omega})$, which in Eqs.~\eqref{eq::EtaNoHFS} and \eqref{eq::RhoNoHFS} are given in the reference system of the problem, can be found in Sect.~5.11 of LL04. 

The functions $H$ and $L$ are the Voigt profile and the associated dispersion profile, respectively. We have also introduced the reduced frequency $u$ and the reduced frequency shift $u_{j_u M_u, j_\ell M_\ell}$, for the transition between the upper state $\ket{j_u M_u}$ and the lower state $\ket{j_\ell M_\ell}$, as
\begin{equation*}
 u = \frac{\nu_0 - \nu}{\Delta\nu_D} \, , \quad \quad %
u_{j_u M_u,j_\ell M_\ell} = \frac{\nu_{j_u M_u, j_\ell M_\ell} - \nu_0}{\Delta\nu_D} \, , 
\end{equation*}
where $\nu_0$ is a reference frequency taken equal to the baricenter of all transitions between the upper and lower states. 
The Doppler widths in frequency and wavelength units are given as $\Delta\nu_D$ and $\Delta\lambda_D$, respectively, and their expressions can be found in Appendix~\ref{sec::AppBroad}. The damping constant $a$ introduced in Eqs.~\eqref{eq::EtaNoHFS} and~\eqref{eq::RhoNoHFS} depends on the Doppler width and its expression can be found in the same Appendix. The frequency-integrated absorption coefficient $k_M$ entering the expressions for $\eta^\ell_i$ and $\rho^\ell_i$ above is given by 
\begin{equation}
 k_M = \frac{c^2}{8 \pi \nu_0^2} \frac{2 L_u + 1}{2 L_\ell + 1} {N}_\ell A_{u \ell} \, ,
 \label{eq::FreqInt}
\end{equation}  
where $c$ has the usual meaning of speed of light, ${N}_\ell$ is the population of the lower term, and $A_{u \ell} = A(\beta_u L_u S \rightarrow \beta_\ell L_\ell S)$ is the Einstein coefficient for spontaneous emission from the upper to the lower term. 

\subsection[Sect2p4]{Emission coefficients for a two-term atom}
\label{sec::sEmis}
The line emission coefficients -- also referred to as emissivities or the components of the emission vector -- directly depend on the populations and quantum coherence between the various states of the upper term. 
These quantities, as well as $N_\ell$ (see Sect.~\ref{sec::sPropMatExp}), can be determined through the statistical equilibrium equations, which usually require a numerical solution. However, in certain cases there exists a closed analytical solution, for instance when considering the two-term atomic model with an unpolarized lower term. In these cases, the line emission coefficients can be related to the incoming radiation field through the redistribution matrix \citep{DomkeHubeny88,Bommier97b} according to
\begin{align}
 \varepsilon^\ell_i(\nu,\mathbf{\Omega}) = & \, k_M \int \! \mathrm{d}\nu^\prime \oint \mathrm{d}\mathbf{\Omega}^\prime 
 \sum_{j = 0}^3 \bigl[\bm{\mathcal{R}}(\nu^\prime,\mathbf{\Omega}^\prime, \nu, \mathbf{\Omega})\bigr]_{i j} \, I_j(\nu^\prime,\mathbf{\Omega}^\prime) \notag \\ 
  & + \varepsilon^{\ell,\mbox{\scriptsize th}}_i(\nu,\mathbf{\Omega}) \, . 
  \label{eq::LinEmis}
\end{align} 
We follow the convention that the primed and unprimed quantities refer to the incident and scattered radiation, respectively. The quantity $\varepsilon^{\ell,\mbox{\scriptsize th}}_i$ is the thermal contribution, which arises from collisionally excited atoms. Throughout this work we assume collisions to be isotropic and neglect the influence of the magnetic field on this contribution. Thus, we take it to be equal to the term appearing in the last line of Eq.~(68) in \cite{Belluzzi+13}. This implies that only the Stokes $I$ component receives a nonzero contribution. 

The first term in the right hand side of Eq.~\eqref{eq::LinEmis} describes the contribution to the emission vector from scattering processes. The redistribution matrix element $[\bm{\mathcal{R}}(\nu^\prime,\mathbf{\Omega}^\prime, \nu, \mathbf{\Omega})]_{i j}$ relates the $j$-th Stokes parameter of the radiation at frequency $\nu^\prime$ and in propagation direction $\mathbf{\Omega}^\prime$ that illuminates the atom to the emissivity in the $i$-th Stokes parameter at frequency $\nu$ and in direction $\mathbf{\Omega}$. 
 
We consider the redistribution matrix presented in \cite{Bommier18}, which is suitable for a two-term atom with an unpolarized and infinitely sharp lower term. This redistribution matrix allows accounting for 
the frequency redistribution effects of elastic collisions and the impact of a magnetic field of {an} arbitrary strength and orientation. 
Its expression can be separated into two matrices, generally indicated as $\bm{\mathcal{R}}_{\mbox{\sc ii}}$ and $\bm{\mathcal{R}}_{\mbox{\sc iii}}$, following the notation introduced by \cite{Hummer62}. 
The matrix $\bm{\mathcal{R}}_{\mbox{\sc{ii}}}$ characterizes the scattering processes that are coherent in the atomic frame and $\bm{\mathcal{R}}_{\mbox{\sc{iii}}}$ characterizes those that are fully noncoherent (i.e., the frequencies of incoming and scattered radiation are uncorrelated) in the atomic frame, due to the redistribution effect of elastic collisions. 
In the observer's frame, Doppler redistribution induces a complex coupling between the angular and frequency dependencies of these matrices. 
In this work we consider the angle-averaged approximation \citep{ReesSaliba82} for the processes quantified by $\bm{\mathcal{R}}_{\mbox{\sc{ii}}}$ and the approximation that CRD occurs in the observer's reference frame for those quantified by $\bm{{\mathcal R}}_{\mbox{\sc{iii}}}$ \citep[see Sect.~10.3 of][]{BHubenyMihalas15}. 
Under such approximations, the expressions of both matrices in the observer's frame can be decomposed into frequency- and angle-dependent components. 
Working in the formalism of the irreducible spherical tensors for polarimetry (see Sect.~5.11 of LL04) we can write 
\begin{equation}
 \bigl[\bm{\mathcal{R}}_{\mbox{\sc{x}}}(\nu^\prime,\mathbf{\Omega}^\prime, \nu, \mathbf{\Omega})\bigr]_{i j} = \!\!\!\! 
 \sum_{K, K^{\prime} = 0}^2 \sum_{Q = -K_\mathrm{min}}^{K_\mathrm{min}} 
 \!\!\!\! \bigl[R_{\mbox{\sc{x}}}\bigr]^{K K^{\prime}}_Q\!\!(\nu^\prime,\nu) \,
 \bm{\mathcal{P}}^{K K^{\prime}}_Q(\mathbf{\Omega}^\prime,\mathbf{\Omega})_{i j} \, ,  
 \label{eq::RedisDecomp}
\end{equation}
where $X = \{ \mbox{\sc{ii}}, \mbox{\sc{iii}} \}$ and $K_\mathrm{min} = \min(K,K^{\prime})$. 
We consider the approximation for $\bm{\mathcal{R}}_{\mbox{\sc{iii}}}$ to be suitable for modeling the scattering polarization of the strong resonance lines considered in this work \citep[e.g.,][]{Sampoorna+17}, for which the contribution from $\bm{\mathcal{R}}_{\mbox{\sc{ii}}}$ is substantially greater. The emergent radiation in the line core region originates from atmospheric depths where the density is low and only a small fraction of scattering processes are perturbed by elastic collisions. Moreover, although the wing photons originate from deeper regions of higher density, it can be shown that the large-amplitude wing scattering polarization signals mainly arise from processes that are not perturbed by elastic collisions \citep[see][]{AlsinaBallester+18}. 
The frequency-dependent components of ${\bm{\mathcal R}}_{\mbox{\sc{ii}}}$ and $\bm{\mathcal R}_{\mbox{\sc{iii}}}$ are given by 
\begin{align}
 \bigl[R_{\mbox{\sc{ii}}}&\bigr]^{K K^{\prime}}_Q(\nu^\prime, \nu) = 
 \sum_{\substack{j_u j_u^\prime \\ M_u M_u^\prime}} 
  \sum_{\substack{j_\ell j_\ell^\prime \\ M_\ell M_\ell^\prime}} 
  \mathcal{A}^{K K^{\prime}}_Q(j_u M_u,j_u^\prime M_u^\prime, j_\ell M_\ell,j_\ell^\prime M_\ell^\prime) \notag \\
 & \times \frac{1}{\pi \Delta\nu_D^2} \frac{\Gamma_R}{\Gamma_R + \Gamma_I + \Gamma_E + 2 \pi \, \mathrm{i}\, {\Delta\nu_D \, u_{j_u M_u,j_u^\prime M_u^\prime}}} \notag \\ 
 & \times \frac{1}{2} \int_0^\pi 
 \!\!\mathrm{d}\Theta\,\,\mathrm{exp}\Biggl[-\,\biggl(\frac{u - u^\prime + u_{j_\ell M_\ell, j_\ell^\prime M_\ell^\prime}}{2 \sin\Theta/2} \biggr)^2 \Biggr] \notag \\ 
 & \; \times \frac{1}{2} \, \Biggl[
   W\,\Biggl(\frac{a}{\cos\Theta/2},\frac{u + u^\prime + u_{j_u^\prime M_u^\prime,j_\ell M_\ell}
  + u_{j_u^\prime M_u^\prime,j_\ell^\prime M_\ell^\prime}}{2 \cos\Theta/2} \Biggr) \, + \notag \\
  & \quad\;\, + W\,\Biggl(\frac{a}{\cos\Theta/2},\frac{u + u^\prime + u_{j_u M_u,j_\ell M_\ell} + u_{j_u M_u,j_\ell^\prime M_\ell^\prime}}{2 \cos\Theta/2} \Biggr)^\ast \Biggr] 
\, ,
 \label{eq::R2ObsnHFS}
\end{align}
and 
\begin{align}
 \bigl[R_{\mbox{\sc{iii}}}&\bigr]^{K K^{\prime}}_Q(\nu^\prime, \nu) = 
  \sum_{\substack{j_u j_u^\prime \\ M_u M_u^\prime}} 
  \sum_{\substack{j_\ell j_\ell^\prime \\ M_\ell M_\ell^\prime}} 
  \mathcal{A}^{K K^{\prime}}_Q(j_u M_u, j_u^\prime M_u^\prime, j_\ell M_\ell, j_\ell^\prime M_\ell^\prime) \notag \\
 & \; \times  
  \frac{1}{\pi \Delta\nu_D^2} \Biggl[\frac{\Gamma_R}{\Gamma_R + \Gamma_I +  2 \pi \, \mathrm{i} \, {\Delta\nu_D \, u_{j_u M_u, j_u^\prime M_u^\prime}}} \notag \\
 & \; \; \; - \frac{\Gamma_R}{\Gamma_R + \Gamma_I + \Gamma_E + 2 \pi \, \mathrm{i} \, {\Delta\nu_D \, u_{j_u M_u,j_u^\prime M_u^\prime}}} \Biggr] \notag \\
 & \; \times \frac{1}{2} \, \Biggl[ W\,\Bigl(a, u^\prime + u_{j_u^\prime M_u^\prime, j_\ell M_\ell}\Bigr)  
 + W\,\Bigl(a, u^\prime + u_{j_u M_u, j_\ell M_\ell}\Bigr)^\ast \Biggr] \notag \\
 & \; \times \frac{1}{2} \, \Biggl[W\,\Bigl(a,u + u_{j_u^\prime M_u^\prime, j_\ell^\prime M_\ell^\prime}\Bigr) 
 + W\,\Bigl(a, u + u_{j_u M_u, j_\ell^\prime M_\ell^\prime}\Bigr)^\ast \Biggr] \, .  
 \label{eq::R3ObsnHFS}
\end{align}
The analogous expressions for atomic systems with HFS can be found in Appendix~\ref{sec::AppRedis2TrmHFS}. 
The function $W(a,x) = H(a,x) + \mathrm{i} \, L(a,x)$ can be equivalently expressed in terms of the complex variable $z = x + \mathrm{i} \, a$; it is a complex analytical function related to the complementary error function (e.g., Sect.~5.4 of LL04). 
 We also introduce the reduced frequency splittings between states pertaining to the same term\footnote{It can be easily verified that
 $u_{j_u M_u, j_u^\prime M_u^\prime} = u_{j_u M_u, j_\ell M_\ell} - u_{j_u^\prime M_u^\prime, j_\ell M_\ell}$ for any lower state $\ket{j_\ell M_\ell}$ and $u_{j_\ell M_\ell, j^\prime_\ell M^\prime_\ell} = u_{j_u M_u, j_\ell^\prime M_\ell^\prime} - u_{j_u M_u, j_\ell M_\ell}$ for any upper state $\ket{j_u M_u}$.}  
\begin{equation*}
  u_{j_u M_u, j_u^\prime M_u^\prime} = \frac{\nu_{j_u M_u, j_u^\prime M_u^\prime}}{\Delta \nu_D} \, \quad
  {\mathrm{and} \quad
  u_{j_\ell M_\ell, j_\ell^\prime M_\ell^\prime} = \frac{\nu_{j_\ell M_\ell, j_\ell^\prime M_\ell^\prime}}{\Delta \nu_D} \, .}
\end{equation*} 
The quantities ${\mathcal A}$ entering Eqs.~\eqref{eq::R2ObsnHFS} and \eqref{eq::R3ObsnHFS} depend on a series of quantum numbers specific to each atomic state. Their expressions can be found in Eq.~\eqref{eqapp::AquantnoHFS}. 
The line broadening constants due to radiative processes ($\Gamma_R$) and due to elastic ($\Gamma_E$) and inelastic  ($\Gamma_I$) collisions have also been introduced in the previous equations. In this work, these constants are determined as explained in Appendix~\ref{sec::AppAtomMod}. In the reference system of the problem the angle-dependent part of the redistribution matrix in Eq.~\eqref{eq::RedisDecomp} is given as 
\begin{align}
    \bm{\mathcal{P}}^{K K^{\prime}}_Q (\mathbf{\Omega}^\prime,\mathbf{\Omega})_{i j} & =
    \sum_{Q^\prime=-K'}^{K'} \sum_{Q^{\prime \prime} = -K}^K (-1)^{Q^{\prime \prime}}
    \mathcal{T}^{K^{\prime}}_{Q^{\prime}}(i,\mathbf{\Omega})
    \mathcal{T}^{K}_{-Q^{\prime \prime}}(j,\mathbf{\Omega}^\prime) \notag \\
    & \times \mathcal{D}^{K}_{Q Q^{\prime \prime}}(R_B) 
    \mathcal{D}^{K^{\prime}}_{Q Q^{\prime}}(R_B)^\ast \, . 
\label{eq::PfcaVertFrame}
\end{align} 
For clarity, we hereafter refer to the calculations that rely on the expressions presented in this subsection without any additional approximations as reference PRD calculations. 

\subsection[Sect2p5]{Continuum contribution to radiative transfer coefficients}
\label{sec::sContinuum} 
Under conditions encountered in the solar atmosphere, it is generally a good approximation to neglect the continuum contribution to the off-diagonal elements of the propagation matrix (see Sect.~9.1 of LL04). The continuum contribution to the absorption coefficient is given by 
\begin{equation}
 \eta^c_I(\nu) = \kappa_c(\nu) + \sigma_c(\nu) \, ,
 \label{eq::ContAbs}
\end{equation}
where $\kappa_c$ is the extinction coefficient due to thermal processes, or ``true'' absorption, and $\sigma_c$ 
is the extinction coefficient due to continuum scattering. For the latter, we account for contributions from Rayleigh scattering due to neutral hydrogen and helium atoms and from Thomson scattering due to free electrons \citep[e.g.,][]{TrujilloBuenoShchukina09}.  

The emission coefficient also receives a contribution from continuum processes, which can be decomposed into a scattering term and a thermal term, or $\varepsilon^c_i = \varepsilon^{c,\mbox{\scriptsize sc}}_i + \varepsilon^{c,\mbox{\scriptsize th}}_i$. The scattering term is given by
\begin{equation}
 \varepsilon^{c,\mbox{\scriptsize sc}}_i(\nu,\mathbf{\Omega}) = \sigma_c(\nu) \sum_{K = 0}^2 \sum_{Q=-K}^K (-1)^Q 
  J^K_{-Q}(\nu) \, \mathcal{T}^K_Q(i,\mathbf{\Omega}) \, \, , 
 \label{eq::ContScatEmis}
 \end{equation}
where $J^K_Q$ are the multipolar components of the radiation field tensor (see Sect.~5.11 of LL04) 
\begin{equation}
 J^K_Q(\nu) = \oint\!\frac{\mathrm{d}\mathbf{\Omega}}{4\pi} \, \sum_{i=0}^3\mathcal{T}^K_Q(i,\mathbf{\Omega}) {I_i(\nu,\vec{\Omega})} \, . 
 \label{eq::RadfieldTens}
\end{equation}

\subsection[Sect2p6]{The iterative scheme}
\label{sec::sIterative}
The general {non-}LTE problem consists in finding a self-consistent solution of the RTE, which yields the Stokes vector at all spatial points, frequencies, and directions (see Eq.~(\ref{eq::RTEq})), and of the statistical equilibrium equations, from which the coefficients of the RTE are obtained. Efficient operator-splitting iterative methods exist to solve this problem, both in the absence \citep[e.g.,][]{Cannon73,Olson+86,TrujilloBuenoFabianiBendicho95} and in the presence \citep[e.g.,][]{Faurobert+97,Nagendra+98,TrujilloBuenoMansoSainz99} of polarization. 
Reviews on such methods can be found in \cite{Hubeny92}, \cite{Nagendra03}, and \cite{TrujilloBueno03}. 
More recently, iterative approaches based on Krylov methods, which could yield better performances than the aforementioned operator-splitting methods, have been proposed \citep[e.g.,][]{PaletouAnterrieu09,Anusha+09,Benedusi+21}. 

\begin{figure*}[!h]
  \centering 
   \includegraphics[width = 0.975\textwidth]{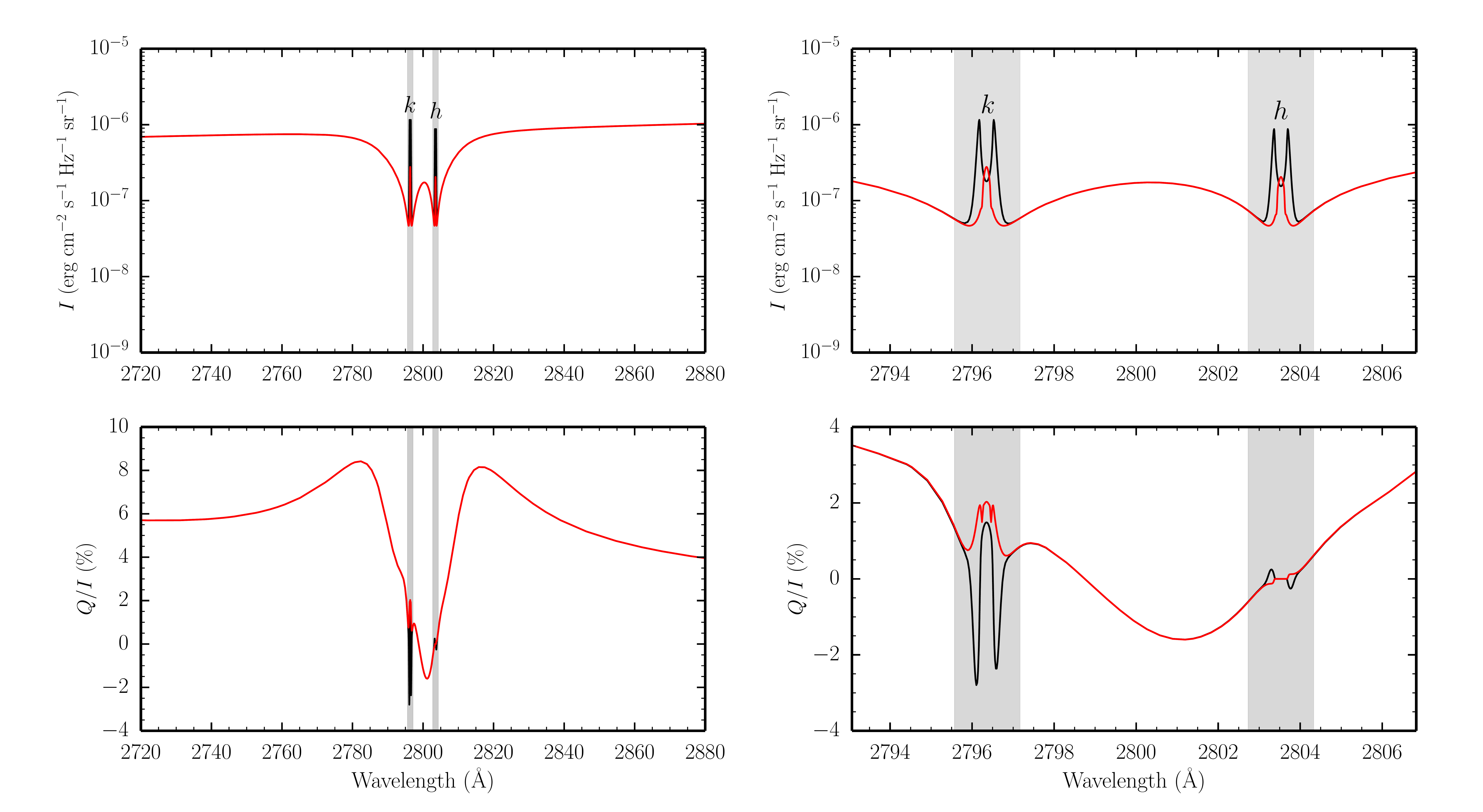}
   \caption{Stokes $I$ (\textit{upper panels}) and $Q/I$ (\textit{lower panels}) profiles as a function of wavelength for the Mg~{\sc{ii}} $h$ and $k$ lines for the wider (\textit{left panels}) and narrower (\textit{right panels}) spectral intervals discussed in the text. 
   Vacuum wavelengths are considered for all figures. The calculations were carried out in the absence of a magnetic field. 
   The solid black curves correspond to the reference PRD calculations (see text), whereas the solid red curves represent the results of calculations taking the approximate redistribution matrix $\bm{\hat{{\mathcal R}}}_{\mbox{\sc{ii}}}$ (see text), in which scattering processes are treated as coherent in the reference frame of the observer. 
 The shaded areas in the figures of this section are centered around the $h$ \& $k$ lines, have a width of $1.6$~\AA , and indicate the spectral region where this approximation is not expected to be suitable. For all the figures presented in this section, the calculations are carried out  considering the atmospheric model C of \cite{Fontenla+93}. All profiles are presented for an LOS with $\mu = 0.1$ and taking the direction for positive Stokes $Q$ parallel to the limb.}  
  \label{Fig::Fig1} 
 \end{figure*} 
 
In our iterative approach the population of the lower term $\mathcal{N}_\ell$ (needed to calculate the frequency-integrated absorption coefficient $k_M$ of Eq.~(\ref{eq::FreqInt})) is kept fixed. 
We generally obtain it from a self-consistent multi-level calculation in the unpolarized case using the RH code of \citet{Uitenbroek01}. Hereafter, we refer to this part of the calculation as step 1. 
This step also provides the continuum quantities $\kappa_c$, $\sigma_c$, and $\varepsilon^{c,\mbox{\scriptsize{th}}}$,
the rates of both elastic and inelastic collisions, and an initial estimate for the unpolarized radiation field.
{We point out that, in principle, these quantities could be obtained from other sources. For instance the population for the lower term of the  H~{\sc{i}} Lyman-$\alpha$ line (see Appendix~\ref{sec::AppLymanAlpha}) can be taken directly from the considered atmospheric model.  

Once the input quantities are obtained, we iteratively solve the  RT problem in the polarized case for a two-term atomic model. We refer to this as step 2 (or the main part) of the calculation, for which we use the RT code that we developed through the implementation of the expressions introduced above. 
This step relies on the approximation that $N_\ell$ is kept fixed\footnote{The population of the upper term $N_u$ is allowed to vary. Because $N_u \ll N_\ell$, this does not introduce significant inconsistencies and it allows us to take into account the impact of polarization on the ratio $N_u / N_\ell$.} during the iterative process. Although this strategy is not fully consistent, it gives a population of the lower term that is much more realistic than the one that would be obtained from a two-term atomic model. Furthermore, by fixing the lower level population the elements of the propagation matrix are known \textit{a priori} and the considered RT problem is linear. 

The step-2 {non-}LTE problem thus consists in finding a self-consistent solution of Eqs.~\eqref{eq::RTEq} and \eqref{eq::LinEmis}, with the latter equation relating the line emission coefficient to the incident radiation. The considered Jacobi-based iterative method represents a generalization of the approach presented in \citet{AlsinaBallester+17} to the case of a two-term atomic model. The RTE is solved through the DELOPAR short-characteristics formal solver of \citet{TrujilloBueno03}. %
Formal solvers of the Diagonal Element Linear Operator (DELO) family are nowadays a popular choice for solving RT problems accounting for polarization \cite[e.g.,][]{AlsinaBallester+17,Sampoorna+17,delaCruz+19} 
because they {can} account for the action of the various elements of the propagation matrix in a straightforward manner. 
Compared to other frequently used methods, such as high-order Runge-Kutta solvers, DELO solvers are particularly well suited to avoiding potential numerical instabilities \citep[see][]{JanettPaganini18}. 

We study the convergence of the iterative procedure in step 2 by introducing the following decomposition of the line scattering contribution to the emission coefficients
\begin{equation}
 \varepsilon^{\ell, \mbox{\scriptsize{sc}}}_i(\nu,\mathbf{\Omega}) = \sum_{K = 0}^2 \sum_{Q=-K}^K \mathcal{E}^K_Q(\nu) \mathcal{T}^K_Q(i,\vec{\Omega}) \, .
 \label{eq::EpsDecomp}
\end{equation} 
At a given iteration, the maximum relative change of the $\mathcal{E}^K_Q$ components, over all spatial and frequency points, is 
\begin{equation}
 R_c\bigl(\mathcal{E}^K_Q\bigr) = \mathrm{max}\left(\frac{\bigl|\mathcal{E}^K_Q(\nu; \mathbf{r})^{\mathrm{new}} - \mathcal{E}^K_Q(\nu; \mathbf{r})^{\mathrm{old}} \bigr|}
 {\mathcal{E}^K_Q(\nu; \mathbf{r})^{\mathrm{new}}} \right) \, ,
 \label{eq::MrelCh}
\end{equation}
where the labels ``$\mathrm{new}$'' and ``$\mathrm{old}$'' refer to the values of the $\mathcal{E}^K_Q$ component at the current and previous iteration, respectively. 

In the applications presented below, convergence is considered to have been achieved when the maximum relative change for the component with $K = Q = 0$ falls below a predetermined value. 
The most computationally demanding task in this RT problem is the evaluation of the redistribution matrix, which is calculated at the first iteration and then stored. Subsequent iterations therefore entail a far lower computational cost. This justifies our choice to take the very strict convergence criterion of $R_c( \mathcal{E}^0_0 ) \le {10^{-10}}$, even though it calls for more iterations than if a higher, but still acceptable, threshold value were taken. For the reference PRD calculations, we also impose that $R_c(\mathcal{E}^2_0 ) \le 10^{-6}$ in order for convergence to be achieved. Due to the possible presence of oscillations in $R_c( \mathcal{E}^2_0 )$ during the initial iterations, these convergence criteria are only applied when more than $30$ iterations have been performed. 
When convergence is reached, the Stokes profiles for the emergent radiation are obtained by solving the RTE for a predetermined set of lines of sight (LOS) and, finally, the results are stored. 

\section[Sect3]{Illustrative results for the Mg~{\sc{ii}} doublet}
\label{sec::Illustrative}
In this section we present a series of calculations carried out with the RT code described above. 
We also quantitatively evaluate the suitability of several approximations of interest for modeling the scattering polarization in the wings of 
the Mg~{\sc{ii}} $h$ \& $k$ lines. We have modeled these resonance lines considering a two-term atomic system with an infinitely sharp and unpolarized lower term without HFS. A discussion on the suitability of a two-term atom and further details of this modeling are presented in Appendix~\ref{sec::AppAtomMod}. The analogous study considering the H~{\sc{i}} Lyman-$\alpha$ line is shown in Appendix~\ref{sec::AppLymanAlpha}. 

We recall that, in all the RT calculations considered in this section, the approximation of CRD in the reference frame of the observer is made for $\bm{\mathcal{R}}_{\mbox{\sc{iii}}}$. We refer to the calculations considering the angle-averaged $\bm{\mathcal{R}}_{\mbox{\sc{ii}}}$, without making the approximations described in Sects.~\ref{sec::sCSObs} and \ref{Sec::PurelyCS}, as reference PRD calculations. 

For all calculations, the semi-empirical 1D atmospheric model C of \cite{Fontenla+93}, hereafter FAL-C, is considered. 
For the figures presented in this section we consider an LOS with $\cos\theta = 0.1$, where $\theta$ is the inclination with respect to the local vertical (see Fig.~\ref{Fig::geometry}).  
We selected an LOS with a large inclination in order to study scattering polarization signals with large amplitudes. Throughout this work, the reference direction for positive Stokes $Q$ is taken parallel to the limb. When present, magnetic fields are taken to be horizontal ($\theta_B = 90^\circ$) with azimuth $\chi_B = 0^\circ$ (i.e., parallel to the $X$-axis in Fig.~\ref{Fig::geometry}) and with a strength of $20$~G at all spatial grid points of the problem. 

We focus our initial discussion on the results of the reference PRD calculations in the absence of a magnetic field. The intensity and linear polarization ($Q/I$) profiles obtained from these calculations are represented by the black curves in Fig.~\ref{Fig::Fig1}, for the two spectral intervals that will be considered throughout this section, given in vacuum wavelengths. The wider of the two intervals was selected to highlight the far-wing behavior of the doublet and ranges from $\sim\!\!75$~\AA\ to the blue of the $k$ line center to $\sim\!\!75$~\AA\ to the red of the $h$ line center. The narrower one focuses on the behavior close to the core region of the lines and corresponds almost exactly to the spectral interval covered by the spectrograph of CLASP2 \citep{Ishikawa+21}, bounded between $\sim\!\!3.3$~\AA\ to the blue of the $k$ line center and $\sim\!\!3.3$~\AA\ to the red of the $h$ line center. 
We verified the accuracy of these calculations by comparing the results with those of existing codes; the results close to the core region of the $k$ line agree with those obtained with the two-level code introduced in \cite{AlsinaBallester+17} and, in the entire interval considered in this work, they coincide with the results obtained with the two-term code that was used in \cite{delPinoAleman+16,delPinoAleman+18}. 
 
In the intensity profiles, when going from the far wings into the core of either line, we find a deep absorption feature where the values become roughly one order of magnitude smaller than those of the far wings. In the region between the two lines, where their extended wings overlap, the intensity is also much lower than in the far wings. On the other hand, emission peaks are found closer to the center of the two lines. The $k$ line in particular presents two peaks (k${}_2$) that reach an intensity greater than that in the far wings. 
Regarding the scattering polarization, we find far-wing signals with  substantial amplitudes, which are known to be strongly influenced by $J$-state interference \citep{BelluzziTrujilloBueno12}. Closer to the core of the two lines, where the absorption features are appreciable in intensity, the $Q/I$ signals fall below the continuum levels. Also in agreement with \cite{BelluzziTrujilloBueno12}, we observe a triple-peak structure around the $k$ line with a positive peak in the center and negative peaks in the near wings, an antisymmetric profile around the $h$ line, and a negative feature between them. 

It must be acknowledged that the intensity and polarization patterns of strong resonance lines are strongly influenced by parameters including the micro-turbulent velocity and the van der Waals (vdW) broadening, which is related to $\Gamma_E$. 
The clear impact of enhancements of the micro-turbulent velocity and the vdW broadening in the intensity wings of the Mg~{\sc{ii}} resonance lines is illustrated in Fig.~4 of \cite{MilkeyMihalas74}. The researcher interested in modeling spectro-polarimetric observations such as those of the CLASP-2 missions should pay special attention to the selection of these parameters. However, for the main purpose of this work -- namely to study the reliability of a number of approximations through comparisons with the reference PRD calculations -- we find it reasonable to take the micro-turbulent velocities that were determined semi-empirically as specified in \cite{Fontenla+91} and to calculate the vdW broadening as discussed in Appendix~\ref{sec::AppAtomMod} without any enhancements. 

In Sects.~\ref{sec::sCSObs} through~\ref{Sec::MagProf} we introduce the approximations investigated throughout this work and evaluate their suitability. In Sect.~\ref{sec::BRs}, we analyze the influence of the branching ratios, especially those that enter the terms of the redistribution matrix that quantify the interference between different upper states. 

\begin{figure*}[]
 \centering
   \includegraphics[width = 0.975\textwidth]{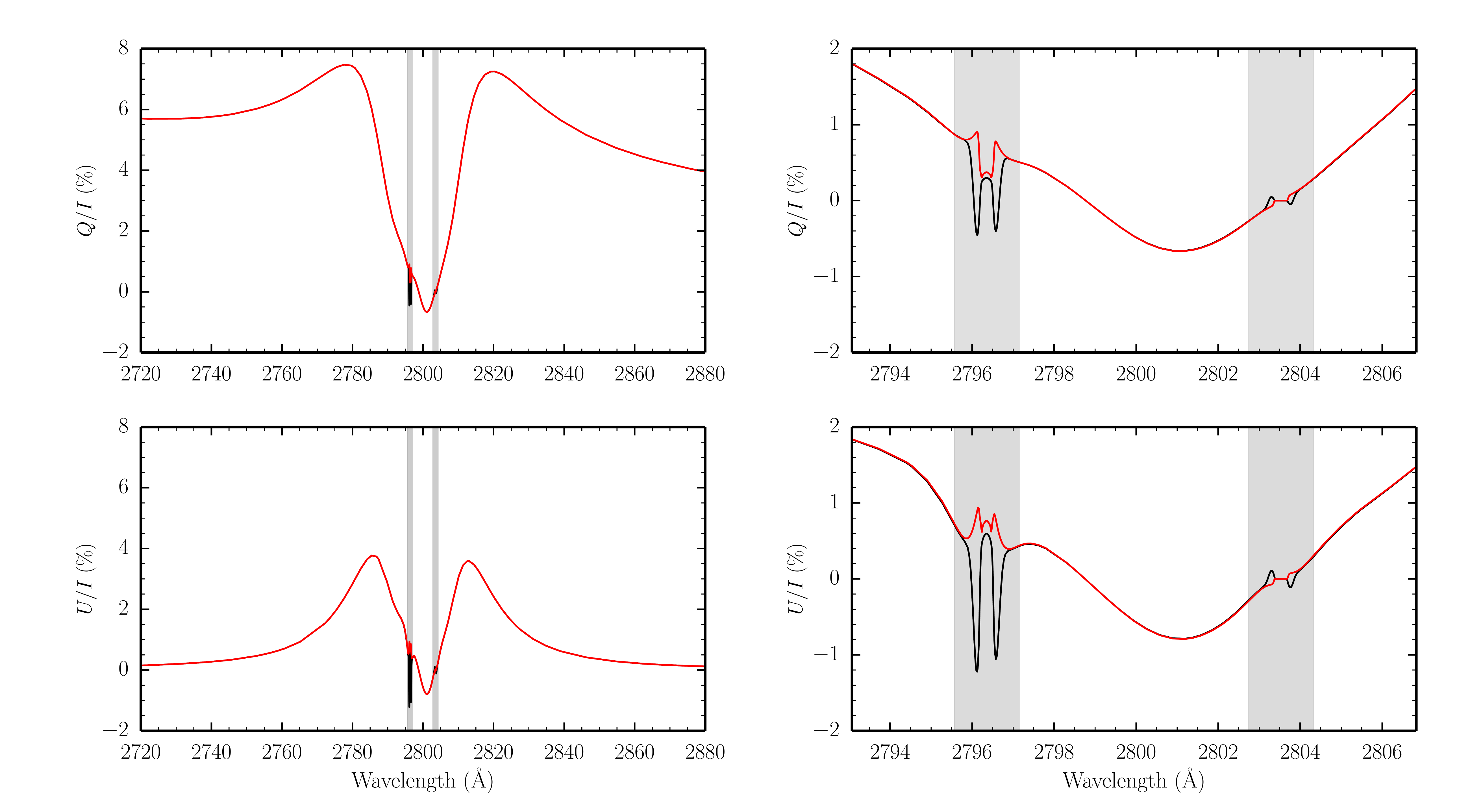}
        \caption{Stokes $Q/I$ (\textit{upper panels}) and $U/I$ (\textit{lower panels}) profiles as a function of wavelength. The black and red curves represent calculations similar to those presented in Fig.~\ref{Fig::Fig1}, but in the presence of $20$~G horizontal ($\theta_B = 90^\circ$) magnetic fields with $\chi_B = 0^\circ$.}
        \label{Fig::Fig2} 
 \end{figure*}

\subsection[Sect3p1]{Coherent scattering in the observer's frame for $\bm{\mathcal R}_{\mbox{\sc{ii}}}$} 
\label{sec::sCSObs} 
In this subsection we investigate an approximation that is applied to the scattering processes that are unperturbed by elastic collisions, quantified by $\bm{\mathcal{R}}_{\mbox{\sc{ii}}}$. 
In particular, we treat such processes as coherent in the reference frame of the observer rather than in that of the atom. In other words, Doppler redistribution is neglected and as a result no frequency-angle coupling is introduced in the ensuing approximate redistribution matrix, which we hereafter refer to as $\bm{\hat{\mathcal{R}}}_{\mbox{\sc{ii}}}$, and whose expression can be found in Eq.~\eqref{eqapp::RedisNoHFSCSObserv}. 
{We note that} we do not neglect the redistribution due to elastic collisions and, for the corresponding scattering processes (quantified by $\bm{\mathcal{R}}_{\mbox{\sc{iii}}}$), we still make the assumption that CRD occurs in the reference frame of the observer (see Sect.~\ref{sec::sEmis}). The proposed approximation greatly reduces the numerical cost of computing the contribution of $\bm{\mathcal{R}}_{\mbox{\sc{ii}}}$ to the emissivity, which 
is the most computationally intensive contribution in the case of the reference PRD calculations. 
The contribution from $\bm{\hat{\mathcal{R}}}_{\mbox{\sc{ii}}}$ to the emission coefficient at any given frequency only requires accounting for the incident radiation at a small number of spectral points determined by the emission frequency and by the energy differences between  
the initial and final lower states. For the considered atomic system, whose lower term only has one FS level, these energy differences only arise in the presence of magnetic fields (see Eq.~\eqref{eqapp::RedisNoHFSCSObserv}).  

The impact of this approximation, relative to the reference PRD calculations, can be seen in the Stokes profiles in Fig.~\ref{Fig::Fig1}. The shaded regions in the figure span $0.8$~\AA\ from the center of the $h$ and $k$ lines, which corresponds to their Doppler width $\Delta\lambda_D$ at the height for which their line-center $\tau$ is equal to unity for $\mu = 0.1$ in the FAL-C model. Within these spectral regions, Doppler redistribution plays a dominant role \citep[e.g.,][]{Thomas57}, and thus the approximation discussed here should not be expected to be suitable. The error incurred by this approximation is only appreciable well within the shaded areas and no farther than $0.5$~\AA\ from the center of the $h$ and $k$ lines, both in intensity and $Q/I$. Far from the shaded core regions of these lines, an excellent agreement is found between the results of this approximation and reference PRD calculations. 

\begin{figure*}[]
 \centering
 \includegraphics[width = 0.975\textwidth]{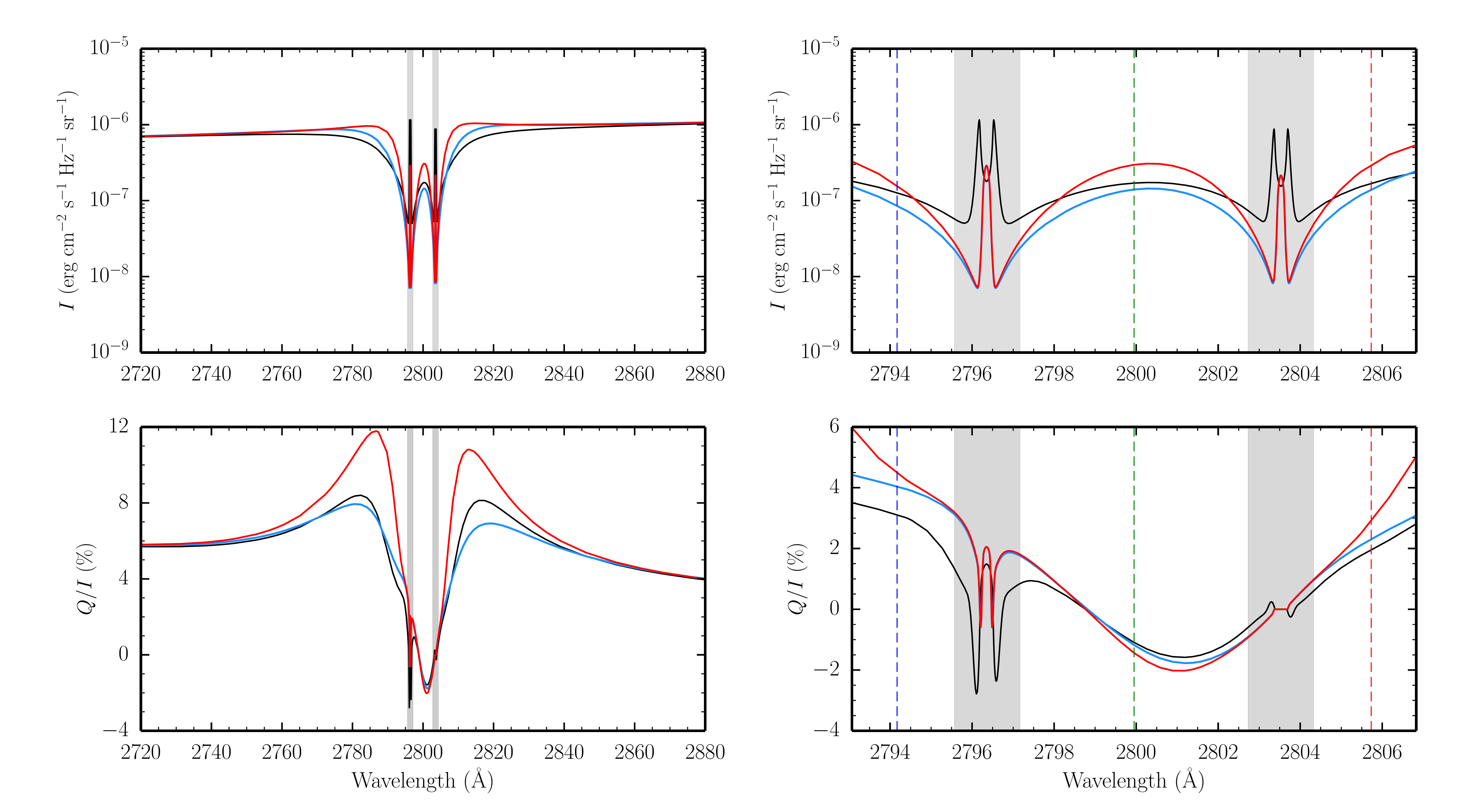}
        \caption{Stokes $I$ (\textit{upper panels}) and $Q/I$ (\textit{lower panels}) profiles as a function of wavelength  
        obtained through the RT calculations described in the text, in the absence of a magnetic field. 
    Reference PRD calculations (black curves) are compared to those in which 
    Doppler and collisional redistribution are neglected in scattering processes (see text), both accounting for the contribution from elastic collisions to the damping constant (blue curves) and neglecting it (red curves). 
    The vertical dashed colored lines indicate the spectral positions discussed in the main text, namely $2.2$~\AA\ to the blue of the $k$ line center (blue), equidistant between the $k$ and $h$ line centers (green), and $2.2$~\AA\ to the red of the $h$ line center (red).} 
        \label{Fig::Fig5} 
 \end{figure*}

We have also evaluated the suitability of using the approximate $\bm{\hat{\mathcal{R}}}_{\mbox{\sc{ii}}}$ in the presence of a $20$~G horizontal magnetic field. The comparison between the $Q/I$ and $U/I$ profiles obtained under this approximation and through the  reference PRD calculations is shown in Fig.~\ref{Fig::Fig2}. 
In the presence of a $20$~G magnetic field, the approximation is also found to be suitable beyond $0.5$~\AA\ from the center of the lines of the magnesium doublet. The intensity profiles are not shown in this figure because they are not meaningfully impacted by fields of such strength and coincide with those shown in Fig.~\ref{Fig::Fig1}. 

\subsection[Sect3p2]{The treatment of noncoherent scattering processes}
\label{Sec::PurelyCS} 
The approximation analyzed in the previous subsection performs very well for modeling the intensity and scattering polarization signals beyond a Doppler width from the centers of the Mg~{\sc{ii}} $h$ \& $k$ lines. In addition, it allows for a considerable decrease in the overall computational cost of the non-LTE 
 RT calculations. Under this approximation, the contribution from the $\bm{\hat{{\mathcal R}}}_{\mbox{\sc{ii}}}$ redistribution matrix may no longer be the most computationally intensive part of the emissivity calculation. 
In this case, it may be of interest to reduce the computational cost of the contribution from $\bm{\mathcal R}_{\mbox{\sc{iii}}}$, which quantifies the scattering processes that are noncoherent due to collisions. 

First we discuss the influence of artificially setting to zero all the frequency-dependent components of $\bm{\mathcal R}_{\mbox{\sc{iii}}}$ (see Eq.~\eqref{eq::RedisDecomp}) except for $K = K^\prime = Q = 0$, which typically represents by far the greatest contribution to $\varepsilon_I$ but has no impact on the other Stokes components of the emission vector. Under this approximation, we calculated the intensity and $Q/I$ profiles both using $\bm{\mathcal{R}}_{\mbox{\sc{ii}}}$ and the approximate $\bm{\hat{\mathcal{R}}}_{\mbox{\sc{ii}}}$.
The profiles obtained through these calculations, which were carried out in the absence of a magnetic field, fully coincide with the corresponding ones shown in Fig.~\ref{Fig::Fig1}, in which all components of $\bm{\mathcal R}_{\mbox{\sc{iii}}}$ were considered. 
Although the scattering processes quantified by $\bm{\mathcal R}_{\mbox{\sc{iii}}}$ have a direct impact on the intensity 
profile,\footnote{We verified that, by setting to zero all the components of $\bm{\mathcal R}_{\mbox{\sc{iii}}}$ including $K = K^\prime = Q = 0$, the intensity profile is severely underestimated, especially in the wings.} 
these comparisons reveal that the same processes only affect the wings of the scattering polarization profile indirectly, by reducing the contribution from the coherent scattering processes through a decrease in the branching ratios in $\bm{\mathcal{R}}_{\mbox{\sc{ii}}}$ or 
$\bm{\hat{\mathcal{R}}}_{\mbox{\sc{ii}}}$ \citep[see also the discussion in][which is focused on the Ca~{\sc{i}} resonance line at $4227$~\AA ]{AlsinaBallester+18}. 
For the rest of the calculations presented in this section, all the components of $\bm{\mathcal R}_{\mbox{\sc{iii}}}$ are taken into account unless otherwise noted. 

A further consequence of this finding is that the scattering polarization profiles are not sensitive to the depolarizing effect of the collisions responsible for noncoherent scattering. In the absence of magnetic fields, the code used in this work can also take into account the depolarizing effect of the collisions that induce transitions between states of the same FS level (see Eqs.~\eqref{eqapp::RedisnoHFSNoB}). We verified that the inclusion of depolarizing collisional rates, computed as detailed in the Appendix of \cite{MansoSainz+14}, has no appreciable impact on the scattering polarization profile compared to the case in which they are neglected. 

When considering scenarios that are especially demanding from the numerical point of view, such as 3D radiative transfer, it may be of interest to further reduce the cost of computing the emissivity. We now go a step beyond the previous approximations and neglect the frequency redistribution due to both the Doppler effect and elastic collisions, so that all scattering process are treated as coherent in the observer's frame and are thus quantified by $\bm{\hat{\mathcal{R}}}_{\mbox{\sc{ii}}}$. 
In the numerical framework presented in this work, this can achieved simply by setting to zero the line broadening constant due to elastic collisions $\Gamma_E$ that enters the branching ratios of the redistribution matrices (see Eq.~\eqref{eqapp::RedisNoHFSCSObserv}). Following this approach, the contribution from $\Gamma_E$ should also be neglected in the computation of the damping constant. Otherwise an inconsistency is introduced because the magnetic splitting in the emission profiles is no longer fully canceled by the  
denominators of the branching ratios (see Appendix~\ref{sec::AppCRDCSApprox}), which results in an artificial magnetic sensitivity in the far wings \citep[see Appendix B of][]{AlsinaBallester+18}. In the following subsection, an approximation through which this spurious sensitivity can be avoided will be discussed. 

Because the contribution from $\bm{\mathcal{R}}_{\mbox{\sc{iii}}}$ is effectively zero in this case, the entire emission vector at any given frequency can be computed accounting only for the radiation field at the small number of frequency points discussed in Sect.~\ref{sec::sCSObs}, which thus allows for an even lighter calculation. 
The intensity and $Q/I$ profiles obtained in the absence of magnetic fields following this approach, both accounting for the contribution from $\Gamma_E$ to $a$ and neglecting it, are shown in Fig.~\ref{Fig::Fig5}, where they are also compared to the reference PRD calculations. Both in the near and far wings, a better agreement is found when accounting for the contribution from elastic collisions to the damping constant.

At wing wavelengths roughly $10$~\AA\ from the center of either line, the reference PRD results are reproduced reasonably well when the contribution of elastic collisions to the damping constant is taken into account, although the intensity and the $Q/I$ signals are slightly overestimated. The agreement is significantly worse when neglecting this contribution, in which case the approximate calculation further overestimates the intensity signal and substantially overestimates the $Q/I$ amplitude. 

Closer to the center of the lines, we do not find a good agreement with the reference PRD intensity profile until going farther than $1$~\AA\ into the wings, in contrast to the results obtained when the redistribution effects of elastic collisions were taken into account (see Sect.~\ref{sec::sCSObs}). In the spectral region a few \AA\ from the center of the $h$ or $k$ lines, the inclusion of $\Gamma_E$ in the damping constant yields a better agreement for both intensity and $Q/I$, although the disparity between calculations including and neglecting this contribution is not as great as in regions farther into the wings. In the regions roughly $2.2$~\AA\ to the blue of the $k$ line center, $2.2$~\AA\ to the red of the $h$ line center, and between the two lines of the doublet close to $2800$~\AA\ (indicated with vertical colored lines in Fig.~\ref{Fig::Fig5}), the intensity is slightly underestimated when including $\Gamma_E$ in $a$ and slightly overestimated when neglecting it. 
The amplitude of the $Q/I$ signal is overestimated in these regions, especially when the damping constant does not include the contribution from elastic collisions. 

 \begin{figure*}[]
 \centering
\includegraphics[width = 0.975\textwidth]{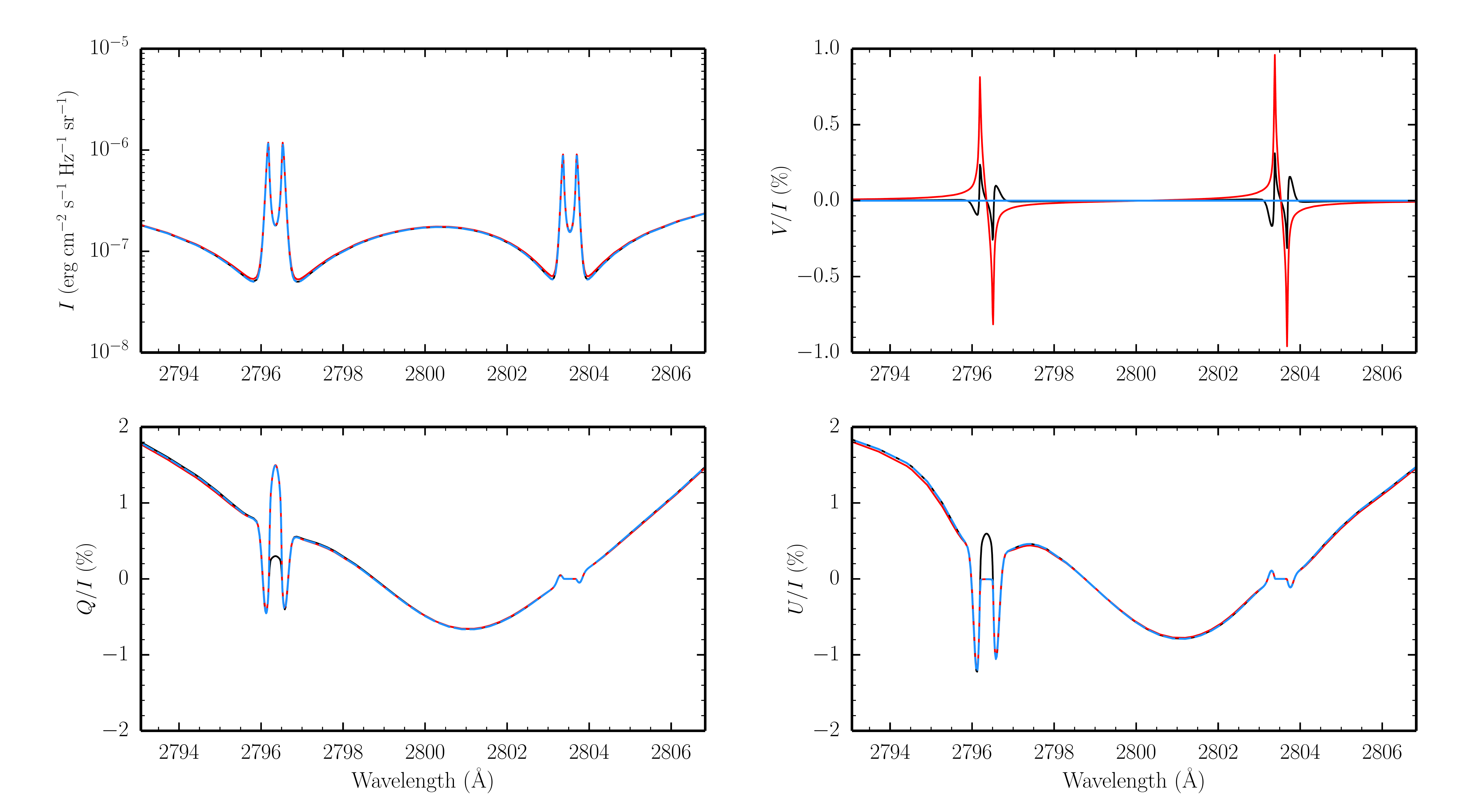}
 \caption{Intensity (\textit{upper left panel}), $V/I$ (\textit{upper right panel}), $Q/I$ (\textit{lower left panel}), and $U/I$ (\textit{lower right panel}) profiles of the Mg~{\sc{ii}} $h$ \& $k$ lines as a function of wavelength. The various colored curves show the results of reference PRD calculations in which the impact of the magnetic field is fully taken into account in all RT coefficients (black curves), neglected only in the computation of the emission coefficient (red curves), and only taken into account in $\eta_I$ and $\rho_V$ (blue curves). The magnetic field is horizontal, with $\chi_B = 0^\circ$ and a strength of $20$~G.} 
         \label{Fig::Fig4_MGHK}
 \end{figure*} 

  
\subsection[Sect3p3]{Neglecting the magnetic field in the redistribution matrix} 
\label{Sec::MagProf}

Here we investigate an approximation for modeling the scattering polarization signals in the wings of the aforementioned lines that is  focused specifically on the treatment of magnetic fields. For field strengths typical of quiet regions of the solar atmosphere, theoretical arguments support that the line scattering emission coefficients should not be sensitive to the magnetic field for wavelengths outside the core regions \citep[see Sect.~10.4 of LL04; also Appendix B of][]{AlsinaBallester+18}. 
Outside the line core, we thus expect the magnetic sensitivity of the linear polarization pattern to be controlled only by the elements of the propagation matrix and particularly by the $\rho_V$ coefficient, which quantifies Faraday rotation.  
If the influence of the magnetic field is neglected in the redistribution matrices, many of the energy states of the atomic system remain degenerate. In this case, far fewer transitions need to be considered (see the expressions in Eq.~\eqref{eqapp::RedisnoHFSNoB}), which considerably reduces the cost of computing the emission vector. 
We recall that 
this is usually the most costly computation in the 
considered RT problem; reducing their complexity would thus bring down the overall computational requirements. 

To study the suitability of this approximation for modeling the scattering polarization signals in the line wings, we have carried out a series of RT calculations in the presence of $20$~G horizontal magnetic fields. 
The four Stokes profiles are shown in Fig.~\ref{Fig::Fig4_MGHK}, comparing the results obtained when the magnetic splitting is (i) taken into account in all the RT coefficients (represented by the black curves), (ii) neglected in the emission coefficients but taken into account in all elements of the propagation matrix $\mathbf{K}$ (red curves), and (iii) only accounted for in the computation of the $\eta_I$ and $\rho_V$ coefficients (blue curves). 
Because we are considering an atomic system with an unpolarized lower term, in case (iii) all elements of $\mathbf{K}$ other than $\eta_I$ and $\rho_V$ are zero (see Eqs.~\eqref{eq::EtaNoHFS} and \eqref{eq::RhoNoHFS}). As expected, these approximations are not suitable to model the circular polarization signals found close to the center of $h$ and $k$, because the Stokes $V$ component of the emissivity is sensitive to the presence of the magnetic field through the Zeeman effect. 
In addition, the Stokes $V/I$ signals of the emergent radiation are shown to be strongly impacted by the dichroism effects quantified by $\eta_V$. 

\begin{figure*}[]
  \centering 
  \includegraphics[width=0.975\textwidth]{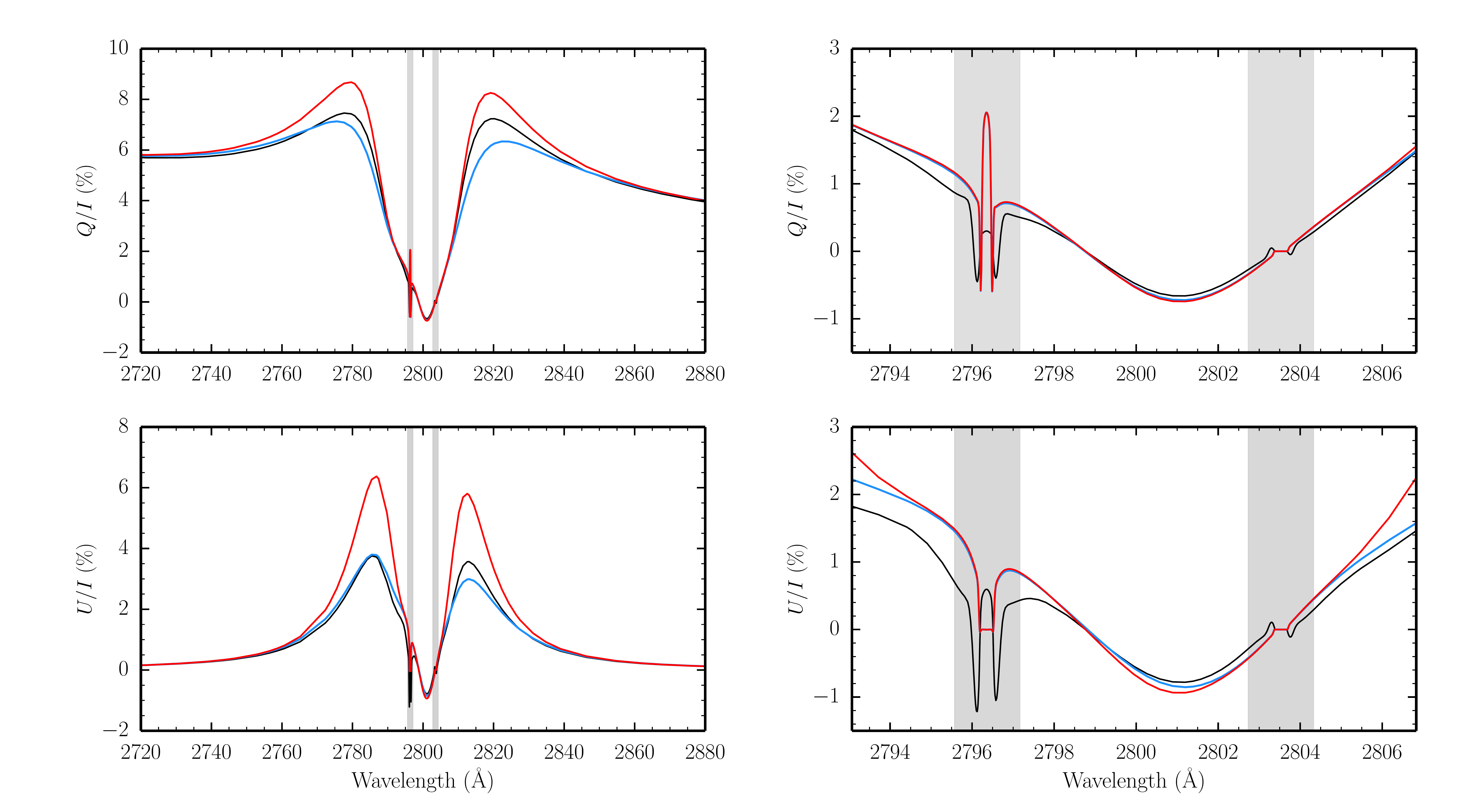}
        \caption{Stokes $Q/I$ (\textit{upper panels}) and $U/I$ (\textit{lower panels}) profiles as a function of wavelength obtained from RT calculations in the presence of a $20$~G magnetic field. Reference PRD calculations accounting for the magnetic field in all RT coefficients (black curves) are compared to those in which Doppler and collisional redistribution are neglected in scattering processes while the magnetic splitting is only accounted for in $\eta_I$ and $\rho_V$ (see text), both including the contribution from elastic collisions in the damping constant (blue curves) and neglecting it (red curves).}
    \label{Fig::Fig6_CSobs_onlyrhov}
  \end{figure*}

\begin{figure*}[!h]
 \centering 
\includegraphics[width = 0.975\textwidth]{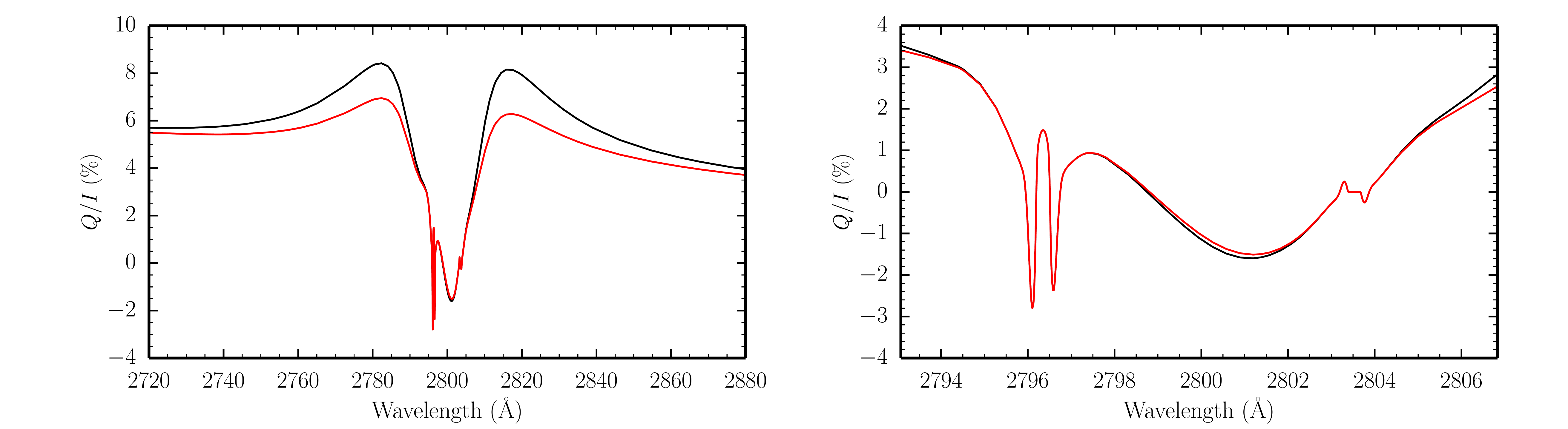}
\caption{$Q/I$ profiles as a function of wavelength obtained from PRD calculations in the absence of a magnetic field, considering the wider (\textit{left panel}) and narrower (\textit{right panel}) spectral ranges discussed in the text. The black and red curves represent, respectively, the calculations carried out using the redistribution matrices given in \cite{Bommier17} and those given in Eq.~\eqref{eqapp::RedisModBR}, for which the main difference concerns the branching ratios.} 
 \label{Fig::Fig6_BR}
 \end{figure*}
In any case, these calculations firmly establish that, for the strong resonance lines considered here, the magnetic sensitivity of the wing linear polarization profiles can be well reproduced by accounting for the magnetic splitting only in $\rho_V$ and $\eta_I$ and neglecting it in all other RT coefficients entering Eq.~\eqref{eq::RTEq}. 
The figure only shows the results in the narrower of the spectral ranges discussed above, but we verified that this approximation is also perfectly suitable for the scattering polarization signals farther from the line wings.  
We also verified that this approximation performs similarly well when considering magnetic fields with different orientations or with field strengths up to one order of magnitude larger than the ones considered here. On the other hand, when neglecting the magnetic splitting in the calculation of the emissivity but taking it into account in all elements of the propagation matrix, slight errors are incurred in the intensities and linear polarization signals in the near wings of both lines. 
We attribute these errors to accounting for the circular polarization dichroism term $\eta_V$ while neglecting the emission in circular polarization $\varepsilon_V$ (a more in-depth discussion in the context of the Lyman-$\alpha$ line can be found in Appendix~\ref{sec::AppLymanAlpha}). Thus, researchers interested in modeling the wing scattering polarization are recommended to account for the magnetic splitting only in $\eta_I$ and $\rho_V$ when making the approximation of neglecting it in the emission coefficient. 

This approximation can be used in combination with the ones discussed in previous subsections, allowing for a further decrease in the computational cost of the RT calculation. Indeed, we have carried out identical calculations to the ones presented in this subsection, but neglecting Doppler redistribution for the scattering processes quantified by $\bm{\mathcal{R}}_{\mbox{\sc{ii}}}$ as in Sect.~\ref{sec::sCSObs}. The results are not shown in the present paper, but we verified that the scattering polarization signals in the wings are also well reproduced by only accounting for the $\rho_V$ and $\eta_I$  coefficients in this case, confirming the suitability of combining the two approximations. 

We also considered combining the approximation of accounting for the magnetic splitting only in $\rho_V$ and $\eta_I$ 
with that of neglecting Doppler and collisional redistribution (see Sect.~\ref{Sec::PurelyCS}). In this case the artificial magnetic sensitivity 
produced when including the contribution from elastic collisions to the damping constant while neglecting it in the branching ratios of the redistribution matrices is of course avoided. The comparison between the scattering polarization profiles obtained under these approximations and through the reference PRD calculations while fully accounting for the magnetic field is shown in Fig.~\ref{Fig::Fig6_CSobs_onlyrhov} for a $20$~G horizontal magnetic field. The level of agreement is similar to that of the nonmagnetic case discussed in Sect.~\ref{Sec::PurelyCS}. Again, the approximation of neglecting Doppler and collisional redistribution performs considerably better when accounting for the contribution from $\Gamma_E$ to the damping constant. 

\subsection[Sect3p4]{The influence of the branching ratios in PRD modeling}
\label{sec::BRs} 
As noted above, the redistribution matrices used for the reference PRD calculations (see Eqs.~\eqref{eq::R2ObsnHFS} and \eqref{eq::R3ObsnHFS}) were derived from the rigorous theoretical framework presented in \cite{Bommier17}. RT calculations in the polarized case for a two-term atom had already been considered in the past, for instance by 
\citet{BelluzziTrujilloBueno12,BelluzziTrujilloBueno14} in the absence of magnetic fields. These authors used a heuristic $\bm{\mathcal{R}}_{\mbox{\sc{iii}}}$, as well as heuristic branching ratios. The RT code introduced in the present paper can consider the latter through the implementation of the expressions given in Eq.~\eqref{eqapp::RedisModBR}. 
More details can be found in Appendix~\ref{sec::AppRedisNoHFS}.  
We verified numerically that we can reproduce the intensity and linear polarization profiles of the Mg~{\sc{ii}} doublet 
obtained by \citet{BelluzziTrujilloBueno12}.  
The main difference between the redistribution matrices for the reference PRD calculations and those given in Eq.~\eqref{eqapp::RedisModBR} is found in the branching ratios of the terms that quantify the interference between different states of the upper level, which contain an imaginary quantity that is proportional to the energy difference of such states.

We now quantitatively evaluate the error introduced in the RT calculations following the approach of \citet{BelluzziTrujilloBueno12,BelluzziTrujilloBueno14}, by comparing with the reference PRD solutions. 
The intensity profiles do not change appreciably, and thus are not shown. However, we find a clear difference in the $Q/I$ profiles, as shown in Fig.~\ref{Fig::Fig6_BR}. This is especially apparent in the far wings of the lines, in which the amplitude of the scattering polarization signals is strongly enhanced by the quantum interference between the upper FS levels of the atomic system.  
As such, these far-wing signals are the most strongly impacted by the branching ratios of the interference terms. Within the narrower spectral interval that represents the range covered by the spectrograph of CLASP2, the two calculations present a much better agreement. Indeed, slight errors are appreciable only close to the blue and red boundaries of the interval and in the negative trough between the two lines, where $J$-state interference is known to play a key role.  

 \section[Sect4]{Conclusions}
 \label{sec::Conclusions} 
In this work, we have presented the numerical approach that we have applied to develop a non-LTE RT code for modeling the intensity and polarization of resonance lines. It is based on the theoretical framework presented in \cite{Bommier17,Bommier18} and represents an extension to the numerical approach of \cite{AlsinaBallester+17}. 
It is suitable to investigate spectral lines arising from a two-term atomic system including HFS, and it allows one to jointly account for the effects of PRD, $J$-state interference,
 and the presence of magnetic fields of an arbitrary strength and orientation. An optimized version of this code will be made publicly available in the near future.  

We applied this RT code to the Mg~{\sc{ii}} doublet and the H~{\sc{i}} Lyman-$\alpha$ line, which are strong resonance lines of interest for diagnostics of magnetic fields in the outer solar atmosphere. Through such RT calculations, we analyzed the 
reliability of a number of approximations for modeling the wings of the intensity and linear polarization profiles of strong resonance lines at a reduced computational cost, which would be particularly useful for the numerically intensive case of 3D radiative transfer. 
The suitability of such approximations for the Mg~{\sc{ii}} doublet is studied in the main body of the paper, and the details of the analogous investigation applied to the stronger H~{\sc{i}} Lyman-$\alpha$ line can be found in Appendix~\ref{sec::AppLymanAlpha}. The approximations we considered are the following. 

 \begin{itemize}
\item \textbf{Doppler redistribution is neglected in coherent scattering processes} and the approximate redistribution matrix $\hat{\mathbf{\mathcal{R}}_{\mbox{\sc{ii}}}}$ is used. 
In the Mg~{\sc{ii}} $h$ \& $k$ lines, this approximation is found to be suitable for modeling the intensity and linear polarization wings at spectral distances larger than a Doppler width from their centers. For the stronger H~{\sc{i}} Lyman-$\alpha$ line, we find that this approximation is suitable outside the core region, but it only yields a perfect agreement in the far wings.  
 
\item \textbf{Only the intensity component of the emission vector receives a contribution from} $\bm{\mathcal{R}}_{\mbox{\sc{iii}}}$, in which all components other than $K = K^\prime = Q = 0$ are neglected. 
We verified that the other components of $\bm{\mathcal R}_{\mbox{\sc{iii}}}$ (i.e., the matrix that quantifies the scattering processes perturbed by elastic collisions) do not appreciably contribute to either the intensity or the scattering polarization profiles of the considered lines, which indicates that this is a very good approximation. Relatedly, even when accounting for all components of $\bm{\mathcal R}_{\mbox{\sc{iii}}}$, the same profiles are found to be insensitive to the depolarizing effect of collisions. 
 
\item \textbf{The redistribution due to {both} the Doppler effect and elastic collisions is neglected}. 
This approximation builds upon the one listed above for $\bm{\hat{\mathcal R}}_{\mbox{\sc{ii}}}$, but making the stronger assumption that all scattering processes are coherent in the observer's frame, which allows for a further decrease in the computational cost.  
The $I$ and $Q/I$ profiles obtained in the wings of the Mg~{\sc{ii}} $h$ \& $k$ lines represent a rough yet suitable approximation of the reference PRD calculations, provided that the contribution from elastic collisions is included in the damping constant. 
This approximation is not suitable for reproducing the wings of the H~{\sc{i}} Lyman-$\alpha$ line because it cannot account for the influence of the strong emission peak in the line core via redistribution effects. 
Moreover, accounting for elastic collisions in the damping constant, while neglecting them in the branching ratios of the redistribution matrices, could in principle produce an artificial magnetic sensitivity in the emission coefficient, which would be appreciable in the wing 
linear polarization pattern. 
 
\item \textbf{The magnetic splitting is neglected in the computation of the emissivity}. 
This allows for a significant reduction in the computational cost of the overall problem. By accounting for the magnetic splitting in only the $\eta_I$ and $\rho_V$ coefficients of the propagation matrix, the scattering polarization wing profiles can be reproduced remarkably well, although we find that additionally accounting for the $\eta_V$ dichroism term introduces some spurious magnetic sensitivity, especially in the H~{\sc{i}} Lyman-$\alpha$ line. This approximation can be used in combination with the others studied in this work, allowing for a further reduction in the computational cost. Indeed, this approximation allows one to avoid the artificial magnetic sensitivity arising in the emission coefficient discussed in the previous point.  

\end{itemize}

 The numerical approach and the resulting RT code presented in this article allow for valuable investigations regarding the polarization of other solar spectral lines of diagnostic interest. 
 For instance, it was already used to account for all the relevant physics required to model the enigmatic linear polarization observed in the core of the sodium D${}_1$ line, which provides a satisfactory {explanation for the appearance} of this signal in the magnetized solar chromosphere \citep{AlsinaBallester+21}.
 Another ongoing investigation studies which physical processes are relevant in shaping the scattering polarization signals of the K~{\sc{i}} D${}_1$ (at $7698.96$~\AA) and D${}_2$ (at $7664.90$~\AA) lines, which the third flight of the SUNRISE mission aims to observe. 
 In another forthcoming publication, RT calculations using the same code will be carried out to further investigate the wing scattering polarization signals of the H~{{\sc i}} Lyman-$\alpha$ line by modeling their frequency-integrated signals and their magnetic sensitivity. These results will provide valuable insights regarding the filter-polarimetric observations obtained with the CLASP missions and, thereby, the 
inference of magnetic fields in the solar chromosphere.  

\begin{acknowledgements}
We thank Petr Heinzel (Astronomical Institute, ASCR) for carefully reading the manuscript and for constructive suggestions. 
We acknowledge the funding received from the Swiss National Science Foundation (SNSF) through Grant 200021\_175997 and from the European Research Council (ERC) under the European Union's Horizon 2020 research and innovation programme (ERC Advanced Grant agreement No. 742265). 
\end{acknowledgements}

\bibliographystyle{aa}
\bibliography{approx}

\begin{thebibliography}{77}
\expandafter\ifx\csname natexlab\endcsname\relax\def\natexlab#1{#1}\fi

\bibitem[{{Alsina Ballester} {et~al.}(2016){Alsina Ballester}, {Belluzzi}, \&
  {Trujillo Bueno}}]{AlsinaBallester+16}
{Alsina Ballester}, E., {Belluzzi}, L., \& {Trujillo Bueno}, J. 2016, \apjl,
  831, L15

\bibitem[{{Alsina Ballester} {et~al.}(2017){Alsina Ballester}, {Belluzzi}, \&
  {Trujillo Bueno}}]{AlsinaBallester+17}
{Alsina Ballester}, E., {Belluzzi}, L., \& {Trujillo Bueno}, J. 2017, \apj,
  836, 6

\bibitem[{{Alsina Ballester} {et~al.}(2018){Alsina Ballester}, {Belluzzi}, \&
  {Trujillo Bueno}}]{AlsinaBallester+18}
{Alsina Ballester}, E., {Belluzzi}, L., \& {Trujillo Bueno}, J. 2018, \apj,
  854, 150

\bibitem[{{Alsina Ballester} {et~al.}(2019){Alsina Ballester}, {Belluzzi}, \&
  {Trujillo Bueno}}]{AlsinaBallester+19}
{Alsina Ballester}, E., {Belluzzi}, L., \& {Trujillo Bueno}, J. 2019, \apj,
  880, 85

\bibitem[{{Alsina Ballester} {et~al.}(2021){Alsina Ballester}, {Belluzzi}, \&
  {Trujillo Bueno}}]{AlsinaBallester+21}
{Alsina Ballester}, E., {Belluzzi}, L., \& {Trujillo Bueno}, J. 2021, \prl,
  127, 081101

\bibitem[{{Anusha} {et~al.}(2009){Anusha}, {Nagendra}, {Paletou}, \&
  {L{\'e}ger}}]{Anusha+09}
{Anusha}, L.~S., {Nagendra}, K.~N., {Paletou}, F., \& {L{\'e}ger}, L. 2009,
  \apj, 704, 661

\bibitem[{{Belluzzi} {et~al.}(2013){Belluzzi}, {Landi Degl'Innocenti}, \&
  {Trujillo Bueno}}]{Belluzzi+13}
{Belluzzi}, L., {Landi Degl'Innocenti}, E., \& {Trujillo Bueno}, J. 2013, \aap,
  551, A84

\bibitem[{{Belluzzi} \& {Trujillo Bueno}(2011)}]{BelluzziTrujilloBueno11}
{Belluzzi}, L. \& {Trujillo Bueno}, J. 2011, \apj, 743, 3

\bibitem[{{Belluzzi} \& {Trujillo Bueno}(2012)}]{BelluzziTrujilloBueno12}
{Belluzzi}, L. \& {Trujillo Bueno}, J. 2012, \apjl, 750, L11

\bibitem[{{Belluzzi} \& {Trujillo Bueno}(2014)}]{BelluzziTrujilloBueno14}
{Belluzzi}, L. \& {Trujillo Bueno}, J. 2014, \aap, 564, A16

\bibitem[{{Belluzzi} {et~al.}(2012){Belluzzi}, {Trujillo Bueno}, \&
  {{\v{S}}t{\v{e}}p{\'a}n}}]{Belluzzi+12}
{Belluzzi}, L., {Trujillo Bueno}, J., \& {{\v{S}}t{\v{e}}p{\'a}n}, J. 2012,
  \apjl, 755, L2

\bibitem[{{Benedusi} {et~al.}(2021){Benedusi}, {Janett}, {Belluzzi}, \&
  {Krause}}]{Benedusi+21}
{Benedusi}, P., {Janett}, G., {Belluzzi}, L., \& {Krause}, R. 2021, \aap, 655,
  A88

\bibitem[{{Bommier}(1997{\natexlab{a}})}]{Bommier97a}
{Bommier}, V. 1997{\natexlab{a}}, \aap, 328, 706

\bibitem[{{Bommier}(1997{\natexlab{b}})}]{Bommier97b}
{Bommier}, V. 1997{\natexlab{b}}, \aap, 328, 726

\bibitem[{{Bommier}(2017)}]{Bommier17}
{Bommier}, V. 2017, \aap, 607, A50

\bibitem[{{Bommier}(2018)}]{Bommier18}
{Bommier}, V. 2018, \aap, 619, C1

\bibitem[{{Cannon}(1973)}]{Cannon73}
{Cannon}, C.~J. 1973, \apj, 185, 621

\bibitem[{{Casini} {et~al.}(2017{\natexlab{a}}){Casini}, {del Pino Alem{\'a}n},
  \& {Manso Sainz}}]{Casini+17a}
{Casini}, R., {del Pino Alem{\'a}n}, T., \& {Manso Sainz}, R.
  2017{\natexlab{a}}, \apj, 835, 114

\bibitem[{{Casini} {et~al.}(2017{\natexlab{b}}){Casini}, {del Pino Alem{\'a}n},
  \& {Manso Sainz}}]{Casini+17b}
{Casini}, R., {del Pino Alem{\'a}n}, T., \& {Manso Sainz}, R.
  2017{\natexlab{b}}, \apj, 848, 99

\bibitem[{{Casini} \& {Landi
  Degl'Innocenti}(2008)}]{CasiniLandiDeglInnocenti08}
{Casini}, R. \& {Landi Degl'Innocenti}, E. 2008, in Plasma Polarization
  Spectroscopy, ed. T.~{Fujimoto} \& A.~{Iwamae}, Vol.~44 (Berlin: Springer),
  247

\bibitem[{{Casini} {et~al.}(2014){Casini}, {Landi Degl'Innocenti}, {Manso
  Sainz}, {Landi Degl'Innocenti}, \& {Landolfi}}]{Casini+14}
{Casini}, R., {Landi Degl'Innocenti}, M., {Manso Sainz}, R., {Landi
  Degl'Innocenti}, E., \& {Landolfi}, M. 2014, \apj, 791, 94

\bibitem[{{Cooper} {et~al.}(1989){Cooper}, {Ballagh}, \& {Hubeny}}]{Cooper+89}
{Cooper}, J., {Ballagh}, R.~J., \& {Hubeny}, I. 1989, \apj, 344, 949

\bibitem[{{de la Cruz Rodr{\'\i}guez} {et~al.}(2019){de la Cruz
  Rodr{\'\i}guez}, {Leenaarts}, {Danilovic}, \& {Uitenbroek}}]{delaCruz+19}
{de la Cruz Rodr{\'\i}guez}, J., {Leenaarts}, J., {Danilovic}, S., \&
  {Uitenbroek}, H. 2019, \aap, 623, A74

\bibitem[{{del Pino Alem{\'a}n} {et~al.}(2016){del Pino Alem{\'a}n}, {Casini},
  \& {Manso Sainz}}]{delPinoAleman+16}
{del Pino Alem{\'a}n}, T., {Casini}, R., \& {Manso Sainz}, R. 2016, \apjl, 830,
  L24

\bibitem[{{del Pino Alem{\'a}n} {et~al.}(2020){del Pino Alem{\'a}n}, {Trujillo
  Bueno}, {Casini}, \& {Manso Sainz}}]{delPinoAleman+20}
{del Pino Alem{\'a}n}, T., {Trujillo Bueno}, J., {Casini}, R., \& {Manso
  Sainz}, R. 2020, \apj, 891, 91

\bibitem[{{del Pino Alem{\'a}n} {et~al.}(2018){del Pino Alem{\'a}n}, {Trujillo
  Bueno}, {{\v{S}}t{\v{e}}p{\'a}n}, \& {Shchukina}}]{delPinoAleman+18}
{del Pino Alem{\'a}n}, T., {Trujillo Bueno}, J., {{\v{S}}t{\v{e}}p{\'a}n}, J.,
  \& {Shchukina}, N. 2018, \apj, 863, 164

\bibitem[{{Domke} \& {Hubeny}(1988)}]{DomkeHubeny88}
{Domke}, H. \& {Hubeny}, I. 1988, \apj, 334, 527

\bibitem[{{Dumont}(1967)}]{Dumont67}
{Dumont}, S. 1967, Annales d'Astrophysique, 30, 861

\bibitem[{{Faurobert-Scholl} {et~al.}(1997){Faurobert-Scholl}, {Frisch}, \&
  {Nagendra}}]{Faurobert+97}
{Faurobert-Scholl}, M., {Frisch}, H., \& {Nagendra}, K.~N. 1997, \aap, 322, 896

\bibitem[{{Fontenla} {et~al.}(1991){Fontenla}, {Avrett}, \&
  {Loeser}}]{Fontenla+91}
{Fontenla}, J.~M., {Avrett}, E.~H., \& {Loeser}, R. 1991, \apj, 377, 712

\bibitem[{{Fontenla} {et~al.}(1993){Fontenla}, {Avrett}, \&
  {Loeser}}]{Fontenla+93}
{Fontenla}, J.~M., {Avrett}, E.~H., \& {Loeser}, R. 1993, \apj, 406, 319

\bibitem[{{Grevesse} \& {Anders}(1991)}]{GrevesseAnders91}
{Grevesse}, N. \& {Anders}, E. 1991, {Solar element abundances} (Tucson:
  University of Arizona Press), 1227--1234

\bibitem[{{Hubeny}(1992)}]{Hubeny92}
{Hubeny}, I. 1992, in The Atmospheres of Early-Type Stars, ed. U.~{Heber} \&
  C.~S. {Jeffery}, Vol. 401, 377

\bibitem[{{Hubeny} \& {Lites}(1995)}]{HubenyLites95}
{Hubeny}, I. \& {Lites}, B.~W. 1995, \apj, 455, 376

\bibitem[{{Hubeny} \& {Mihalas}({2015})}]{BHubenyMihalas15}
{Hubeny}, I. \& {Mihalas}, D. {2015}, {Theory of Stellar Atmospheres: An
  Introduction to Astrophysical Non-equilibrium Quantitative Spectroscopic
  Analysis} ({Princeton}: {Princeton University Press})

\bibitem[{{Hummer}(1962)}]{Hummer62}
{Hummer}, D.~G. 1962, \mnras, 125, 21

\bibitem[{{Ishikawa} {et~al.}(2021){Ishikawa}, {Trujillo Bueno}, {del Pino
  Aleman}, {Okamoto}, {McKenzie}, {Auchere}, {Kano}, {Song}, {Yoshida},
  {Rachmeler}, {Kobayashi}, {Hara}, {Kubo}, {Narukage}, {Sakao}, {Shimizu},
  {Suematsu}, {Bethge}, {De Pontieu}, {Sainz Dalda}, {Vigil}, {Winebarger},
  {Alsina Ballester}, {Belluzzi}, {Stepan}, {Asensio Ramos}, {Carlsson}, \&
  {Leenaarts}}]{Ishikawa+21}
{Ishikawa}, R., {Trujillo Bueno}, J., {del Pino Aleman}, T., {et~al.} 2021,
  Science Adv., 7, eabe8406

\bibitem[{{Janett} {et~al.}(2017){Janett}, {Carlin}, {Steiner}, \&
  {Belluzzi}}]{Janett+17}
{Janett}, G., {Carlin}, E.~S., {Steiner}, O., \& {Belluzzi}, L. 2017, \apj,
  840, 107

\bibitem[{{Janett} \& {Paganini}(2018)}]{JanettPaganini18}
{Janett}, G. \& {Paganini}, A. 2018, \apj, 857, 91

\bibitem[{{Kano} {et~al.}(2017){Kano}, {Trujillo Bueno}, {Winebarger},
  {Auch{\`e}re}, {Narukage}, {Ishikawa}, {Kobayashi}, {Bando}, {Katsukawa},
  {Kubo}, {Ishikawa}, {Giono}, {Hara}, {Suematsu}, {Shimizu}, {Sakao},
  {Tsuneta}, {Ichimoto}, {Goto}, {Belluzzi}, {{\v{S}}t{\v{e}}p{\'a}n}, {Asensio
  Ramos}, {Manso Sainz}, {Champey}, {Cirtain}, {De Pontieu}, {Casini}, \&
  {Carlsson}}]{Kano+17}
{Kano}, R., {Trujillo Bueno}, J., {Winebarger}, A., {et~al.} 2017, \apjl, 839,
  L10

\bibitem[{{Kramida} {et~al.}(2021){Kramida}, {Ralchenko}, {Reader}, \& {NIST
  ASD Team}}]{NIST_ASD}
{Kramida}, A., {Ralchenko}, Y., {Reader}, J., \& {NIST ASD Team}. 2021, {NIST
  Atomic Spectra Database (version 5.9), Gaithersburg, MD, National Institute
  of Standards and Technology [Online]. Available:
  {\url{https://physics.nist.gov/}} }

\bibitem[{{Landi Degl'Innocenti} {et~al.}(1997){Landi Degl'Innocenti}, {Landi
  Degl'Innocenti}, \& {Landolfi}}]{LandiDeglInnocenti+97}
{Landi Degl'Innocenti}, E., {Landi Degl'Innocenti}, M., \& {Landolfi}, M. 1997,
  in THEMIS Forum: Science with THEMIS, ed. N.~{Mein} \& S.~{Sahal-Br\'{e}chot}
  (Paris: Observatoire de Paris-Meudon), 59

\bibitem[{{Landi Degl'Innocenti} \& {Landolfi}(2004)}]{BLandiLandolfi04}
{Landi Degl'Innocenti}, E. \& {Landolfi}, M. 2004, {Polarization in Spectral
  Lines}, Vol. 307 (Dordrecht: Kluwer)

\bibitem[{{Leenaarts} {et~al.}(2013){Leenaarts}, {Pereira}, {Carlsson},
  {Uitenbroek}, \& {De Pontieu}}]{Leenaarts+13}
{Leenaarts}, J., {Pereira}, T.~M.~D., {Carlsson}, M., {Uitenbroek}, H., \& {De
  Pontieu}, B. 2013, \apj, 772, 89

\bibitem[{{Manso Sainz} {et~al.}(2014){Manso Sainz}, {Roncero}, {Sanz-Sanz},
  {Aguado}, {Asensio Ramos}, \& {Trujillo Bueno}}]{MansoSainz+14}
{Manso Sainz}, R., {Roncero}, O., {Sanz-Sanz}, C., {et~al.} 2014, \apj, 788,
  118

\bibitem[{{Milkey} \& {Mihalas}(1973{\natexlab{a}})}]{MilkeyMihalas73b}
{Milkey}, R.~W. \& {Mihalas}, D. 1973{\natexlab{a}}, \solphys, 32, 361

\bibitem[{{Milkey} \& {Mihalas}(1973{\natexlab{b}})}]{MilkeyMihalas73}
{Milkey}, R.~W. \& {Mihalas}, D. 1973{\natexlab{b}}, \apj, 185, 709

\bibitem[{{Milkey} \& {Mihalas}(1974)}]{MilkeyMihalas74}
{Milkey}, R.~W. \& {Mihalas}, D. 1974, \apj, 192, 769

\bibitem[{{Nagendra}(2003)}]{Nagendra03}
{Nagendra}, K.~N. 2003, in Astronomical Society of the Pacific Conference
  Series, Vol. 288, Stellar Atmosphere Modeling, ed. I.~{Hubeny}, D.~{Mihalas},
  \& K.~{Werner}, 583

\bibitem[{{Nagendra} {et~al.}(1998){Nagendra}, {Frisch}, \&
  {Faurobert-Scholl}}]{Nagendra+98}
{Nagendra}, K.~N., {Frisch}, H., \& {Faurobert-Scholl}, M. 1998, \aap, 332, 610

\bibitem[{{Olson} {et~al.}(1986){Olson}, {Auer}, \& {Buchler}}]{Olson+86}
{Olson}, G.~L., {Auer}, L.~H., \& {Buchler}, J.~R. 1986, \jqsrt, 35, 431

\bibitem[{{Paganini} {et~al.}(2021){Paganini}, {Hashemi}, {Alsina Ballester},
  \& {Belluzzi}}]{Paganini+21}
{Paganini}, A., {Hashemi}, B., {Alsina Ballester}, E., \& {Belluzzi}, L. 2021,
  \aap, 645, A4

\bibitem[{{Paletou} \& {Anterrieu}(2009)}]{PaletouAnterrieu09}
{Paletou}, F. \& {Anterrieu}, E. 2009, \aap, 507, 1815

\bibitem[{{Pereira} {et~al.}(2015){Pereira}, {Carlsson}, {De Pontieu}, \&
  {Hansteen}}]{Pereira+15}
{Pereira}, T. M.~D., {Carlsson}, M., {De Pontieu}, B., \& {Hansteen}, V. 2015,
  \apj, 806, 14

\bibitem[{{Przybilla} \& {Butler}(2004)}]{PrzybillaButler04}
{Przybilla}, N. \& {Butler}, K. 2004, \apj, 609, 1181

\bibitem[{{Racah}(1942)}]{Racah42}
{Racah}, G. 1942, Physical Review, 62, 438

\bibitem[{{Rees} \& {Saliba}(1982)}]{ReesSaliba82}
{Rees}, D.~E. \& {Saliba}, G.~J. 1982, \aap, 115, 1

\bibitem[{{Sampoorna} {et~al.}(2017){Sampoorna}, {Nagendra}, \&
  {Stenflo}}]{Sampoorna+17}
{Sampoorna}, M., {Nagendra}, K.~N., \& {Stenflo}, J.~O. 2017, \apj, 844, 97

\bibitem[{{Sigut} \& {Pradhan}(1995)}]{SigutPradhan95}
{Sigut}, T.~A.~A. \& {Pradhan}, A.~K. 1995, Journal of Physics B Atomic
  Molecular Physics, 28, 4879

\bibitem[{{Sowmya} {et~al.}(2015){Sowmya}, {Nagendra}, {Sampoorna}, \&
  {Stenflo}}]{Sowmya+15}
{Sowmya}, K., {Nagendra}, K.~N., {Sampoorna}, M., \& {Stenflo}, J.~O. 2015,
  \apj, 814, 127

\bibitem[{{Sowmya} {et~al.}(2014){Sowmya}, {Nagendra}, {Stenflo}, \&
  {Sampoorna}}]{Sowmya+14}
{Sowmya}, K., {Nagendra}, K.~N., {Stenflo}, J.~O., \& {Sampoorna}, M. 2014,
  \apj, 786, 150

\bibitem[{{Stenflo}(1994)}]{BStenflo94}
{Stenflo}, J. 1994, {Solar Magnetic Fields: Polarized Radiation Diagnostics},
  Vol. 189 (Dordrecht: Springer)

\bibitem[{{Sutton}(1978)}]{Sutton78}
{Sutton}, K. 1978, \jqsrt, 20, 333

\bibitem[{{Thomas}(1957)}]{Thomas57}
{Thomas}, R.~N. 1957, \apj, 125, 260

\bibitem[{Traving(1960)}]{BTraving60}
Traving, G. 1960, {\"U}ber die Theorie der Druckverbreiterung von
  Spektrallinien (Hamburg: Astronomische Gesellschaft)

\bibitem[{{Trujillo Bueno}(2001)}]{TrujilloBueno01}
{Trujillo Bueno}, J. 2001, in Astronomical Society of the Pacific Conference
  Series, Vol. 236, Advanced Solar Polarimetry -- Theory, Observation, and
  Instrumentation, ed. M.~{Sigwarth}, 161

\bibitem[{{Trujillo Bueno}(2003)}]{TrujilloBueno03}
{Trujillo Bueno}, J. 2003, in Astronomical Society of the Pacific Conference
  Series, Vol. 288, Stellar Atmosphere Modeling, ed. I.~{Hubeny}, D.~{Mihalas},
  \& K.~{Werner} (San Francisco: Astronomical Society of the Pacific), 551

\bibitem[{{Trujillo Bueno} \& {Fabiani
  Bendicho}(1995)}]{TrujilloBuenoFabianiBendicho95}
{Trujillo Bueno}, J. \& {Fabiani Bendicho}, P. 1995, \apj, 455, 646

\bibitem[{{Trujillo Bueno} {et~al.}(2017){Trujillo Bueno}, {Landi
  Degl'Innocenti}, \& {Belluzzi}}]{TrujilloBueno+17}
{Trujillo Bueno}, J., {Landi Degl'Innocenti}, E., \& {Belluzzi}, L. 2017, \ssr,
  210, 183

\bibitem[{{Trujillo Bueno} \& {Manso Sainz}(1999)}]{TrujilloBuenoMansoSainz99}
{Trujillo Bueno}, J. \& {Manso Sainz}, R. 1999, \apj, 516, 436

\bibitem[{{Trujillo Bueno} \& {Shchukina}(2009)}]{TrujilloBuenoShchukina09}
{Trujillo Bueno}, J. \& {Shchukina}, N. 2009, \apj, 694, 1364

\bibitem[{{Trujillo Bueno} {et~al.}(2012){Trujillo Bueno},
  {{\v{S}}t{\v{e}}p{\'a}n}, \& {Belluzzi}}]{TrujilloBueno+12}
{Trujillo Bueno}, J., {{\v{S}}t{\v{e}}p{\'a}n}, J., \& {Belluzzi}, L. 2012,
  \apjl, 746, L9

\bibitem[{{Trujillo Bueno} {et~al.}(2018){Trujillo Bueno},
  {{\v{S}}t{\v{e}}p{\'a}n}, {Belluzzi}, {Asensio Ramos}, {Manso Sainz}, {del
  Pino Alem{\'a}n}, {Casini}, {Ishikawa}, {Kano}, {Winebarger}, {Auch{\`e}re},
  {Narukage}, {Kobayashi}, {Bando}, {Katsukawa}, {Kubo}, {Ishikawa}, {Giono},
  {Hara}, {Suematsu}, {Shimizu}, {Sakao}, {Tsuneta}, {Ichimoto}, {Cirtain},
  {Champey}, {De Pontieu}, \& {Carlsson}}]{TrujilloBueno+18}
{Trujillo Bueno}, J., {{\v{S}}t{\v{e}}p{\'a}n}, J., {Belluzzi}, L., {et~al.}
  2018, \apjl, 866, L15

\bibitem[{{Trujillo Bueno} {et~al.}(2011){Trujillo Bueno},
  {{\v{S}}t{\v{e}}p{\'a}n}, \& {Casini}}]{TrujilloBueno+11}
{Trujillo Bueno}, J., {{\v{S}}t{\v{e}}p{\'a}n}, J., \& {Casini}, R. 2011,
  \apjl, 738, L11

\bibitem[{{Uitenbroek}(2001)}]{Uitenbroek01}
{Uitenbroek}, H. 2001, \apj, 557, 389

\bibitem[{{Uns\"old}(1955)}]{Unsold55}
{Uns\"old}, A. 1955, {Physik der Sternatmosph\"aren, mit besonderer
  Ber\"ucksichtigung der Sonne.} ({Berlin}: {Springer})

\bibitem[{{Vernazza} {et~al.}(1973){Vernazza}, {Avrett}, \&
  {Loeser}}]{Vernazza+73}
{Vernazza}, J.~E., {Avrett}, E.~H., \& {Loeser}, R. 1973, \apj, 184, 605

\end{thebibliography}

\onecolumn
\appendix 
\section{The atomic model for the Mg~{\sc{ii}} $h$ \& $k$ lines}
\label{sec::AppAtomMod}
\begin{table*}[!h]
 \centering
 \caption{\label{tab:Table1} Atomic parameters for the Mg~{\sc{ii}} $h$ \& $k$ lines}
 \bgroup
\def\arraystretch{1.3}
 \begin{tabular}{c c c c c } \hline \hline
  Upper state & $E_u$(cm${}^{-1}$) & Lower state & $E_\ell$(cm${}^{-1}$) & $A_{u \ell}$ (s${}^{-1}$) \\ \hline 
  $2p^6 3p \; {}^{2}$P${}_{1/2}$ & $35669.31$ & $2p^6 3s \; {}^{2}$S${}_{1/2}$ & $0.00$ & $2.57 \cdot 10^8$ \\ \hline
  $2p^6 3p \; {}^{2}$P${}_{3/2}$ & $35760.88$ & $2p^6 3s \; {}^{2}$S${}_{1/2}$ & $0.00$ & $2.60 \cdot 10^8$ \\ \hline
 \end{tabular}
 \egroup
\end{table*}

The RT calculations for the Mg~{\sc ii} doublet presented in this work were carried out numerically following the two-step approach described in Sect.~\ref{sec::sIterative}. 
In step 1, which relies on the use of the RH code, we consider a relatively simple multi-level model for the Mg~{\sc ii} $h$ \& $k$ lines that consists of 4 levels, namely the ground level of Mg~{\sc{ii}} (which corresponds to the 3$s\,{}^2\mathrm{S}$ term) and the upper levels of the $h$ \& $k$ resonance lines (which belong to the 3$p\,{}^2\mathrm{P}$ term), as well as the ground level of Mg~{\sc{iii}}. The $h$ \& $k$ lines are the only two line transitions considered by this model, for which PRD effects are taken into account \citep[see][]{Uitenbroek01}. In addition, this model includes 3 continuum transitions. 
The results of \cite{Leenaarts+13} support the suitability of this model to calculate the ground level population of Mg~{\sc{ii}}, which is the only population that is required as an input for step 2. 

In step 2 the polarization of radiation and the magnetic field are taken into account. In this step the calculations are carried out using the numerical code described in Sect.~\ref{sec::formulation}. A two-term atomic model with spin $S = 1/2$ is considered, in which the upper term has orbital angular momentum quantum number $L = 1$ and the lower term has $L = 0$. 
The lower term corresponds to the ground level of Mg~{\sc{ii}}. Due to their long lifetime, the states pertaining to this term can be treated as infinitely sharp and, because of the impact of collisions on such long-lived states, the atomic polarization of the term can be safely neglected.\footnote{The only fine structure level of the lower term has total angular momentum $J = 1/2$, and therefore cannot carry atomic alignment by definition. 
We recall that a given fine structure level is said to be oriented when there are population imbalances between two given sublevels with magnetic quantum numbers $M$ and $-M$ and aligned when population imbalances exist between sublevels with different $|M|$. 
Atomic orientation and alignment contribute, respectively, to the circular and linear polarization signals produced via scattering processes.} 

Two-term atoms were considered in prior works to model the intensity and polarization of the Mg~{\sc{ii}} resonance lines \citep[see][]{BelluzziTrujilloBueno12}. 
In the past it was suggested that the inclusion of the 3$d\,{}^2\mathrm{D}$ term in the atomic model could have an impact on the intensity profile of the $h$ \& $k$ lines if a PRD treatment was made for the subordinate lines that couple this term to 3$p\,{}^2\mathrm{P}$ \citep[see][]{MilkeyMihalas74}, although it was already established that this was not the case when the subordinate lines were modeled under CRD \citep{Dumont67}. Subsequent RT calculations by \cite{Pereira+15} confirmed that the same intensity profiles are obtained for the $h$ \& $k$ lines whether PRD or CRD is considered for the subordinate lines. 
This makes a compelling case for the suitability of modeling the Mg~{\sc{ii}} $h$ \& $k$ lines with a two-term atomic system.  
Moreover, a comparison between the calculations for intensity and polarization considering a two-term atomic model \citep[see][]{delPinoAleman+16} and those considering a three-term model \citep[see][]{delPinoAleman+20} suggests that the $h$ \& $k$ lines can be well modeled by considering a two-term system, also when accounting for polarization. However, it must be noted that a CRD treatment was made for the subordinate lines in the latter calculations. 
Clearly, a definitive confirmation of the suitability of a two-term modeling will require a formalism through which polarization and PRD effects can be taken into account in all the lines of a cascade-type three-term system 
such as the one responsible for the resonance and subordinate lines \citep[see][]{delPinoAleman+20}, which is not yet available.  

Relevant parameters for the atomic model considered in step 2 include the upper and lower fine structure (FS) levels, their respective energies $E$ relative to the ground state, and the Einstein coefficient for spontaneous emission from upper level $\ket{\beta_u L_u S J_u}$ to lower level $\ket{\beta_\ell L_\ell S J_\ell}$, given as $A_{u \ell} = A(\beta_u L_u S J_u \to \beta_\ell L_\ell S J_\ell)$. 
For all the calculations considered in the present work, the radiative line broadening parameter $\Gamma_R$ is taken to be equal to $A_{u \ell}$, whose experimental values are given in \cite{NIST_ASD}. These quantities are given in Table~\ref{tab:Table1}.  

The collisional line broadening constants for elastic ($\Gamma_E$) and inelastic ($\Gamma_I$) collisions, which are required as input quantities for step 2, enter the branching ratios of the redistribution matrices and the expression of the damping constant $a$ (see Appendix.~\ref{sec::AppBroad}). In this work, $\Gamma_E$ is obtained accounting only for the contributions due to Van der Waals interaction and the Stark effect. $\Gamma_I$ is taken to be equal to the rate of inelastic collisions responsible for the atomic transitions from states of the upper term to those of the lower term, $C_{u \ell}$. For the Mg~{\sc{ii}} doublet, these quantities are obtained in step 1 directly from the execution of RH. 
The collisional strengths required to compute $C_{u \ell}$ are obtained following \cite{SigutPradhan95}. The Van der Waals contribution to $\Gamma_E$ due to neutral hydrogen and helium atoms is obtained making use of Uns\"{o}ld's approximation \citep{Unsold55}. The quadratic Stark effect contribution, due to electrons and singly charged ions, is taken into account \citep{BTraving60}. 
The chemical abundance of magnesium relative to hydrogen is given by 
\begin{equation}
    A(\mbox{Mg}) = 12 + \log_{10}\Bigl(n(\mbox{Mg})/n(\mbox{H})\Bigr) \, ,
\end{equation}
where $n(\mbox{Mg})$ and $n(\mbox{H})$ are the populations of magnesium and hydrogen, respectively. Throughout this work, we have taken $A(\mbox{Mg}) = 7.58$ \citep[see][]{GrevesseAnders91}. 

\section{Applications to the H~{\sc{i}} Lyman-$\alpha$ line}
\label{sec::AppLymanAlpha}
\begin{table*}[!h]
 \centering
 \caption{\label{tab:Table2} Atomic parameters for the FS components of the H~{\sc{i}} Lyman-$\alpha$ line}
  \bgroup
\def\arraystretch{1.3}
 \begin{tabular}{c c c c c} \hline \hline
  Upper state & $E_u$(cm${}^{-1}$) & Lower state & $E_\ell$(cm${}^{-1}$) & $A_{u \ell}$ (s${}^{-1}$) \\ \hline 
   $2p \; {}^{2}$P${}_{1/2}$ & $82258.92$  & $1s \; {}^{2}$S${}_{1/2}$ & $0.00$ & $6.26 \cdot 10^8$ \\ \hline
  $2p \; {}^{2}$P${}_{3/2}$ & $82259.29$  & $1s \; {}^{2}$S${}_{1/2}$ & $0.00$ & $6.26 \cdot 10^8$ \\ \hline 
 \end{tabular}
 \egroup
\end{table*}
Here we present an analogous investigation to the one presented in Sect.~\ref{sec::Illustrative} for the Mg~{\sc{ii}} $h$ \& $k$ resonance lines, but applied to the H~{\sc{i}} Lyman-$\alpha$ line instead. 
We recall that the Lyman-$\alpha$ line arises from the radiative transitions between the $n = 1$ and $n = 2$ levels of neutral hydrogen. 
In step 1 the calculations (see Sect.~\ref{sec::sIterative}) are carried out with RH, taking an atomic model with 6 levels in which FS is neglected, which includes 5 levels of H~{\sc{i}} with principal quantum numbers between $n = 1$ and $n = 5$ and a level corresponding to the ionized H~{\sc ii}. We account for 10 line transitions and 5 continuum transitions that involve these levels. The contributions from all transitions are computed under the CRD (or flat-spectrum) assumption, except for the one corresponding to Lyman-$\alpha$ itself, which receives a PRD treatment. The suitability of considering PRD only for the latter transition when determining the population of the ground state of hydrogen is supported by the results presented in \cite{HubenyLites95}. 

In step 2 of the problem (see Sect.~\ref{sec::sIterative}) we consider a two-term atomic system. 
Indeed, considering the fine structure of hydrogen and recalling that the 1$s\,{}^2\mathrm{S}$ and 2$s\,{}^2\mathrm{S}$ terms are prohibited by the electric dipole selection rules, we can treat the H~{\sc{i}} Lyman-$\alpha$ line as arising from the transitions between the 1$s\,{}^2\mathrm{S}$ and 2$p\,{}^2\mathrm{P}$ terms. The relevant atomic parameters are provided in Table~\ref{tab:Table2}, in which the energies of the FS levels and the  Einstein coefficients $A_{u \ell}$ were taken from \cite{NIST_ASD}. 
Unlike in the calculations for the Mg~{\sc{ii}} doublet, only the continuum quantities and the initial guess for the unpolarized radiation field are provided from step 1. 
The population of the lower level $N_\ell$ is taken to be equal to the population of the ground level of hydrogen provided by the atmospheric model \citep[see][]{Fontenla+93}. The  
$\Gamma_I$ line broadening constant is taken to be equal to the inelastic collision rates ${\mathcal{C}}_{u \ell}$, which are calculated according to the expressions given in \cite{PrzybillaButler04}. 
Regarding $\Gamma_E$, we consider two contributions. The first is the van der Waals contribution 
due to neutral hydrogen, which is calculated following Sect.~7.13 of LL04. 
The second is the linear (rather than quadratic) Stark broadening contribution, for which we use the formula presented in \cite{Sutton78}. 

We highlight two important differences with respect to the atomic parameters of the Mg~{\sc{ii}} doublet. 
First, the energy separation between the FS levels of the upper term of the H~{\sc{i}} Lyman-$\alpha$ line is considerably smaller and, as a result, its two FS components are blended. 
Second, the Einstein coefficients for spontaneous emission $A_{u \ell}$, which are proportional to the oscillator strength of the FS components of the line, are larger for the Lyman-$\alpha$ line by more than a factor two. Together with the far greater abundance of hydrogen, this implies that  the Lyman-$\alpha$ line is much stronger \citep[i.e., characterized by a stronger line opacity, see][]{MilkeyMihalas73} than the Mg~{\sc{ii}} $h$ or $k$ lines, and thus originates at more external atmospheric regions. 

We carried out RT calculations as detailed above to evaluate the suitability of the same approximations discussed in Sect.~\ref{sec::Illustrative}, but applied here to the H~{\sc{i}} Lyman-$\alpha$ line. 
The FAL-C atmospheric model is used in all calculations. The magnetic field, when present, is taken with the same strength and orientation at all heights. The field is horizontal ($\theta_B = 90^\circ$) with azimuth $\chi_B = 0^\circ$ (see Fig.\ref{Fig::Fig1}). However, here its strength is taken to be $50$~G, which characterizes the onset of the Hanle effect (see Sect.~10.3 of LL04) for this line. 
The profiles are shown for an LOS with $\mu = 0.1$ and taking the reference direction for positive Stokes $Q$ parallel to the limb. 
Vacuum wavelengths are considered in all the figures shown in this Appendix. 
The spectral interval considered in these figures has a width of $30$~\AA\ and is centered at the baricenter of the transitions contributing to this line. The baricenter falls at $1215.67$~\AA , which we hereafter refer to as the line center.

 \begin{figure*}[!h]
  \centering
  \includegraphics[width = 0.975\textwidth]{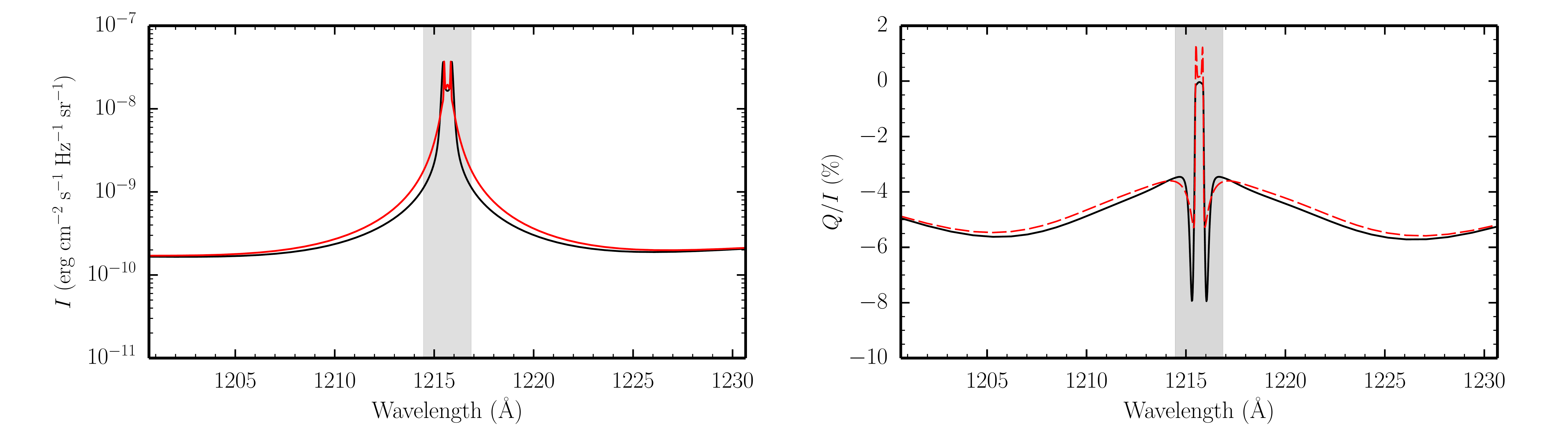}
        \caption{Stokes $I$ (\textit{left panel}) and $Q/I$ (\textit{right panel}) profiles as a function of vacuum wavelength for the H~{\sc{i}} Lyman-$\alpha$ line, obtained in the absence of a magnetic field. The solid black curves correspond to reference PRD calculations, whereas the red curves represent the results of calculations considering the approximate redistribution matrix $\bm{\hat{\mathcal R}}_{\mbox{\sc{ii}}}$ (see main text). In the right panel, the latter curve is dashed to enhance visibility. The shaded areas have a width of $2.4$~\AA\ and indicate the spectral region where the latter approximation is not expected to be suitable. For all the calculations presented in this Appendix, the FAL-C atmospheric model is considered. For the figures in this Appendix we consider an LOS with $\mu = 0.1$ and take the direction for positive Stokes $Q$ parallel to the limb.}  
        \label{Fig::App1} 
\end{figure*}
First we focus on the intensity and linear polarization ($Q/I$) profiles resulting from reference PRD calculations, represented by the black curves in Fig.~\ref{Fig::App1}. Unlike for the Mg~{\sc{ii}} doublet, the two FS components are clearly blended. In the intensity profile we find the most striking difference with respect to the Mg~{\sc{ii}} lines; for this line a much sharper emission profile is produced, with a value within the core region that is roughly two orders of magnitude larger than that in the far wings \citep[in qualitative agreement with the theoretical profiles in][]{MilkeyMihalas73}. We note that we obtained comparable emission profiles when carrying out calculations for the Mg~{\sc{ii}} doublet in which we artificially increased the abundance of magnesium (not shown in this work). This indicates that this remarkable emission feature is a consequence of the fact that the H~{\sc{i}} Lyman-$\alpha$ line is much stronger. The $Q/I$ profile obtained through the reference PRD calculation is negative throughout the considered spectral range, with very broad lobes with large amplitudes in the far wings, sharp peaks in the near wings, and a small amplitude close to the line center. 

\begin{figure*}[!h]
 \centering
 \includegraphics[width = 0.975\textwidth]{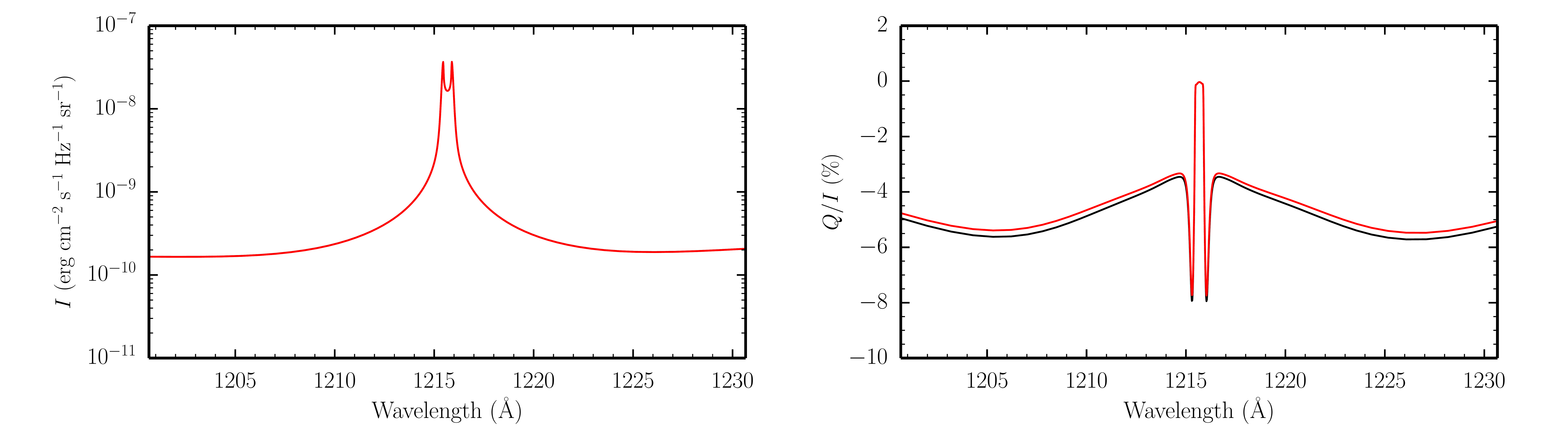}
 \caption{Stokes $I$ (\textit{left panel}) and $Q/I$ (\textit{right panel}) profiles as a function of wavelength obtained from PRD calculations in the absence of a magnetic field. The black and red curves represent the calculations obtained considering the redistribution matrices given in Eqs.~\eqref{eq::R2ObsnHFS} and~\eqref{eq::R3ObsnHFS} and those given in Eqs.~\eqref{eqapp::RedisModBR}, respectively. The latter calculation was considered in order to reproduce the results of \cite{Belluzzi+12}.} 
 \label{Fig::App2} 
\end{figure*} 
We verified that the intensity and linear polarization patterns found in the line wings, taking an LOS with $\mu = 0.3$, perfectly agree with the calculations neglecting FS presented in \cite{AlsinaBallester+19}; this is in accordance with the principle of spectroscopic stability (see Sect.~7.10 of LL04). We also found a perfect agreement with the line-core profiles presented in \cite{Belluzzi+12}. However, a slight discrepancy was found in the line-wing $Q/I$ pattern, in the region where $J$-state interference plays an important role, because in that work the authors made use of redistribution matrices in which the branching ratios and $\bm{\mathcal{R}}_{\mbox{\sc{iii}}}$ were obtained using heuristic arguments 
\citep[see][]{BelluzziTrujilloBueno14}. 
Indeed, we could reproduce these profiles through calculations using the redistribution matrices that correspond to the latter approach, given in Eqs.~\eqref{eqapp::RedisModBR}. A comparison between such calculations (red curves) and the reference PRD calculations (black curves) is shown in Fig.~\ref{Fig::App2} for an LOS with $\mu = 0.1$. 
An error is indeed incurred in the $Q/I$ wings, but it is not as large as in the case of the Mg~{\sc{ii}} $h$ \& $k$ lines (see Sect.~\ref{sec::BRs}) because of the smaller energy separation between FS levels, which thus has a smaller impact on the branching ratios for the interference terms. On the other hand, it can be seen from the figure that the intensity profile is not appreciably impacted by the changes in branching ratios.  

In addition to the reference PRD calculations, Fig.~\ref{Fig::App1} also shows the results obtained when neglecting Doppler redistribution in the $\bm{\hat{\mathcal{R}}}_{\mbox{\sc{ii}}}$ redistribution matrix, so that the scattering processes that it quantifies are treated as coherent in the reference frame of the observer (see Sect.~\ref{sec::sCSObs}). 
The shaded area indicates the spectral range in which this approximation is not expected to be suitable. For this line, it spans $1.2$~{\AA} from the line center, corresponding to the Doppler width at the height where $\tau = 1$, at line center and for $\mu = 0.1$. 
Unlike for the Mg~{\sc{ii}} doublet, discrepancies in the intensity profile are appreciable even beyond the shaded area, up to several angstroms from line center. This can be explained because of the exceptionally prominent emission feature of this line. Through redistribution effects, the large intensity signal in the core region strongly impacts the intensity in the line wings, and this impact is of course underestimated when Doppler redistribution is neglected. In the linear polarization profiles, discrepancies are also appreciable at similar distances from line center. 
\begin{figure*}[!h]
  \centering
  \includegraphics[width = 0.975\textwidth]{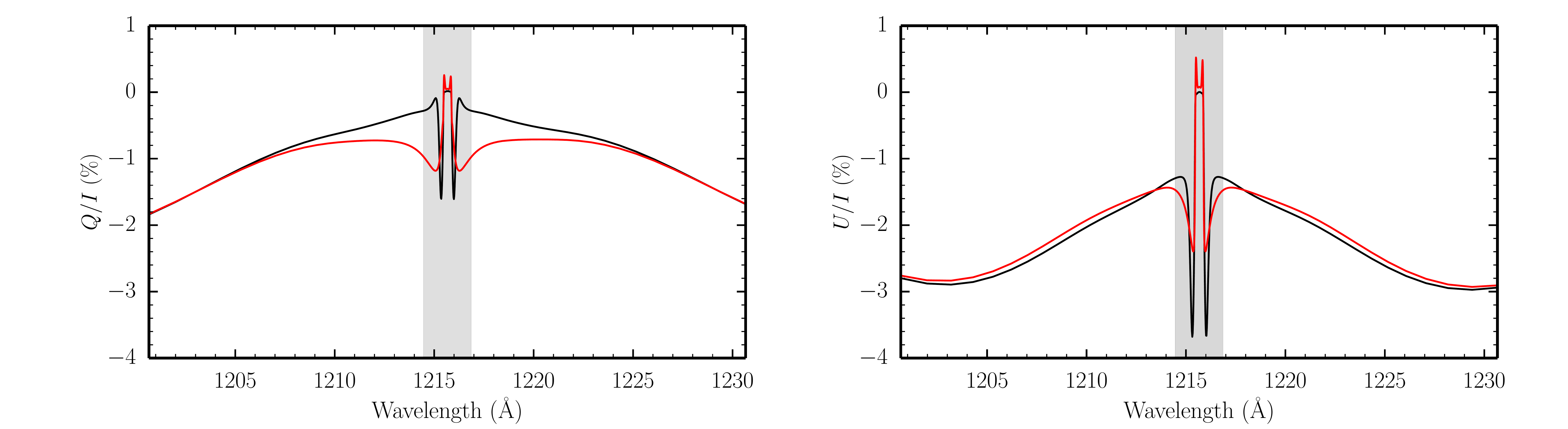}
        \caption{Stokes $Q/I$ (\textit{left panel}) and $U/I$ (\textit{right panel}) profiles for the H~{\sc{i}} Lyman-$\alpha$ line as a function of wavelength. The black and red curves represent calculations   {similar to}  those presented in Fig.~\ref{Fig::App1}, but in the presence of $50$~G horizontal ($\theta_B = 90^\circ$) magnetic fields with $\chi_B = 0^\circ$.}  
        \label{Fig::App3} 
\end{figure*}

In the presence of a magnetic field, discrepancies between the reference PRD calculations and the ones obtained under the abovementioned approximation are also found beyond the shaded area, as can be seen in Fig.~\ref{Fig::App3}. Indeed, a few angstroms from the line center, the discrepancies in $Q/I$ are substantially larger than in case without magnetic fields.  
We recall that the magnetic sensitivity of the  wing linear polarization patterns is mainly controlled by the MO effects quantified by $\rho_V$ \citep{AlsinaBallester+16,AlsinaBallester+19}, which act on the radiation emitted within the medium. %
This suggests that the approximate treatment of coherent scattering processes has a considerable impact on the linear polarization components of the emission vector %
several angstroms from line center even in the absence of a magnetic field.\ 
{However,} in that case the impact on the emergent radiation {was found to be} modest, possibly because of transfer effects. 
In the presence of magnetic fields, the profiles found in the line wings also coincide with those reported in \cite{AlsinaBallester+19}.
\begin{figure*}[!h]
  \centering 
  \includegraphics[width = 0.975\textwidth]{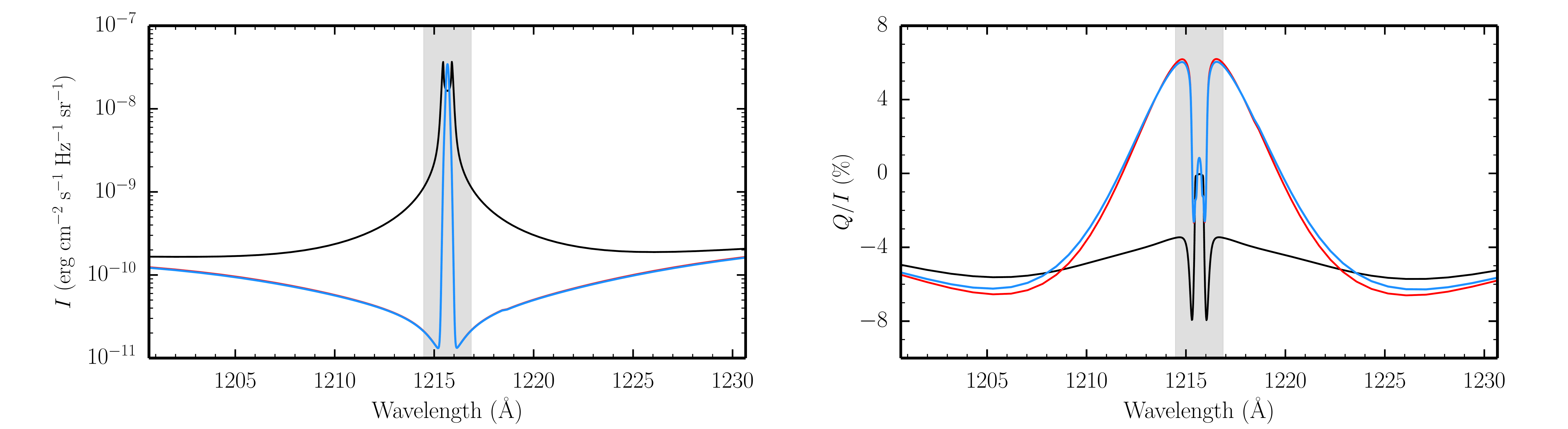}
        \caption{Stokes $I$ (\textit{left panel}) and $Q/I$ (\textit{right panel}) profiles as a function of wavelength obtained through the RT calculations described in the text in the absence of a magnetic field. 
    Reference PRD calculations (black curves) are compared to those in which scattering is treated as fully coherent in the observer's reference frame (neglecting Doppler and collisional redistribution; see text), both accounting for the contribution from elastic collisions to the damping constant (blue curves) and neglecting it (red curves).}  
        \label{Fig::App4} 
\end{figure*}

\begin{figure*}[!h]
  \centering 
 \includegraphics[width = 0.975\textwidth]{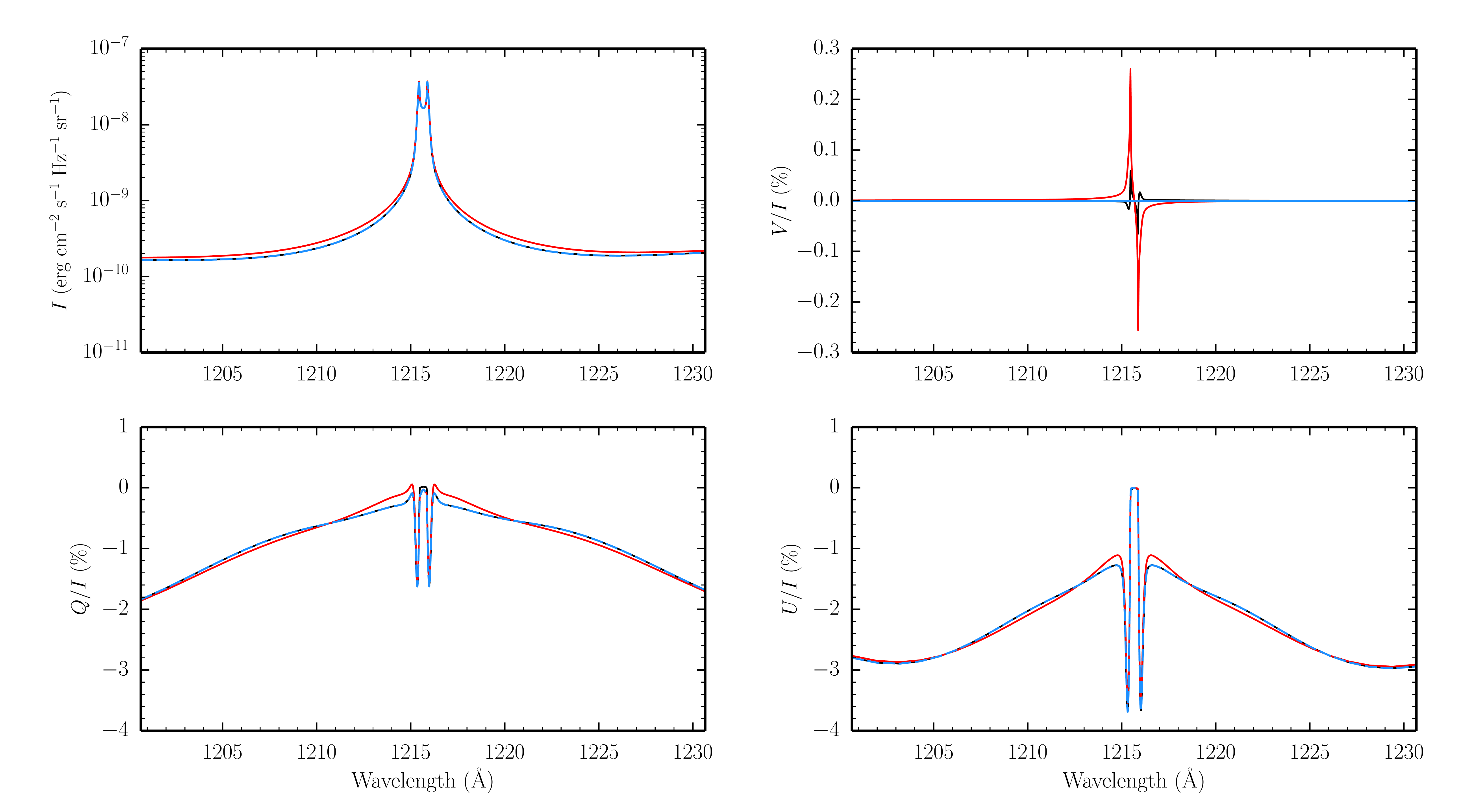}
        \caption{Intensity (\textit{upper left panel}), $V/I$ (\textit{upper right panel}), $Q/I$ (\textit{lower left panel}), and $U/I$ (\textit{lower right panel}) profiles as a function of wavelength. The various colored curves show the results of reference PRD calculations in which the magnetic field is fully taken into account in all RT coefficients (black curves), neglecting its impact only in the computation of the emission coefficient (red curves), and accounting for it only for $\eta_I$ and $\rho_V$ (blue curves). The magnetic field is horizontal ($\theta_B = 90^\circ$) with $\chi_B = 0^\circ$ and has a strength of $50$~G.}  
        \label{Fig::App5} 
\end{figure*}

Although we are not showing the results here, we verified that there is no appreciable impact on the intensity and $Q/I$ profiles when artificially setting to zero all the components of $\bm{\mathcal{R}}_{\mbox{\sc{iii}}}$ except for $K = K^\prime = Q = 0$ (see Eq.~\eqref{eq::RedisDecomp}), as was found also for the Mg~{\sc{ii}} doublet in Sect.~\ref{Sec::PurelyCS}. 
In Fig.~\ref{Fig::App4} a comparison is presented between the results of reference PRD calculations and calculations in which Doppler and collisional redistribution are neglected (see Sect.~\ref{Sec::PurelyCS} for details). For the latter calculations, we consider both the case in which the contribution from $\Gamma_E$ to the damping constant is taken into account (blue curves) and in which it is neglected (red curves). Such approximations perform very poorly for this line until wavelengths many Doppler widths from the line center are considered, independent of whether the contribution from elastic collisions is included in the damping constant. The intensity profiles present considerable discrepancies until roughly $10$~\AA\ from the line center. Although the $Q/I$ signals begin to suitably approximate the reference PRD signals somewhat closer to the center, the discrepancies are striking within a range of a few Doppler widths from the line center, even presenting a different sign. We again attribute the unsuitability of this approximation for modeling the H~{\sc{i}} Lyman-$\alpha$ line wings to its particularly strong intensity emission peak paired with redistribution effects. 

Finally, Fig.~\ref{Fig::App5} shows a comparison between reference PRD calculations, applied to the H~{\sc{i}} Lyman-$\alpha$ line, in which the magnetic splitting is taken into account (i) for all RT coefficients, (ii) for all elements of the propagation matrix but not the emission vector, and (iii) only in $\eta_I$ and $\rho_V$ (see Sect.~\ref{Sec::MagProf} for details). It is noteworthy that, when the magnetic splitting is neglected in the computation of the emission vector, the intensity and the scattering polarization signals in the wings depend appreciably on whether the splitting is taken into account in all elements of the propagation matrix or only in $\eta_I$ and $\rho_V$ (see Fig.~\ref{Fig::App5}). 
Further numerical tests revealed that the RT coefficient responsible for this difference is the $\eta_V$ coefficient for dichroism. By accounting for the impact of the magnetic field in the calculation of $\eta_V$ but not in that of $\varepsilon_V$, a spurious $V/I$ signal is produced. 
As can be seen from the RTE given in Eq.~\eqref{eq::RTEq}, this impacts the intensity profile close to the line core. Because of the particularly strong emission profile of Lyman-$\alpha$, this in turn has a substantial impact on the $I$, $Q/I$, and $U/I$ signals in the line wings, again due to redistribution effects. This impact was much smaller for the Mg~{\sc{ii}} $h$ \& $k$ lines (see Sect.~\ref{Sec::MagProf}) and was appreciable only in the very near wings, because they present far weaker emission features. In any case, the results presented in this Appendix do not invalidate the suitability of the approximation of neglecting the magnetic splitting in the redistribution matrix, and thereby the emissivity, for modeling the wing scattering polarization signals. These findings only serve to emphasize that, under this approximation, the magnetic splitting should be taken into account only in $\eta_I$ and $\rho_V$ but not in the other elements of the propagation matrix. 

In summary, certain approximations that were suitable for the Mg~{\sc{ii}} doublet are not applicable for the H~{\sc{i}} Lyman-$\alpha$ line because of the exceptionally strong emission feature in its line core, which is related to the strength of this line. Neglecting Doppler redistribution in the scattering processes quantified by $\bm{\hat{{\mathcal R}}}_{\mbox{\sc{ii}}}$ represents a rough approximation that is suitable in the far wings. The noncoherent scattering processes quantified by $\bm{\mathcal R}_{\mbox{\sc{iii}}}$ do not appreciably impact the scattering polarization. On the other hand, the approximation of fully neglecting Doppler and collisional redistribution is found to be unsuitable in this case.  
Finally, the approximation of neglecting the magnetic splitting in the emissitivity remains reliable even in the near wings, as long as the splitting is only taken into account in $\eta_I$ and $\rho_V$. 

\section{Analytical expressions for the numerical framework}
\label{sec::AppendixExpression} 
The {non-}LTE RT code used in this work, based on the numerical framework discussed in the main text (see Sect.~\ref{sec::formulation}),  
is suitable for modeling the generation and transfer of polarized radiation in spectral lines that arise from two-term atomic systems with an unpolarized and infinitely sharp lower term. 
Although it is not considered in the main text, it also allows for the inclusion of 
HFS in the atomic system. In this Appendix, we present the expressions involved in the RT problem for this more general case (see Appendices~\ref{sec::AppendixHamilton}--\ref{sec::AppRedis2TrmHFS}). Subsequently, we discuss how to implement the approximations of fully coherent scattering and of fully noncoherent scattering, also referred to as the CRD approximation (see Appendix~\ref{sec::AppCRDCSApprox}). Then, we show how the RT coefficients change when considering a magnetic field whose orientation changes at small scales (see Appendix~\ref{sec::AppMicro}). In Appendix~\ref{sec::AppRedisZero} we show simplified expressions that are suitable for the case without magnetic fields, and which can additionally take into account the depolarizing effect of collisions that induce transitions within the same HFS level. Finally, in Appendix~\ref{sec::AppRedisNoHFS} we present the simplified expressions for atomic systems without HFS, which are used in the main text. The same Appendix also contains the expressions obtained by applying the approximations investigated in the main text. 
We point out that the framework presented here can also be trivially extended to the case in which the considered spectral line arises from multiple isotopes of the same atomic species, simply by computing the contributions from each isotope to the various RT coefficients separately and then summing them, weighted by the relative abundances of the isotopes.  
\subsection{The atomic Hamiltonian}
\label{sec::AppendixHamilton}
In the $L$-$S$ coupling scheme, a given term is specified by $(\beta, L, S)$, where $\beta$ is a set of inner quantum numbers, $L$ is the quantum number for the orbital angular momentum operator $\vec{L}$, and $S$ is the quantum number for the electronic spin $\vec{S}$. 
The various fine structure (FS) levels that pertain to a given term are determined by the spin-orbit Hamiltonian $H_{\mbox{\scriptsize{FS}}}$, whose eigenvectors are of the form $\ket{\beta L S J M}$, where $J$ is the quantum number for the total electronic angular momentum $\vec{J} = \vec{L} + \vec{S}$, and $M$ is its projection along the quantization axis. 
If the nuclear spin of the atom $I$ is nonzero, it interacts with the electron cloud; through this interaction, which is much weaker than the spin-orbit one, a given fine structure $J$-level splits into various HFS levels. 
These levels are determined by the HFS Hamiltonian, which can be expressed as an infinite series of electric and magnetic multipoles. It is generally a good approximation to only consider the first two terms of the series, namely the magnetic dipole and the electric quadrupole terms (see Sect.~3.5 of LL04). 
The eigenvectors of the HFS Hamiltonian are of the form $\ket{\beta L S J F f}$, where $F$ is the quantum number for the total (electronic plus nuclear) angular momentum $\vec{F} = \vec{J} + \vec{I}$, and $f$ its projection along the quantization axis. 
The matrix elements of the unperturbed Hamiltonian $H_0 = H_{\mbox{\scriptsize{FS}}} + H_{\mbox{\scriptsize{HFS}}}$ are given by 
\begin{align}
        & \bra{\beta L S I J F f}\!| H_{\mbox{\scriptsize{FS}}} |\! 
        \ket{\beta L S I J^\prime F^\prime f^\prime} = 
        \delta_{J, J^\prime} \delta_{F, F^\prime} \delta_{f, f^\prime}
        E_{\beta L S}(J) \, , \notag \\
        & \bra{\beta L S I J F f}\!| H_{\mbox{\scriptsize{HFS}}} |\! 
        \ket{\beta L S I J^\prime F^\prime f^\prime} = 
        \delta_{J, J^\prime} \delta_{F, F^\prime} \delta_{f, f^\prime} 
        \left[ \frac{1}{2} \mathcal{A}(\beta,L,S,J,I) K 
        + \mathcal{B}(\beta, L, S, J, I) 
        \frac{3 K (K + 1) - 4 I (I + 1) J (J + 1)}{8 I (2 I - 1) J (2 J - 1)} 
        \right] \, ,
 \label{eqapp::FSHFSHamil}
\end{align}
where $E_{\beta L S}(J)$ is the energy of the considered FS $J$-level, $\mathcal{A}(\beta,L,S,J,I)$ and $\mathcal{B}(\beta,L,S,J,I)$ are the  magnetic dipole and electric quadrupole constants for HFS, and 
\begin{equation*}
 K = F (F + 1) - I (I + 1) - J (J + 1) \, .
\end{equation*}
In the presence of a magnetic field, the energy eigenvectors and eigenvalues of the atomic system have to be calculated by diagonalizing 
the total Hamiltonian $H = H_{\mbox{\scriptsize{FS}}} + H_{\mbox{\scriptsize{HFS}}} + H_B$, where $H_B$ is the magnetic Hamiltonian, 
given by 
\begin{equation}
        H_B = \mu_0 \vec{B} \cdot \bigl( \vec{J} + \vec{S} \bigr) \, 
\end{equation}
where $\mu_0$ is the Bohr magneton and $\vec{B} = B \vec{b}$ is the magnetic field, with $B$ the magnetic field strength and $\vec{b}$ the unit vector for the direction of the field. Through the application of the Wigner-Eckart theorem and its corollaries (see Sect.~2.8 of LL04), the matrix elements of the magnetic Hamiltonian are found to be
\begin{align}
 \bra{\beta L S I J F f}\!| H_B & |\!\ket{\beta L S I J^\prime F^\prime f^\prime} = \delta_{f, f^\prime} 
 \mu_0 B \sqrt{(2 F + 1) (2 F^\prime + 1) (2 J + 1)} 
 \left( 
 \begin{array}{c c c}
  F & F^\prime & 1 \\
  -f & f & 0
 \end{array}
 \right)
 \left\{
 \begin{array}{c c c}
  F & F^\prime & 1 \\
  J^\prime & J & I
 \end{array}
 \right\} \notag \\
 & \times (-1)^{J + I - f}
 \Biggl[ \delta_{J, J^\prime} \sqrt{J (J + 1)}  
  + (-1)^{L + S + J + 1} \sqrt{(2 J^\prime + 1) S (S + 1) (2 S + 1)} 
  \left\{   
  \begin{array}{c c c}
   S & S & 1 \\
   J^\prime & J & L
   \end{array}
  \right\}\Biggr] \, . 
  \label{eqapp::MagHamil}
\end{align} 
With some minor changes in notation, this expression coincides with the one presented in \cite{Sowmya+15}. 
We recall that the {$\Tjs$} and {$\Sjs$} symbols couple different angular momenta, and their expressions can be found in Sects.~2.2 and~2.3 of LL04 \citep[see also][]{Racah42}. 
We note that the magnetic Hamiltonian is still diagonal in $f$, as well as in $\beta$, $L$, $S$, and $I$. However, it introduces a coupling between states with different $F$ and $J$, which therefore are no longer good quantum numbers for the system. We thus label the eigenstates of the atomic Hamiltonian as $\ket{\beta L S I \mu f}$, which can be given as linear combinations of the eigenstates of $\vec{J}$ and $\vec{F}$ as
\begin{equation}
 \ket{\beta L S I \mu f} = 
 \sum_{J} \sum_{F} 
 C^{J F}_\mu(f) \, \ket{\beta L S I J F f} \, ,
\end{equation}
where the label $\mu$ is introduced to distinguish between different states with the same quantum numbers $\beta$, $L$, $S$, $I$, and $f$. 
The sums over $F$ and $J$ range from $|J - I|$ to $(J + I)$ and from $|L - S|$ to $(L + S)$, respectively. For simplicity of notation these limits will not be shown explicitly in the following sections of this Appendix. 
The values of the coupling coefficients $C^{J F}_\mu(f)$ are determined through the diagonalization of the atomic (unperturbed plus magnetic) Hamiltonian. 
Of course, in the absence of a magnetic field there is no state mixing and $C^{J F}_\mu(f) = \delta_{\mu, J F}$. 

Hereafter, we write the eigenstates $\ket{\beta L S I \mu_i f_i}$ and eigenvalues (i.e., the energies) $E_{\beta L S I}(\mu_i f_i)$ using the shorthand notation $\ket{\mu_i f_i}$ and $ E_{\mu_i f_i}$, respectively. For the two-term system considered in this work, we write $u$ or $\ell$ in place of the label $i$, depending on whether it corresponds to a state of the upper or lower term, respectively. 
The frequency of the transition between the state of the upper term $\ket{\mu_u f_u}$ and the state of the lower term $\ket{\mu_\ell f_\ell}$ is given by  
\begin{equation*}
 \nu_{\mu_u f_u, \mu_\ell f_\ell} = 
 (E_{\mu_u f_u} - E_{\mu_\ell f_\ell})/h . 
\end{equation*} 
\subsection{Doppler width and damping constant}
 \label{sec::AppBroad} 
 
In the expressions presented below, we consider the reference frame of the observer rather than the rest frame of the atom. As such, the Doppler frequency shifts due to the small-scale motions of the atmospheric medium are taken into account. These effects cause the spectral line to be further broadened according to the Doppler width, which is given as $\Delta\nu_D$ in frequency units or as $\Delta\lambda_D$ in wavelength units 
 \begin{equation}
 \Delta\nu_D = \frac{\nu_0}{c} \sqrt{\frac{2 k_B T}{m} + \mathrm{v}_m^2} \, \;\;
\mbox{and} \,\;\, \Delta\lambda_D = \frac{\lambda_0^2}{c} \Delta\nu_D \, , 
 \label{eqapp::Doppler}
\end{equation}
where $\nu_0$ is the reference frequency defined as the baricenter of the frequencies of all the transitions between states of the upper and lower term and $\lambda_0$ is the corresponding vacuum wavelength. In this expression the Boltzmann constant $k_B$ and the temperature of the medium $T$ are also introduced. 
The Doppler width receives a thermal contribution, for which we assume a Maxwellian distribution of velocities, and a nonthermal contribution, which depends on the so-called micro-turbulent velocity $\mathrm{v}_m$. We also define the damping constant in terms of the Doppler width as 
\begin{equation}
 a = \frac{\Gamma}{4 \pi \Delta\nu_D} \, ,  
 \label{eqapp::Damping}
\end{equation}
where $\Gamma$ is the line broadening parameter, which is given as the sum of the contribution from radiative processes $\Gamma_R$, elastic collisions $\Gamma_E$, and inelastic collisions $\Gamma_I$. For the applications to the Mg~{\sc{ii}} lines presented in the main text, these contributions are obtained as discussed in Appendix~\ref{sec::AppAtomMod}. For applications to the H~{\sc{i}} Lyman-$\alpha$ line, they are obtained as explained in Appendix~\ref{sec::AppLymanAlpha}. 

\subsection{The elements of the propagation matrix}
\label{sec::AppProp} 
Here we present the analytical expressions for the quantities that enter the RT equation. Unless otherwise noted, the expressions in this Appendix are given in the reference frame of the observer, and taking the quantization axis along which the quantum number $f$ is measured to be parallel to the local vertical (hereafter the vertical reference system). 
The geometrical information of these expressions is fully contained in the geometrical tensors ${\mathcal T}^K_Q$ %
(see LL04). The reference system in which the quantization axis is parallel to the direction of the magnetic field (hereafter the magnetic reference system) and the vertical reference system are related \citep[e.g.,][]{AlsinaBallester+17} as follows 
\begin{align}
 & {\mathcal T}^K_Q(i,\mathbf{\Omega})\bigl|_{B} = \sum_{Q^\prime = -K}^K {\mathcal D}^K_{Q Q^\prime}(R_B)^\ast \, 
 {\mathcal T}^K_{Q^\prime}(i,\mathbf{\Omega}) \bigl|_{\mathrm{v}} \, , \notag \\
 & {\mathcal T}^K_Q(i,\mathbf{\Omega})\bigl|_{\mathrm{v}} = \sum_{Q^\prime = -K}^K {\mathcal D}^K_{Q^\prime Q}(R_B) \, 
 {\mathcal T}^K_{Q^\prime}(i,\mathbf{\Omega}) \bigl|_{B} \, , 
 \label{eqapp::GeomTransK}
\end{align}
where the label $B$ indicates the magnetic reference system and $\mathrm{v}$ indicates the vertical one. 
The quantities $D^K_{Q Q^\prime}(R_B)$ are the elements of the so-called rotation matrices, and $R_B = (0,-\theta_B,-\chi_B)$ are the Euler angles through which the system is rotated from the magnetic frame into the vertical one. 

For the considered two-term atomic model with HFS and an infinitely sharp and unpolarized lower term, the elements of the propagation matrix appearing in Eqs.~\eqref{eq::RTEq} and \eqref{eq::Kmat} may be given in the vertical reference system as
\begin{subequations}
 \begin{align}
    \eta_i(\nu,\mathbf{\Omega}) = {\frac{k_M}{\sqrt{\pi} \Delta\nu_D}} & \sum_{K=0}^2 \sum_{Q=-K}^K  \sqrt{\frac{2 K + 1}{3}}
    \mathcal{D}^K_{0 Q}(R_B)^\ast \, 
    \mathcal{T}^K_Q(i,\mathbf{\Omega})  
     \sum_{q=-1}^1 (-1)^{1+q} 
      \left(\begin{array}{c c c}
            1 & 1 & K\\
            q & -q & 0
         \end{array} \right)  
  \sum_{\mu_u f_u \mu_\ell f_\ell} 
    S^{\mu_\ell f_\ell, \mu_u f_u}_q \, 
    H\bigl(a, u + u_{\mu_u f_u, \mu_\ell f_\ell} \bigr) \, , \\
   \rho_i(\nu,\mathbf{\Omega}) ={\frac{k_M}{\sqrt{\pi} \Delta\nu_D}} & \sum_{K=0}^2 \sum_{Q=-K}^K \sqrt{\frac{2 K + 1}{3}}
   \mathcal{D}^K_{0 Q}(R_B)^\ast \, 
   \mathcal{T}^K_Q(i,\mathbf{\Omega}) 
    \sum_{q=-1}^1 (-1)^{1+q} 
      \left(\begin{array}{c c c}
            1 & 1 & K\\
            q & -q & 0
         \end{array} \right)  
   \sum_{\mu_u f_u \mu_\ell f_\ell} 
     S^{\mu_\ell f_\ell, \mu_u f_u}_q \, 
    L\bigl(a, u + u_{\mu_u f_u, \mu_\ell f_\ell} \bigr) \, . 
 \end{align} 
  \label{eqapp::PropLineTransformedHFS}
  \end{subequations}
The functions $H$ and $L$ are the Voigt and associated dispersion profiles, respectively. They arise from the convolution of the Lorentzian and associated dispersion profiles (which enter the expressions in the atomic rest frame), respectively, with the Gaussian profile that quantifies the Doppler broadening. The summation limits for $f$ depend on the quantum numbers $S\!$, $I$, and $L$ of the considered term, and the number of distinct $\mu$ states depends on $S\!$, $I$, $L$, and $f$. For simplicity of notation, these limits are not shown here.  
The reduced frequency $u$ is introduced in Sect.~\ref{sec::sPropMatExp}. The reduced frequency shift between upper state $\ket{\mu_u f_u}$ and lower state $\ket{\mu_\ell f_\ell}$ is given by 
\begin{equation*}
u_{\mu_u f_u, \mu_\ell f_\ell} = \frac{\nu_{\mu_u f_u, \mu_\ell f_\ell} - \nu_0}{\Delta\nu_D} \, .
\end{equation*} 
The expressions for the elements of the propagation matrix shown above are given in terms of the so-called transition strengths 
\begin{align}
  S^{\mu_\ell f_\ell, \mu_u f_u}_q & = \frac{3}{(2 S + 1)(2 I + 1)} 
 \sum_{\substack{J_u J_u^\prime \\F_u F_u^\prime}} C_{\mu_u}^{J_u F_u}(f_u) \, C_{\mu_u}^{J_u^\prime F_u^\prime}(f_u)
 \sum_{\substack{J_\ell J_\ell^\prime \\F_\ell F_\ell^\prime}} C_{\mu_\ell}^{J_\ell F_\ell}(f_\ell) \, 
 C_{\mu_\ell}^{J_\ell^\prime F_\ell^\prime}(f_\ell) \notag \\
& \times (-1)^{J_u + J_u^\prime + J_\ell + J_\ell^\prime} 
\sqrt{(2 J_u + 1) (2 J_u^\prime + 1) (2 F_u + 1) (2 F_u^\prime + 1) \vphantom{(2 J_\ell^\prime + 1)}} 
\sqrt{(2 J_\ell + 1) (2 J_\ell^\prime + 1) (2 F_\ell + 1) (2 F_\ell^\prime + 1)} \notag \\
& \times \left\{
 \begin{array}{c c c}
  J_u  & J_\ell & 1\\
  L_\ell & L_u  & S
 \end{array}
 \right\} 
  \left\{
 \begin{array}{c c c}
  J_u^\prime  & J_\ell^\prime & 1\\
  L_\ell & L_u  & S
 \end{array}
 \right\}
  \left\{
 \begin{array}{c c c}
  F_u  & F_\ell & 1\\
  J_\ell & J_u  & I
 \end{array}
 \right\}
  \left\{
 \begin{array}{c c c}
  F_u^\prime  & F_\ell^\prime & 1\\
  J_\ell^\prime & J_u^\prime & I
 \end{array}
 \right\}
 \left(
  \begin{array}{c c c}
  F_u  & F_\ell & 1\\
  -f_u & f_\ell & -q
 \end{array}
 \right)
  \left(
  \begin{array}{c c c}
  F_u^\prime  & F_\ell^\prime & 1\\
  -f_u & f_\ell & -q
 \end{array}
 \right) \, , 
 \label{eqapp::TranStr}
\end{align}
where the quantum number $q = f_\ell - f_u$ takes integer values $-1$, $0$ or $1$. These transition strengths can be obtained making use of the definition provided in Sect.~3.4 of LL04, which can be straightforwardly generalized to an atomic system with HFS. Moreover, they fulfill the normalization condition 
\begin{equation}
  \sum_{\mu_\ell f_\ell \mu_u f_u} S^{\mu_\ell f_\ell, \mu_u f_u}_q = 1, \quad (q = -1, 0, 1). 
\end{equation}  
\subsection{The redistribution matrices (two-term atomic model with HFS)}
\label{sec::AppRedis2TrmHFS} 
The line scattering emission coefficient is related to the incident radiation field following Eq.~\eqref{eq::LinEmis}. 
The redistribution matrix entering this expression can be separated into $\bm{\mathcal R}_{\mbox{\sc{ii}}}$ and $\bm{\mathcal R}_{\mbox{\sc{iii}}}$, which characterize scattering processes that are coherent and noncoherent in the atomic frame, respectively (see Sect.~\ref{sec::sEmis}). For a two-term atomic model with an unpolarized lower term and HFS, we have made use of the expression given in Eq.~(A.6) of \cite{Bommier18}. 
For $\bm{\mathcal R}_{\mbox{\sc{ii}}}$ we carried out an integration over velocities assuming a Maxwellian distribution (including also a nonthermal contribution) and accounting for the Doppler frequency shifts as detailed in Sect.~10.3 of \cite{BHubenyMihalas15}, making also use of the angle-averaged approximation \citep[see][]{ReesSaliba82}. As for $\bm{\mathcal R}_{\mbox{\sc{iii}}}$, we have made the assumption that complete frequency redistribution (CRD) occurs in the observer's frame, accounting for Doppler redistribution by substituting each of the Lorentzian and associated dispersion profiles within it by the corresponding $H$ and $L$ functions, also taking into account a normalization factor $\!\sqrt{\pi} \Delta\nu_D$. 
Under these approximations, we can perform the decomposition into frequency and angle-dependent components shown in Eq.~\eqref{eq::RedisDecomp}. The angle-dependent factors are of the form shown in Eq.~\eqref{eq::PfcaVertFrame}, where the transformation into the vertical reference system is performed according to the discussion in Appendix~\ref{sec::AppRedis2TrmHFS}. 
The frequency-dependent factors for $\bm{\mathcal R}_{\mbox{\sc{ii}}}$ and $\bm{\mathcal R}_{\mbox{\sc{iii}}}$ are given by 
\begin{subequations}
\begin{align}
 \bigl[R_{\mbox{\sc{ii}}}&\bigr]^{K K^{\prime}}_Q(\nu^\prime, \nu) = 
 \sum_{\substack{\mu_u \mu_u^\prime \\ f_u f_u^\prime}} 
 \sum_{\substack{\mu_\ell \mu_\ell^\prime \\ f_\ell f_\ell^\prime}} 
 \mathcal{A}^{K K^{\prime}}_Q(\mu_u f_u,\mu_u^\prime f_u^\prime, \mu_\ell f_\ell,\mu_\ell^\prime f_\ell^\prime) \notag \\[-8pt]
 & \times \frac{1}{\pi \Delta\nu_D^2}\frac{\Gamma_R}{\Gamma_R + \Gamma_I + \Gamma_E + 2 \pi \, \mathrm{i}\, {\Delta\nu_D \,u_{\mu_u f_u, \mu_u^\prime f_u^\prime}}} 
  \frac{1}{2} \int_0^\pi\!\!\mathrm{d}\Theta\,\exp\Biggl[-\,\biggl(\frac{u - u^\prime + u_{\mu_\ell f_\ell, \mu_\ell^\prime f_\ell^\prime}}{2 \sin\Theta/2} \biggr)^2 \Biggr] \notag \\
 & \;\times \Biggl[ \frac{1}{2} W\,\biggl(\frac{a}{\cos\Theta/2},\frac{u + u^\prime + u_{\mu_u^\prime f_u^\prime, \mu_\ell f_\ell}
 + u_{\mu_u^\prime f_u^\prime, \mu_\ell^\prime f_\ell^\prime}}{2 \cos\Theta/2} \biggr)
 + \frac{1}{2} W\,\biggl(\frac{a}{\cos\Theta/2},\frac{u + u^\prime + u_{\mu_u f_u, \mu_\ell f_\ell} + 
 u_{\mu_u f_u, \mu_\ell^\prime f_\ell^\prime}}{2 \cos\Theta/2} \biggr)^\ast \Biggr] \, ,
  \label{eqapp::R2HFSObserv} \\[10pt]
 \bigl[R_{\mbox{\sc{iii}}}&\bigr]^{K K^{\prime}}_Q(\nu^\prime, \nu) = 
  \sum_{\substack{\mu_u \mu_u^\prime \\ f_u f_u^\prime}} 
 \sum_{\substack{\mu_\ell \mu_\ell^\prime \\ f_\ell f_\ell^\prime}} 
 \mathcal{A}^{K K^{\prime}}_Q(\mu_u f_u, \mu_u^\prime f_u^\prime, \mu_\ell f_\ell,\mu_\ell^\prime f_\ell^\prime) \notag \\[-8pt]
  & \times \frac{1}{\pi \Delta\nu_D^2} \Biggl[\frac{\Gamma_R}{\Gamma_R + \Gamma_I +  2 \pi \, \mathrm{i} \, {\Delta\nu_D \, u_{\mu_u f_u, \mu_u^\prime f_u^\prime}}} 
  - \frac{\Gamma_R}{\Gamma_R + \Gamma_I + \Gamma_E + 2 \pi \, \mathrm{i} \, {\Delta\nu_D \, u_{\mu_u f_u, \mu_u^\prime f_u^\prime}}} \Biggr] \notag \\ 
  & \times \Biggl[\frac{1}{2} W\,\Bigl(a, u^\prime + u_{\mu_u^\prime f_u^\prime, \mu_\ell f_\ell} \Bigr)
  + \frac{1}{2} W\,\Bigl(a, u^\prime + u_{\mu_u f_u, \mu_\ell f_\ell}\Bigr)^\ast \Biggr] 
  \, \,\Biggl[\frac{1}{2} W\,\Bigl(a, u + u_{\mu_u^\prime f_u^\prime, \mu_\ell^\prime f_\ell^\prime}\Bigr) 
  + \frac{1}{2} W\,\Bigl(a, u + u_{\mu_u f_u, \mu_\ell^\prime f_\ell^\prime}\Bigr)^\ast \Biggr] \, . 
 \label{eqapp::R3HFSObserv}
\end{align} 
\label{eqapp::RedisHFSObserv}
\end{subequations}
Notable changes in notation with respect to the expressions given in \cite{Bommier18} include the exchange of the indices $K$ and $K^\prime$ and that the quantum number $M$ in that paper corresponds to the quantum number $f$ used here. 
The reduced frequency splitting between states that belong to the same term is given by  
\begin{equation*} 
  u_{\mu_u f_u, \mu_u^\prime f_u^\prime} = \frac{\nu_{\mu_u f_u, \mu_u^\prime f_u^\prime}}{\Delta \nu_D} \, \quad
  \mathrm{and} \quad
  u_{\mu_\ell f_\ell, \mu_\ell^\prime f_\ell^\prime} = \frac{\nu_{\mu_\ell f_\ell, \mu_\ell^\prime f_\ell^\prime}}{\Delta \nu_D} \, .  
\end{equation*} 
The expressions for ${R}_{\mbox{\sc{ii}}}$ and $R_{\mbox{\sc{iii}}}$ contain the same quantity $\mathcal{A}$, which depends on the quantum numbers of the states involved in each of the terms of the redistribution matrix, and the coupling between them. This quantity is given by 
\begin{align}
& {\mathcal A}^{K K^{\prime}}_Q(\mu_u f_u,\mu^\prime_u f^\prime_u,\mu_\ell f_\ell, \mu^\prime_\ell f^\prime_\ell) =
\frac{3 (2 L_u + 1)}{(2 I + 1) (2 S + 1)} 
\sqrt{(2 K + 1) (2 K^\prime + 1)} \notag \\
& \times (-1)^{f_\ell - f^\prime_\ell} \sum_{\substack{J_u {\bar J}_u J^\prime_u {\bar{J}^\prime_u}\\ F_u \bar{F}_u F^\prime_u{\bar{F}^\prime_u}}} 
C_{{\mu_u}}^{{J_u} {F_u}} ({f_u}) \, C_{{\mu_u}}^{{\bar{J}_u} {\bar{F}_u}} ({f_u}) 
\, C_{\mu^\prime_u}^{J^\prime_u F^\prime_u} (f^\prime_u) \, C_{\mu^\prime_u}^{{\bar{J}^\prime_u} {\bar{F}^\prime_u}} (f^\prime_u) 
\sum_{\substack{J_\ell \bar{J}_\ell J^\prime_\ell {\bar{J}^\prime_\ell} \\ F_\ell \bar{F}_\ell F^\prime_\ell {\bar{F}^\prime_\ell}}}
\,  {C_{{\mu_\ell}}^{{J_\ell} {F_\ell}} ({f_\ell}) \, C_{{\mu_\ell}}^{{\bar{J}_\ell} {\bar{F}_\ell}} ({f_\ell})
\, C_{\mu^\prime_\ell}^{J^\prime_\ell F^\prime_\ell} (f^\prime_\ell) \,  C_{\mu^\prime_\ell}^{\bar{J}^\prime_\ell {\bar{F}^\prime_\ell}} (f^\prime_\ell)} \notag \\
& \times 
{(-1)^{J_u + J^\prime_u + \bar{J}_u + {\bar{J}^\prime_u}}
(-1)^{J_\ell + J^\prime_\ell + \bar{J}_\ell + {\bar{J}^\prime_\ell}}} 
\sqrt{(2 F_u + 1) (2 F^\prime_u + 1) (2 \bar{F}_u + 1) (2 {\bar{F}^\prime_u} + 1) 
  (2 J_u + 1) (2 J^\prime_u + 1) (2 \bar{J}_u + 1) (2 {\bar{J}^\prime_u} + 1)}  \notag \\
& \times \sqrt{(2 F_\ell + 1) (2 F^\prime_\ell + 1) (2 \bar{F}_\ell + 1) (2 {\bar{F}^\prime_\ell} + 1) 
 (2 J_\ell + 1) (2 J^\prime_\ell + 1) (2 \bar{J}_\ell + 1) (2 {\bar{J}^\prime_\ell} + 1)} \notag \\ 
     & \times {\left\{\begin{array}{c c c}
     F_u & F_\ell & 1 \\
     J_\ell & J_u & I
    \end{array}\right\}
     \left\{\begin{array}{c c c}
     F^\prime_u & \bar{F}_\ell & 1 \\
     \bar{J}_\ell & J^\prime_u & I
    \end{array}\right\}
     \left\{\begin{array}{c c c}
     {\bar{F}^\prime_u} & {\bar{F}^\prime_\ell} & 1 \\
     {\bar{J}^\prime_\ell} & {\bar{J}^\prime_u} & I
    \end{array}\right\} 
    \left\{\begin{array}{c c c}
     \bar{F}_u & F^\prime_\ell & 1 \\
     J^\prime_\ell & \bar{J}_u & I
    \end{array}\right\} }
 \left\{\begin{array}{c c c}
     J_u & J_\ell & 1 \\
     L_\ell & L_u & S
    \end{array}\right\}
     \left\{\begin{array}{c c c}
     J^\prime_u & \bar{J}_\ell & 1 \\
     L_\ell & L_u & S
    \end{array}\right\} \notag \\
  & \times \left\{\begin{array}{c c c}
     {\bar{J}^\prime_u} & {\bar{J}^\prime_\ell} & 1 \\
     L_\ell & L_u & S
    \end{array}\right\}
     \left\{\begin{array}{c c c}
     \bar{J}_u & J^\prime_\ell & 1 \\
     L_\ell & L_u & S
    \end{array}\right\} 
   \sum_{p p^\prime p^{\prime \prime} p^{\prime \prime \prime} = -1}^1 
     \left(\begin{array}{c c c}
      1 & 1 & K \\
      -p & p^\prime & Q \end{array}\right)
     \left(\begin{array}{c c c}
      1 & 1 & K^{\prime} \\
      -p^{\prime \prime \prime} & p^{\prime \prime} &  Q \end{array}\right) \notag \\
 & \times 
  \left(\begin{array}{c c c}
     F_u & F_\ell & 1 \\
     -f_u & f_\ell & p
     \end{array} \right)
   \left(\begin{array}{c c c}
     F_u^\prime & \bar{F}_\ell & 1 \\
     -f_u^\prime & f_\ell & p^\prime
     \end{array} \right) 
    \left(\begin{array}{c c c}
     {\bar{F}^\prime_u} & {\bar{F}^\prime_\ell} & 1 \\
     -f^\prime_u & f^\prime_\ell & p^{\prime \prime}
     \end{array} \right)  
   \left(\begin{array}{c c c}
     \bar{F}_u & {F^\prime_\ell} & 1 \\
     -f_u & f^\prime_\ell & p^{\prime \prime \prime}
     \end{array} \right) \, .
     \label{eqapp::AquanHFS}
\end{align}
We arrive at this expression by making use of the equivalence between the coupling coefficients $C_{\mu}^{F J}(f)$ and the product of the pair $C^J_{J^\ast M (F^\ast f)}$ $C^F_{F^{\ast \ast}(J M)f}$ given in \cite{Bommier18}, in which the perturbed Hamiltonians for FS and HFS are diagonalized separately.
Here the quantity $M$ is the value of $\vec{J}$ measured along the selected quantization axis, and it is referred to as $M_J$ in \cite{Bommier18}. We also changed the set of labels $J^\ast F^\ast$ to the equivalent $\mu$, as shown in Eqs.~\eqref{eqapp::RedisHFSObserv}.
 
\subsection{The limits of CRD and fully coherent scattering}
\label{sec::AppCRDCSApprox}
A theoretical framework that has been extensively used to model spectral line polarization is the one introduced by LL04. It was derived from first principles of quantum electrodynamics and relies on a first-order perturbative treatment of matter-radiation interaction. Thus, scattering processes are treated as a succession of statistically independent absorption and emission processes \citep{CasiniLandiDeglInnocenti08}. This corresponds to the limit of complete frequency redistribution (CRD), in which any frequency coherence in scattering processes is effectively lost. As discussed in LL04, this approach is valid under the flat-spectrum approximation, in which the incident radiation field is assumed to be independent of frequency. 

By contrast, the formalism considered in this work \citep[see][]{Bommier17} allows one to take the spectral structure of the incident radiation fully into account. It can account for scattering processes that are coherent in frequency (i.e., that preserve the frequency correlation between incoming and outgoing radiation) and, the same time, it accounts for loss of such coherence due to elastic collisions. As can be seen from the redistribution matrices shown in Appendix~\ref{sec::AppRedis2TrmHFS}, a higher elastic collisional rate $\Gamma_E$ implies smaller branching ratios in $\bm{\mathcal R}_{\mbox{\sc{ii}}}$ (which quantify coherent scattering processes) and larger ones in $\bm{\mathcal R}_{\mbox{\sc{iii}}}$ (noncoherent processes). 
The CRD limit can in principle be reached by artificially setting a very large value of $\Gamma_E$ (i.e., $\Gamma_E \gg \Gamma_R, \Gamma_I$) in the branching ratios so that the contribution from $\bm{\mathcal R}_{\mbox{\sc{iii}}}$ dominates over that from $\bm{\mathcal R}_{\mbox{\sc{ii}}}$. 
However, one must bear in mind that $\Gamma_E$ also enters the definition of the damping constant $a$ (which could introduce inconsistencies) and that elastic collisions may in principle relax atomic polarization (so that polarization phenomena could become negligible in a highly collisional regime). We also observe that the $\bm{\mathcal R}_{\mbox{\sc{iii}}}$ considered in this framework was not obtained under the flat-spectrum assumption and could potentially introduce a sensitivity to the spectral structure of the radiation field. 
Conversely, the limit of fully coherent scattering (CS) can by reached by setting $\Gamma_E = 0$ 
so that the contribution from $\bm{\mathcal R}_{\mbox{\sc{iii}}}$ vanishes. In this work, we have considered 
calculations that correspond to this limit both accounting for the contribution from $\Gamma_E$ to the damping constant and neglecting it (see Sect~\ref{sec::AppRedisNoHFS} for the approximation of additionally neglecting Doppler redistribution).  
We note that accounting for this contribution in the damping constant gives rise to an inconsistency, because the emission profiles contained in the redistribution matrices no longer cancel perfectly with the branching ratios in the far wings, which could introduce an artificial magnetic sensitivity \citep[see Sect.~10.4 of LL04;][]{AlsinaBallester+18}. 

 \subsection{Micro-structured magnetic fields}
 \label{sec::AppMicro} 
The expressions presented above are suitable for the computation of the RT coefficients when the external magnetic field has a fixed strength and orientation at any given spatial point. Hereafter, such fields are referred to as deterministic. 
However, in stellar atmospheres it is common for magnetic fields to be tangled at small spatial scales, so that its orientation changes multiple times across the mean free path of the line's photons. Making the assumption that the strength of the magnetic field remains constant over these scales (i.e., the magnetic field is unimodal), the expressions for the RT coefficients in the vertical reference system can be found through an analytical integration over the considered distribution of magnetic field orientations. 
The dependency on the magnetic field orientation is fully contained in the rotation matrices, both in the elements of the propagation matrix and in the line scattering emission coefficient (see Eqs.~\eqref{eq::PropLineTransformed} and~\eqref{eq::PfcaVertFrame}). Thus, these matrices are the only quantities entering the integration. Below, we consider two distinct scenarios for the distribution of magnetic field orientations and show the corresponding expressions. 

\paragraph{\textit{Isotropic distribution of magnetic fields:}} The analytical integration is carried out over all possible inclinations $\theta_B$ and azimuths $\chi_B$.
\begin{align*}
 & \frac{1}{4\pi} \int_0^{2 \pi}\!\mathrm{d}\chi_B \int_0^\pi\!\mathrm{d}\theta_B \sin\theta_B \, 
 {\mathcal D}^K_{0 Q}(R_B)^\ast = \delta_{K, 0} \, \delta_{Q, 0} \, , \\ 
 & \frac{1}{4\pi} \int_0^{2 \pi}\!\mathrm{d}\chi_B \int_0^\pi\!\mathrm{d}\theta_B \sin\theta_B \,
 {\mathcal D}^{K}_{Q Q^{\prime \prime}}(R_B) \, 
 {\mathcal D}^{K^{\prime}}_{Q Q^{\prime}}(R_B)^\ast = \frac{1}{2 K + 1}
 \delta_{K, K^{\prime}} \delta_{Q^\prime, Q^{\prime \prime}} \, .  
\end{align*}
\noindent For the latter integration we used the Weyl theorem (see Sect.~2.6 of LL04). Making use of these integrals, the expression for the absorption coefficient becomes
\begin{align}
 \eta_I(\nu,\mathbf{\Omega}) = \frac{k_M}{3 \sqrt{\pi}\Delta\nu_D}  
 \sum_{q} \sum_{\mu_u f_u \mu_\ell f_\ell} 
     S^{\mu_\ell f_\ell, \mu_u f_u }_q \, 
    H\bigl(a,u + u_{\mu_u f_u,\mu_\ell f_\ell} \bigr) \, .
 \label{eqapp::PropIso}   
\end{align}
The propagation matrix elements $\rho_i$ and $\eta_i$ with subscripts other than $I$ are zero and therefore their expressions are not shown. The elements of the scattering phase matrix contained in the redistribution matrices become 
\begin{align}
 \bm{\mathcal P}^{K K^{\prime}}_Q(\mathbf{\Omega}^\prime,\mathbf{\Omega})_{i j} = \delta_{K, K^{\prime}} \frac{1}{2 K + 1} 
 \sum_{Q^\prime = -K}^K (-1)^{Q^\prime} {\mathcal T}^{K}_{Q^\prime}(i,\mathbf{\Omega}) {\mathcal T}^{K}_{-Q^\prime}(j,\mathbf{\Omega}^\prime) \, . 
 \label{eqapp::ScatPhasMatIso}
\end{align}
It is often useful to consider this distribution in order to investigate the small-scale magnetic fields present in the quiescent solar photosphere. 
\paragraph{\textit{Magnetic field with a fixed inclination and a random distribution of azimuths:}} 
In this case, the analytical integration is only carried out over $\chi_B$, while $\theta_B$ is kept fixed, resulting in 
\begin{align*}
 & \frac{1}{2\pi} \int_0^{2 \pi}\!\mathrm{d}\chi_B \, 
 {\mathcal D}^K_{0 Q}(R_B)^\ast = \delta_{Q, 0} \, d^{K}_{0 0}(\theta_B) ,\\ 
 & \frac{1}{2\pi} \int_0^{2 \pi}\!\mathrm{d}\chi_B \, 
 {\mathcal D}^{K}_{Q Q^{\prime \prime}}(R_B) \, 
 {\mathcal D}^{K^{\prime}}_{Q Q^{\prime}}(R_B)^\ast = (-1)^{Q - Q^{\prime \prime}}
 \sum_{\kappa=|K - K^{\prime}|}^{K + K^{\prime}} (2 \kappa + 1) 
 \left( 
  \begin{array}{c c c}
   K & K^{\prime} & \kappa \\
   Q & -Q & 0
  \end{array}
 \right)
 \left(
 \begin{array}{c c c}
  K & K^{\prime} & \kappa \\
  Q^{\prime \prime} & -Q^{\prime} & 0 
  \end{array} 
  \right)
  d^{\kappa}_{0 0}(\theta_B) \, , 
\end{align*}
where $d^K_{0 0}(\theta_B)$ is the so-called reduced rotation matrix (see Sect.~2.6 of LL04). Through these relations we reach %
\begin{subequations}
\begin{align}
    \eta_i(\nu,\mathbf{\Omega}) = {\frac{k_M}{\sqrt{\pi} \Delta\nu_D}} 
    & \sum_{K=0}^2 \sqrt{\frac{2 K + 1}{3}}
    d^K_{0 0}(\theta_B) \, 
    \mathcal{T}^K_0(i,\mathbf{\Omega}) 
     \sum_{q=-1}^1 (-1)^{1+q} 
      \left(\begin{array}{c c c}
            1 & 1 & K\\
            q & -q & 0
         \end{array} \right) 
  \sum_{\mu_u f_u \mu_\ell f_\ell} 
     S^{\mu_\ell f_\ell, \mu_u f_u}_q \, 
    H\bigl(a,u + u_{\mu_u f_u,\mu_\ell f_\ell} \bigr) \, , \label{eqapp::EtaRandazi} \\
   \rho_i(\nu,\mathbf{\Omega}) = {\frac{k_M}{\sqrt{\pi} \Delta\nu_D}} 
   & \sum_{K=0}^2 \sqrt{\frac{2 K + 1}{3}}
   d^K_{0 0}(\theta_B) \, 
   \mathcal{T}^K_0(i,\mathbf{\Omega})  
    \sum_{q = -1}^1 (-1)^{1+q} 
      \left(\begin{array}{c c c}
            1 & 1 & K\\
            q & -q & 0
         \end{array} \right)  
   \sum_{\mu_u f_u \mu_\ell f_\ell} 
    S^{\mu_\ell f_\ell, \mu_u f_u}_q \, 
    L\bigl(a, u + u_{\mu_u f_u, \mu_\ell f_\ell} \bigr) \, , \label{eqapp::RhoRandazi}
 \end{align}
 \label{eqapp::PropLineRandazi}
\end{subequations} 
and
\begin{equation}
 \bm{\mathcal{P}}^{K K^{\prime}}_Q (\mathbf{\Omega}^\prime,\mathbf{\Omega})_{i j} = (-1)^{Q} \!\!
 \sum_{Q^\prime = -K^\prime}^{K^\prime} \sum_{Q^{\prime \prime} = -K}^K \mathcal{T}^{K^{\prime}}_{Q^{\prime}}(i,\mathbf{\Omega}) \, 
 \mathcal{T}^K_{-Q^{\prime \prime}}(j,\mathbf{\Omega}^\prime) 
 \sum_{\kappa=|K - K^{\prime}|}^{K + K^{\prime}} (2 \kappa + 1) 
 \left(
 \begin{array}{c c c}
  K & K^{\prime} & \kappa \\
  Q & -Q & 0
 \end{array}
 \right)
 \left(
 \begin{array}{c c c}
  K & K^{\prime} & \kappa \\
  Q^{\prime \prime} & -Q^{\prime} & 0
 \end{array}
 \right) d^{\kappa}_{0 0}(\theta_B)
 \, , 
 \label{eqapp::ScatPhasMatRandazi}
\end{equation}
It is often useful to consider this distribution in order to investigate the canopy-like magnetic fields in the outer layers of the solar atmosphere.  
\subsection{The redistribution matrices in the absence of a magnetic field, including the depolarizing effect of elastic collisions}
\label{sec::AppRedisZero} 
In the absence of magnetic fields, $J$ and $F$ remain good quantum numbers. The energy eigenstates are of the form $\ket{\beta L S I J F f}$ (or, with shorthand notation, $\ket{J F f}$) and states with different quantum number $f$ but the same numbers $J$ and $F$ remain degenerate. As a result, the number of transitions with different frequencies greatly decreases and the line scattering emission coefficients can be computed at a considerably lower computational cost. 
Simpler expressions for the redistribution matrix, suitable for the nonmagnetic case, are found in Eq.~(40) of \cite{Bommier17}.  
Following the approach described in Appendix~\ref{sec::AppRedis2TrmHFS} and making the same changes in notation we write them as 
\begin{subequations}
 \begin{align}
  \bigl[R_{\mbox{\sc{ii}}}&\bigr]^{K K^{\prime}}_Q(\nu^\prime, \nu) =
 \delta_{K K^{\prime}} 
 \sum_{\substack{J_u J_u^\prime \\ F_u F_u^\prime}} 
 \sum_{\substack{J_\ell J_\ell^\prime \\ F_\ell F_\ell^\prime}} 
 \mathcal{A}^{K K}_Q\bigl(J_u F_u, J_u^\prime F_u^\prime, J_\ell F_\ell, J_\ell^\prime F_\ell^\prime\bigr) \notag \\[-5pt]
 & \;\times \frac{1}{{\pi} \Delta\nu_D^2} \frac{\Gamma_R}{\Gamma_R + \Gamma_I + \Gamma_E + 2 \pi \, \mathrm{i}\, {\Delta\nu_D \, u_{J_u F_u, J_u^\prime F_u^\prime}}} 
  \frac{1}{2} \int_0^\pi\!\!\mathrm{d}\Theta\,
  \exp\Biggl[-\,\biggl(\frac{u - u^\prime + u_{J_\ell F_\ell, J_\ell^\prime F_\ell^\prime}}{2 \sin\Theta/2} \biggr)^2 \Biggr] \notag \\
 & \;\times \Biggl[ \frac{1}{2} W\,\biggl(\frac{a}{\cos\Theta/2},\frac{u + u^\prime + u_{J_u^\prime F_u^\prime, J_\ell F_\ell}
 + u_{J_u^\prime F_u^\prime, J_\ell^\prime F_\ell^\prime}}{2 \cos\Theta/2} \biggr)
 + \frac{1}{2} W\,\biggl(\frac{a}{\cos\Theta/2},\frac{u + u^\prime + u_{J_u F_u, J_\ell F_\ell} + 
 u_{J_u F_u, J_\ell^\prime F_\ell^\prime}}{2 \cos\Theta/2} \biggr)^\ast \Biggr] \, ,
  \label{eqapp::R2RedisHFSNoB} \\[10pt]
 \bigl[R_{\mbox{\sc{iii}}}&\bigr]^{K K^{\prime}}_Q(\nu^\prime, \nu) = 
 \delta_{K K^{\prime}} 
 \sum_{\substack{J_u J_u^\prime \\ F_u F_u^\prime}} 
 \sum_{\substack{J_\ell J_\ell^\prime \\ F_\ell F_\ell^\prime}} 
 \mathcal{A}^{K K}_Q\bigl(J_u F_u, J_u^\prime F_u^\prime, J_\ell F_\ell, J_\ell^\prime F_\ell^\prime\bigr) \notag \\[-8pt]
  & \times \frac{1}{\pi \Delta\nu_D^2}  \Biggl[\frac{\Gamma_R}{\Gamma_R + \Gamma_I + \frac{1}{2}\bigl[D^{(K)}(J_u F_u) + D^{(K)}(J_u^\prime F_u^\prime) \bigr] 
  + 2 \pi \, \mathrm{i} \, {\Delta\nu_D \, u_{J_u F_u, J_u^\prime F_u^\prime}}} 
  - \frac{\Gamma_R}{\Gamma_R + \Gamma_I + \Gamma_E + 2 \pi \, \mathrm{i} \, {\Delta\nu_D \, u_{J_u F_u, J_u^\prime F_u^\prime}}} \Biggr] \notag \\ 
  & \times \Biggl[\frac{1}{2} W\,\Bigl(a, u^\prime + u_{J_u^\prime F_u^\prime, J_\ell F_\ell} \Bigr)
  + \frac{1}{2} W\,\Bigl(a, u^\prime + u_{J_u F_u, J_\ell F_\ell}\Bigr)^\ast \Biggr] 
  \, \,\Biggl[\frac{1}{2} W\,\Bigl(a, u + u_{J_u^\prime F_u^\prime, J_\ell^\prime F_\ell^\prime}\Bigr) 
  + \frac{1}{2} W\,\Bigl(a, u + u_{J_u F_u, J_\ell^\prime F_\ell^\prime}\Bigr)^\ast \Biggr] \, , \label{eqapp::R3RedisHFSNoB}
 \end{align}
\label{eqapp::RedisHFSNoB}
\end{subequations}
where 
\begin{align}
 \mathcal{A}^{K K}_Q \Bigl(J_u F_u, & J_u^\prime F_u^\prime, J_\ell F_\ell, J_\ell^\prime F_\ell^\prime \Bigr) = 
  \frac{3 (2 L_u + 1)}{(2 S + 1)(2 I + 1)} %
  \, (-1)^{F_\ell - F_\ell^\prime} 
  (2 J_u + 1) \, (2 J_u^\prime + 1) \, (2 J_\ell + 1) \, (2 J_\ell^\prime + 1) \notag \\ 
 & \times (2 F_u + 1) \, (2 F_u^\prime + 1) \, (2 F_\ell + 1) \, (2 F_\ell^\prime + 1) 
 \left\{
  \begin{array}{c c c}
  1    &  1          & K \\
  F_u  &  F_u^\prime & F_\ell
 \end{array} \right\} 
 \left\{
  \begin{array}{c c c}
   1   &  1          & K \\
   F_u  &  F_u^\prime & F_\ell^\prime
 \end{array} \right\} \notag \\
 & \times
 \left\{
  \begin{array}{c c c}
  J_u    &  J_\ell &  1 \\
  L_\ell &  L_u    &  S  
 \end{array} \right\}  
 \left\{
  \begin{array}{c c c}
   J_u^\prime    &  J_\ell &  1 \\
   L_\ell &  L_u    &  S  
 \end{array} \right\} 
 \left\{
  \begin{array}{c c c}
   J_u^\prime &  J_\ell^\prime &  1 \\
   L_\ell     &  L_u    &  S  
  \end{array} \right\}  
 \left\{
  \begin{array}{c c c}
   J_u &  J_\ell^\prime &  1 \\
   L_\ell     &  L_u    &  S  
  \end{array} \right\} \notag \\
  & \times
 \left\{
  \begin{array}{c c c}
  F_u    &  F_\ell &  1 \\
  J_\ell &  J_u    &  I  
 \end{array} \right\}  
 \left\{
  \begin{array}{c c c}
   F_u^\prime    &  F_\ell &  1 \\
   J_\ell &  J_u^\prime    &  I  
 \end{array} \right\} 
 \left\{
  \begin{array}{c c c}
   F_u^\prime &  F_\ell^\prime &  1 \\
   J_\ell^\prime     &  J_u^\prime    &  I  
  \end{array} \right\}  
 \left\{
  \begin{array}{c c c}
   F_u &  F_\ell^\prime &  1 \\
   J_\ell^\prime     &  J_u    &  I  
  \end{array} \right\} \, . 
  \label{eqapp::AquanHFSNoB}
 \end{align}
Similar expressions can be obtained directly from those presented in Appendix~\ref{sec::AppRedis2TrmHFS}, observing that in the unmagnetized case $C_\mu^{J F} = \delta_{\mu, J F}$. 
The only notable difference is that the branching ratios in the expressions obtained from \cite{Bommier17} contain $K$-multipole depolarization rates for each HFS level $D^{(K)}(J_u, F_u)$ produced by elastic collisions. In that work the $D^{(K)}$ rates could be included because the master equations for the atomic states allow for a closed analytical solution even when accounting for the collisional transitions between states belonging to the same HFS level, provided there is no external magnetic field. 
We note that, even in the absence of magnetic fields, no such solution can be found when accounting for the collisional transitions between states that belong to the same term but with different $J$ or $F$ quantum numbers, which are therefore not included in the expressions introduced above. 

\paragraph{\textit{Neglecting interference between states of the upper term}} 
The expressions above can be straightforwardly modified to artificially neglect the quantum interference between states belonging to the upper term with different $J$ or $F$, which remain good quantum numbers of the system in the absence of a magnetic field. 
In order to neglect the interference between states with different $J$ (i.e., $J$-state interference), we simply introduce a $\delta_{J_u, J_u^\prime}$ in the quantity ${\mathcal A}$ shown above. Similarly, the interference between states with different quantum number $F$ can be neglected by introducing $\delta_{F_u, F_u^\prime}$ in $\mathcal A$. 

\subsection{The two-term atomic model without HFS}
\label{sec::AppRedisNoHFS}
The Mg~{\sc{ii}} doublet (see main text) and the H~{\sc{i}} Lyman-$\alpha$ line (see Appendix~\ref{sec::AppLymanAlpha}) can be suitably modeled considering atomic systems without HFS. In this case, the expressions shown in the previous sections simplify considerably. 
The matrix elements of the magnetic Hamiltonian also simplify, taking the form shown in Sect.~3.4 of LL04, and the atomic Hamiltonian is given by $H = H_{\mbox{\scriptsize{FS}}} + H_B$. 
For a given $(\beta, L, S)$ term, its eigenstates are specified by quantum number $M$, which is the value of the total electronic angular momentum $\vec{J}$ measured along the specified quantization axis, and by the label $j$. The energies of these states and their couplings to the eigenstates of $\vec{J}$ are given by
 \begin{align*}
  H \ket{\beta L S j M} & = E_{\beta L S} \ket{\beta L S j M} \, ;  \\
  \ket{\beta L S j M} & = \sum_J C^{J}_j(M) \ket{\beta L S J M} \, . 
 \end{align*}  
Hereafter, we use the shorthand labels $\ket{j M}$ and $\ket{J M}$ for states $\ket{\beta L S j M}$ and $\ket{\beta L S J M}$, respectively. In this case, the RT coefficients can be obtained from those corresponding to an atomic system with HFS (see Appendices~\ref{sec::AppProp} and \ref{sec::AppRedis2TrmHFS}), but setting the nuclear spin $I = 0$ and making use of the relation shown in Eq.~(2.36a) of LL04 for the $\Sjs$ symbols. 
This also requires the substitution of labels $\mu$ with $j$, quantum numbers $f$ with $M$, and coupling coefficients $C_\mu^{J F}(f)$ with $C_j^{J}(M)$. The resulting expressions for the elements of the propagation matrix are found in Eqs.~\eqref{eq::EtaNoHFS} and \eqref{eq::RhoNoHFS}, and those for the frequency-dependent terms of the redistribution matrices are found in Eqs.~\eqref{eq::R2ObsnHFS} and \eqref{eq::R3ObsnHFS}. The corresponding quantity $\mathcal{A}$ is given by 
\begin{align} 
\mathcal{A}^{K K^{\prime}}_Q & \Bigl(j_u M_u,j_u^\prime M_u^\prime, j_\ell M_\ell,j_\ell^\prime M_\ell^\prime \Bigr) = 
 \frac{3 (2 L_u + 1)}{(2 S + 1)} \sqrt{(2 K + 1) (2 K^{\prime} + 1)} \notag \\
 & \times \sum_{\substack{J_u \bar{J}_u J_u^\prime \bar{J}_u^\prime}} C^{J_u}_{j_u}(M_u) \, C^{\bar{J}_u}_{j_u}(M_u) \, C^{J_u^\prime}_{j_u^\prime}(M_u^\prime) \, 
 C^{\bar{J}_u^\prime}_{j_u^\prime}(M_u^\prime) \sqrt{(2 J_u + 1)(2 J^\prime_u + 1)(2 \bar{J}_u + 1)(2 \bar{J}_u^\prime + 1)} \notag \\
 & \times \sum_{\substack{J_\ell \bar{J}_\ell J_\ell^\prime \bar{J}_\ell^\prime}} (-1)^{M_\ell - M_\ell^\prime}
 C^{J_\ell}_{j_\ell}(M_\ell) \,  C^{\bar{J}_\ell}_{j_\ell}(M_\ell) \, 
 C^{J_\ell^\prime}_{j_\ell^\prime}(M_\ell^\prime) \, 
 C^{\bar{J}_\ell^\prime}_{j_\ell^\prime}(M_\ell^\prime) \sqrt{(2 J_\ell + 1)(2 J^\prime_\ell + 1)(2 \bar{J}_\ell + 1)(2 \bar{J}_\ell^\prime + 1)} \notag \\
 & \times
 \left\{\begin{array}{c c c}
         J_u & J_\ell & 1 \\
         L_\ell & L_u & S
        \end{array} \right\}
 \left\{\begin{array}{c c c}
         J_u^\prime & \bar{J}_\ell & 1 \\
         L_\ell & L_u & S
        \end{array} \right\}
 \left\{\begin{array}{c c c}
         \bar{J}_u^\prime & \bar{J}_\ell^\prime & 1 \\
         L_\ell & L_u & S
        \end{array} \right\}        
  \left\{\begin{array}{c c c}
         \bar{J}_u & J_\ell^\prime & 1 \\
         L_\ell & L_u & S
        \end{array} \right\} \notag \\
& \times
  \sum_{p p^\prime p^{\prime \prime} p^{\prime \prime \prime} = -1}^1
  \left(\begin{array}{c c c }
         1  & 1  & K \\
         -p & p^\prime & Q
        \end{array}\right)
  \left(\begin{array}{c c c }
         1  & 1  & K^{\prime} \\
         -p^{\prime \prime \prime} & p^{\prime \prime} & Q
        \end{array}\right)      \notag \\
& \times 
\left(\begin{array}{c c c}
       J_u & J_\ell & 1 \\
       - M_u & M_\ell & p
      \end{array} \right)
\left(\begin{array}{c c c}
       J_u^\prime & \bar{J}_\ell & 1 \\
       -M_u^\prime & M_\ell & p^\prime       
      \end{array} \right)
\left(\begin{array}{c c c}
       \bar{J}_u^\prime & \bar{J}_\ell^\prime & 1 \\
       -M_u^\prime & M_\ell^\prime & p^{\prime \prime}       
      \end{array} \right)
\left(\begin{array}{c c c}
       \bar{J}_u & J_\ell^\prime & 1 \\
       -M_u & M_\ell^\prime & p^{\prime \prime \prime}        
      \end{array} \right) \, .   
\label{eqapp::AquantnoHFS}      
\end{align}
We note that these expressions for the redistribution matrix coincide with those that would be obtained from Eq.~(A.1) of  \cite{Bommier18}, making the abovementioned changes in notation and the approximations discussed in Sect.~\ref{sec::sEmis} for the reference PRD calculations. In the rest of this section, we present the expressions for the redistribution matrices under the approximations that are studied in the main text. 

\paragraph{\textit{Coherent scattering in the reference frame of the observer:}}
The first approximation considered in the main text is that of treating the scattering processes quantified by $\bm{\mathcal R}_{\mbox{\sc{ii}}}$ as coherent in the reference frame of the observer, or equivalently, neglecting Doppler redistribution in this matrix.  
Under this approximation we still account for the broadening effect of small-scale atomic motions in the absorption profiles. 
This can be achieved by neglecting the Doppler shifts in the Dirac delta function in Eq.~(A.1) of \cite{Bommier18}. Thus, when integrating over velocities with a Maxwellian distribution, the Lorentzian and associated dispersion profiles in the $R_{\mbox{\sc{ii}}}$ terms are replaced with Voigt and associated dispersion profiles, respectively, multiplied by a $1\!/\!\sqrt{\pi}\Delta\nu_D$ normalization factor. 
Making the abovementioned changes in notation, the frequency-dependent components become  
 \begin{align}
  \bigl[\hat{R}_{\mbox{\sc{ii}}}&\bigr]^{K K^{\prime}}_Q(\nu^\prime, \nu) = \sum_{\substack{j_u j_u^\prime \\ M_u M_u^\prime}} 
 \sum_{\substack{j_\ell j_\ell^\prime \\ M_\ell M_\ell^\prime}} 
  \mathcal{A}^{K K^{\prime}}_Q(j_u M_u,j_u^\prime M_u^\prime, j_\ell M_\ell,j_\ell^\prime M_\ell^\prime) 
 \,\, \frac{\Gamma_R}{\Gamma_R+\Gamma_I+\Gamma_E + 2 \pi \, \mathrm{i}\, {\Delta\nu_D \, u_{j_u M_u, j_u^\prime M_u^\prime}}} \notag \\[-8pt] %
  & \times \frac{1}{\sqrt{\pi} \Delta\nu_D^2} \,\delta\,\Bigl(u - u^\prime + u_{j_\ell M_\ell, j_\ell^\prime M_\ell^\prime} \Bigr) 
 \,\,\, \Biggl[\frac{1}{2} W\,\Bigl(a, u^\prime + u_{j_u^\prime M_u^\prime, j_\ell M_\ell}\Bigr) 
 + \frac{1}{2} W\,\Bigl(a, u^\prime + u_{j_u M_u, j_\ell M_\ell} \Bigr)^\ast \Biggr] \, .
 \label{eqapp::RedisNoHFSCSObserv}
\end{align}
Like in the main text (see Sect.~\ref{sec::sCSObs}), we write the components of this approximate redistribution matrix 
as $\hat{R}_{\mbox{\sc{ii}}}$. The additional $\Delta\nu_D$ in the denominator is obtained when changing the argument of the Dirac delta function from frequency to reduced frequency. It can easily be recognized that, under this approximation, the coupling between the incoming and scattered radiation is contained only in the Dirac delta. Thus, the radiation scattered at reduced frequency $u$ receives contributions only from the incident radiation at the few reduced frequencies $u^\prime =  u + u_{j_\ell M_\ell, j_\ell^\prime M_\ell^\prime}$ determined by the energy difference between the initial and final state. This represents a significant decrease in numerical complexity. 
Indeed, if all states of the lower term are degenerate, then these scattering processes introduce no coupling between the radiation field at different frequencies. We recall that this approximation is only applied to $\bm{\mathcal R}_{\mbox{\sc{ii}}}$, but it can be combined with the approximation of treating all scattering processes as coherent, discussed in Appendix.~\ref{sec::AppCRDCSApprox}. Indeed, this is done in Sect.~\ref{Sec::PurelyCS} of the main text. 

We finally observe that in the line wings, where this approximation is valid, 
\begin{equation*}
 H(a,u) \sim {\sqrt{\pi} \Delta\nu_D} \, \phi(\nu_0 - \nu) \, \quad  
  \mathrm{and} \quad  
 L(a,u) \sim {\sqrt{\pi} \Delta\nu_D} \, \psi(\nu_0 - \nu) \, , 
\end{equation*}
where $\phi$ and $\psi$ are the Lorentzian and associated dispersion profiles (see Sect.~5.3 of LL04). 
Therefore, at these spectral ranges, one should expect the results to be the same as if $\bm{\hat{{\mathcal R}}}_{\mbox{\sc{ii}}}$ were considered directly in the atomic rest frame without accounting for Doppler broadening. 
\paragraph{\textit{Expressions in the absence of a magnetic field:}} 
The expressions for the redistribution matrix in the unmagnetized case, accounting for the depolarizing effect of elastic collisions (see Appendix~\ref{sec::AppRedisZero}) can be further simplified for atomic systems without HFS. In analogy to what is discussed in Appendix~\ref{sec::AppRedisZero}, they can be written as  
\begin{subequations}
 \begin{align}
  \bigl[R_{\mbox{\sc{ii}}}&\bigr]^{K K^{\prime}}_Q(\nu^\prime, \nu) =
 \delta_{K K^{\prime}} 
 \sum_{J_u J_u^\prime} 
 \sum_{J_\ell J_\ell^\prime} 
 \mathcal{A}^{K K}_Q\bigl(J_u, J_u^\prime, J_\ell, J_\ell^\prime \bigr) \notag \\[-5pt]
 & \;\times \frac{1}{\pi \Delta\nu_D^2}\frac{\Gamma_R}{\Gamma_R + \Gamma_I + \Gamma_E + 2 \pi \, \mathrm{i}\, {\Delta\nu_D \, u_{J_u, J_u^\prime}}} 
  \frac{1}{2} \int_0^\pi\!\!\mathrm{d}\Theta\,
  \exp\Biggl[-\,\biggl(\frac{u - u^\prime + u_{J_\ell, J_\ell^\prime}}{2 \sin\Theta/2} \biggr)^2 \Biggr] \notag \\
  & \;\times \Biggl[ \frac{1}{2} W\,\biggl(\frac{a}{\cos\Theta/2},\frac{u + u^\prime + u_{J_u^\prime, J_\ell}
  + u_{J_u^\prime, J_\ell^\prime}}{2 \cos\Theta/2} \biggr)
  + \frac{1}{2} W\,\biggl(\frac{a}{\cos\Theta/2},\frac{u + u^\prime + u_{J_u, J_\ell} + 
  u_{J_u, J_\ell^\prime}}{2 \cos\Theta/2} \biggr)^\ast \Biggr] \, ,
   \label{eqapp::R2RedisnoHFSNoB} \\[10pt]
 \bigl[R_{\mbox{\sc{iii}}}&\bigr]^{K K^{\prime}}_Q(\nu^\prime, \nu) = 
 \delta_{K K^{\prime}} 
 \sum_{J_u J_u^\prime} 
 \sum_{J_\ell J_\ell^\prime} 
 \mathcal{A}^{K K}_Q\bigl(J_u, J_u^\prime, J_\ell, J_\ell^\prime\bigr) \notag \\[-5pt]
&  \times \frac{1}{\pi \Delta\nu_D^2} \Biggl[\frac{\Gamma_R}{\Gamma_R + \Gamma_I + \frac{1}{2}\bigl[D^{(K)}(J_u) + D^{(K)}(J_u^\prime) \bigr] 
  + 2 \pi \, \mathrm{i} \, { \Delta\nu_D \, u_{J_u, J_u^\prime}}} 
 - \frac{\Gamma_R}{\Gamma_R + \Gamma_I + \Gamma_E + 2 \pi \, \mathrm{i} \,
 {\Delta\nu_D \, u_{J_u, J_u^\prime}} } \Biggr] \notag \\[-5pt]
  & \times \Biggl[\frac{1}{2} W\,\Bigl(a, u^\prime + u_{J_u^\prime, J_\ell} \Bigr)
  + \frac{1}{2} W\,\Bigl(a, u^\prime + u_{J_u, J_\ell}\Bigr)^\ast \Biggr] 
  \, \,\Biggl[\frac{1}{2} W\,\Bigl(a, u + u_{J_u^\prime, J_\ell^\prime}\Bigr) 
  + \frac{1}{2} W\,\Bigl(a, u + u_{J_u, J_\ell^\prime}\Bigr)^\ast \Biggr] \, , \label{eqapp::R3RedisnoHFSNoB}
 \end{align}
\label{eqapp::RedisnoHFSNoB}
\end{subequations}
and the ${\mathcal A}$ quantity contained within them is given by
 \begin{align}
  \mathcal{A}^{K K}_Q \Bigl(J_u, J_u^\prime, J_\ell, & J_\ell^\prime \Bigr) = 
  \frac{3 (2 L_u + 1)}{(2 S + 1)} 
  \, (-1)^{J_\ell - J_\ell^\prime} 
  (2 J_u + 1) \, (2 J_u^\prime + 1) \, (2 J_\ell + 1) \, (2 J_\ell^\prime + 1) \notag \\
 &  \times 
 \left\{
  \begin{array}{c c c}
  1    &  1          & K \\
  J_u  &  J_u^\prime & J_\ell
 \end{array} \right\} 
 \left\{
  \begin{array}{c c c}
   1   &  1          & K \\
  J_u  &  J_u^\prime & J_\ell^\prime
 \end{array} \right\}
 \left\{
  \begin{array}{c c c}
  J_u    &  J_\ell &  1 \\
  L_\ell &  L_u    &  S  
 \end{array} \right\}  
    \left\{
  \begin{array}{c c c}
   J_u^\prime    &  J_\ell &  1 \\
   L_\ell &  L_u    &  S  
  \end{array} \right\} 
 \left\{
  \begin{array}{c c c}
   J_u^\prime &  J_\ell^\prime &  1 \\
   L_\ell     &  L_u    &  S  
  \end{array} \right\}  
 \left\{
  \begin{array}{c c c}
   J_u &  J_\ell^\prime &  1 \\
   L_\ell     &  L_u    &  S  
  \end{array} \right\} \, .
  \label{eqapp::AquanNoB}
 \end{align}
These expressions can also be obtained from the redistribution matrix presented in Eq.~(34) of \cite{Bommier17}, making the abovementioned changes in notation and approximations for the reference PRD case (see Sect.~\ref{sec::sEmis}). 
These redistribution matrices were used for the calculations presented in Sect.~\ref{Sec::MagProf}, in which the depolarizing effect of elastic collisions was fully neglected by setting all $D^{(K)}(J_u)$ rates to zero. 

\paragraph{\textit{Reproducing the redistribution matrix of Belluzzi \& Trujillo Bueno (2014):}} These authors proposed a redistribution 
matrix for the case of a two-term atomic model with an unpolarized and infinitely sharp lower term in the absence of a magnetic field. 
The branching ratios and $\bm{\mathcal R}_{\mbox{\sc{iii}}}$ were obtained from heuristic arguments.  
In order to recover this redistribution matrix using the present formalism, we consider the following frequency-dependent factors for $\bm{\mathcal R}_{\mbox{\sc{ii}}}$ and $\bm{\mathcal R}_{\mbox{\sc{iii}}}$ 
\begin{subequations}
 \begin{align}
 \bigl[R_{\mbox{\sc{ii}}}&\bigr]^{K K^{\prime}}_Q(\nu^\prime, \nu) =  \delta_{K K^{\prime}}  \sum_{J_u J_u^\prime} \sum_{J_\ell J_\ell^\prime} 
  \mathcal{A}^{K K}_Q\bigl(J_u, J_u^\prime, J_\ell, J_\ell^\prime\bigl) \notag \\[-5pt]
  &\times \frac{1}{\pi \Delta\nu_D^2} \frac{\Gamma_R + \Gamma_I}{\Gamma_R + \Gamma_I + \Gamma_E} \frac{\Gamma_R}{\Gamma_R+\Gamma_I 
  + 2 \pi \, \mathrm{i}\, {\Delta\nu_D \, u_{J_u,J_u^\prime}}} 
  \frac{1}{2} \int_0^\pi\!\!\mathrm{d}\Theta\,\mathrm{exp} 
  \Biggl[-\,\Biggl(\frac{u - u^\prime + u_{J_\ell, J_\ell^\prime}}{2 \sin\Theta/2} \Biggr)^2 \Biggr] \notag \\
 & \;\;\times \left[
 \frac{1}{2} W\,\Biggl(\frac{a}{\cos\Theta/2},\frac{u + u^\prime + u_{J_u^\prime, J_\ell}
 + u_{J_u^\prime, J_\ell^\prime}}{2 \cos\Theta/2} \Biggr)
 + \frac{1}{2} W\,\Biggl(\frac{a}{\cos\Theta/2},\frac{u + u^\prime + u_{J_u,J_\ell} + u_{J_u, J_\ell^\prime}}{2 \cos\Theta/2} \Biggr)^\ast \right] \, , 
 \label{R2ModBR}\\[10pt] 
  \bigl[R_{\mbox{\sc{iii}}}&\bigr]^{K K^{\prime}}_Q(\nu^\prime, \nu) =  \delta_{K K^{\prime}}  
  \sum_{J_u J_u^\prime} \sum_{J_\ell J_\ell^\prime} 
  \mathcal{A}^{K K}_Q(J_u, J_u^\prime, J_\ell, J_\ell^\prime) \notag \\[-5pt]
  & \times \frac{1}{\pi \Delta\nu_D^2} 
\frac{\Gamma_E}{\Gamma_R + \Gamma_I + \Gamma_E} \frac{\Gamma_R}{\Gamma_R + \Gamma_I +  2 \pi \, \mathrm{i} \, \Delta\nu_D \, u_{J_u, J_u^\prime}}
 \notag \\
  & \times \Biggl[\frac{1}{2} W\,\Bigl(a, u^\prime + u_{J_u^\prime, J_\ell}\Bigr)
  + \frac{1}{2} W\,\Bigl(a, u^\prime + u_{J_u, J_\ell}\Bigr)^\ast \Biggr] 
  \,\, \Biggl[\frac{1}{2} W\,\Bigl(a, u + u_{J_u^\prime, J_\ell^\prime}\Bigr) 
  + \frac{1}{2} W\,\Bigl(a, u + u_{J_u, J_\ell^\prime}\Bigr)^\ast \Biggr] \, , \label{R3ModBR}
 \end{align}
\label{eqapp::RedisModBR}
\end{subequations}
\noindent where the quantity ${\mathcal A}$ is exactly as given in Eq.~\eqref{eqapp::AquanNoB}. 
We observe that the only difference between these expressions and those given in Eq.~\eqref{eqapp::RedisnoHFSNoB} is found in their branching ratios. Neglecting depolarizing collisions in the latter, we note that the branching ratios of the two formulations coincide if (i) the CS limit is taken ($\Gamma_E = 0$), (ii) the CRD limit is taken ($\Gamma_E \gg \Gamma_R, \Gamma_I$), or (iii) the fine structure energy splitting of both the upper and lower terms (contained in $u_{J_u, J_\ell}$ and $u_{J_\ell, J_\ell^\prime}$) is negligible. 

The $\bm{\mathcal R}_{\mbox{\sc{ii}}}$ that results from Eq.~\eqref{R2ModBR} perfectly coincides with the analogous quantity in \cite{BelluzziTrujilloBueno14}, which is the first term on the right-hand side of Eq.~(36) in that paper. On the other hand, discrepancies remain between the $\bm{\mathcal R}_{\mbox{\sc{iii}}}$ resulting from Eq.~\eqref{R3ModBR} and the second term of Eq.~(36) in \cite{BelluzziTrujilloBueno14}. 
These discrepancies are contained in the absorption profiles of the two expressions (i.e., the profiles whose arguments depend on the incident radiation); in the latter work this profile was introduced heuristically. Nevertheless, the two expressions fully coincide if we make the assumption that the incident radiation field is spectrally flat, in which case the frequency shifts in their absorption profiles can be neglected.  
\end{document}